\documentclass[namedreferences]{solarphysics}
\usepackage[optionalrh]{spr-sola-addons} 
\usepackage{graphicx}                    
\usepackage{color}                       
\usepackage{url}                         
\usepackage{ifpdf}

\begin{document}

\begin{article}

\begin{opening}

\title{Geoeffectiveness of Coronal Mass Ejections in the SOHO era}

%
\author{M.~\surname{Dumbovi\'c}$^{1}$\sep
				A.~\surname{Devos}$^{2}$\sep
        B.~\surname{Vr\v snak}$^{1}$\sep
        D.~\surname{Sudar}$^{1}$\sep
        L.~\surname{Rodriguez}$^{2}$\sep
        D.~\surname{Ru\v zdjak}$^{1}$\sep             
        K.~\surname{Leer}$^{3}$\sep
        S.~\surname{Vennerstrom}$^{3}$\sep
        A.~\surname{Veronig}$^{4}$
       }


\runningauthor{Dumbovi\' c et al.}


\institute{$^{1}$ Hvar Observatory, Faculty of Geodesy, University of Zagreb, Zagreb, Croatia;
                     email: \url{mdumbovic@geof.hr}\\ 
           $^{2}$ Royal Observatory of Belgium, Brussels, Belgium\\
					 $^{3}$ Technical University of Denmark, Copenhagen, Denmark\\
					 $^{4}$ IGAM/Institute of Physics, University of Graz, Graz, Austria\\
           }


\begin{abstract}

The main objective of the study is to determine the probability distributions of the geomagnetic $Dst$ index as a function of the coronal mass ejection (CME) and solar flare parameters for the purpose of establishing a probabilistic forecast tool for the geomagnetic storm intensity.
Several CME and flare parameters as well as the effect of successive-CME occurrence in changing the probability for a certain range of $Dst$ index values, were examined.
The results confirm some of already known relationships between remotely-observed properties of solar eruptive events and geomagnetic storms, namely the importance of initial CME speed, apparent width, source position, and the associated solar flare class.
In this paper we quantify these relationships in a form to be used for space weather forecasting in future.
The results of the statistical study are employed to construct an empirical statistical model for predicting the probability of the geomagnetic storm intensity based on remote solar observations of CMEs and flares.
\end{abstract}

%
\end{opening}


\section{Introduction}
				\label{intro}

Interplanetary coronal mass ejections (ICMEs) are the drivers of the most intense geomagnetic storms (e.g., \opencite{gosling90}; \opencite{koskinen06}). They carry enhanced magnetic fields, usually at a speed higher than the background solar wind, both parameters being crucial for generating geomagnetic storms in a process that involves magnetic reconnection with the Earth's magnetosphere. A geomagnetic storm occurs if the topology of the magnetic field in the ICME is favorable for reconnection, \emph{i.e.}, if there is a strong southward component of the magnetic field (\opencite{dungey61}; \opencite{russell74}; \opencite{akasofu81}). The connection between \emph{in situ} properties of ICMEs and geomagnetic storms has been investigated in numerous studies considering different geomagnetic indices (see, \emph{e.g.}, \opencite{huttunen05}; \opencite{srivastava04}; \opencite{zhang07}; \opencite{richardson11b}; \opencite{yermolaev12}; \opencite{verbanac13}, and references therein).
Monitoring of the near-Earth solar wind parameters can give a quite reliable prediction of the potentially harmful events. However, the warnings precede the event only by an hour (for spacecraft located at the Lagrangian point L1), providing very limited "response time" \cite{richardson11b}. Therefore, it would be more useful to predict geoeffectiveness of ICMEs based on the remotely-observed CME/flare parameters.

Numerous studies dealing with the properties of geoeffective CMEs have been carried out including several attempts to construct geomagnetic storm prediction-models based on the remotely-measured properties of CMEs (\opencite{srivastava05}; \opencite{srivastava06}; \opencite{valach09}; \opencite{kim10}; \opencite{uwamahoro12}). The studies led to the conclusion that the geoeffectiveness of CMEs is related to the following solar properties of CMEs and the associated solar flares: CME initial speed (\opencite{srivastava04}; \opencite{gopalswamy07}), apparent angular width (\opencite{zhang03}; \opencite{srivastava04}; \opencite{zhang07}), source region location (\opencite{zhang03}; \opencite{srivastava04}; \opencite{gopalswamy07}; \opencite{zhang07}; \opencite{richardson10}), the intensity of the CME-related flare (\opencite{srivastava04}), occurrence of successive CMEs (\opencite{gopalswamy07}; \opencite{zhang07}). However, most of the above studies did not take account for the false and missing alarms, \emph{i.e.}, CMEs apparently having favorable solar properties, which did not produce geomagnetic storms, and the geomagnetic storms produced by CMEs with apparently non-favorable solar properties, respectively (see, \emph{e.g.}, \opencite{schwenn05}; \opencite{rodriguez09}), since they considered only storm-related CMEs.

The aim of this study is to quantitatively analyze CME/flare parameters and their relationship to the storm intensity \emph{viz.} the Disturbance Storm Time ($Dst$) geomagnetic index, derived from changes of the horizontal component of the geomagnetic field (see, \emph{e.g.}, \opencite{verbanac11} and references therein). To account for both, geomagnetic storms and false alarms, we include a large sample of the events on the Sun which can be associated with the geomagnetic activity at the Earth without considering the interplanetary component. It is based on the data related to CME take-off and therefore is suited for the near real-time advance forecasting and warning. Based on the statistical analysis, we develop a model to construct a probability distribution for geoeffectiveness of an observed CME. However, this model suffers from some limitations regarding the forecast and number of false alarms, as discussed in Section \ref{model} and will be analyzed further in future.


\section{Data and event selection}
				\label{data}

The CME data was taken from the SOHO LASCO CME Catalog (\opencite{yashiro04}, \url{http://cdaw.gsfc.nasa.gov/CME_list/}). The solar flare data was taken from the NOAA X-ray solar flare list (\url{ftp://ftp.ngdc.noaa.gov/STP/space-weather/solar-data/solar-features/solar-flares}).
\linebreak CMEs were associated with solar flares in the time period 10 January 1996 -- 30 June 2011 (hereafter "the SOHO era") using an automated method based on temporal and spatial criteria as described in \inlinecite{vrsnak05}. The temporal criterion is used to associate a CME with all the flares within the $\pm 1$ hour period of the CME liftoff time, where liftoff time is derived by back-extrapolation of the CME height-time plot (available on the CDAW website) to the solar surface assuming a linear speed. The spatial criterion associates a CME with all flares that were located within the opening angle of a CME, where the CME opening angle is a projection of the CME apparent width on the solar disc, centred around the central position angle of the CME obtained from LASCO-catalog. Therefore, the spatial criterion could not be used for halo CMEs (due to their apparent width of 360 degrees) and solar flares for which the location was not reported. Starting with a total of 16824 CMEs and 25907 flares in the SOHO era (reported by LASCO-catalog and NOAA Xray solar flare list, respectively) we first applied a temporal criterion to associate CMEs and flares. Then, the spatial criterion was used for the applicable events, resulting in a sample of 1392 CMEs and 1617 associated flares, meaning that some CMEs were associated with more than one flare. For those cases, the associated flare of the strongest intensity was chosen, resulting in 1392 CME-flare pairs. All but 38 pairs had a source position identified on the visible side of the Sun, meaning that they were front sided events. The remaining 38 CMEs for which the source position were not available, are halo CMEs therefore the association with flares was taken from the HALO CME SOHO LASCO catalog (\url{http://cdaw.gsfc.nasa.gov/CME_list/halo/halo.html}).

For the present analysis, we selected a subsample of the events, consisting of CMEs with speeds larger than 400 $\mathrm{km\,s^{-1}}$. From all the CMEs, we selected 211 events in order to equally cover the range of velocities (from $400$ $\mathrm{km\,s^{-1}}$ to the fastest CMEs, \emph{i.e.}, $v>1500$ $\mathrm{km\,s^{-1}}$). Equal sampling was used due to the fact that 78\% of CMEs in the sample of 1392 CME--flare pairs have speed less than 800 $\mathrm{km\,s^{-1}}$ (53\% of CMEs have speed less than 500 $\mathrm{km\,s^{-1}}$). Further, previous studies have shown that faster CMEs are more geoeffective (\emph{e.g.} \opencite{gopalswamy07}). Therefore, using a random sample would include only a small number of large geomagnetic storms in the sample, \emph{i.e.} most interesting events. For this purpose all fast CMEs ($v>1500$ $\mathrm{km\,s^{-1}}$) were taken, including a total of 53 events, whereas for CMEs with 400 $\mathrm{km\,s^{-1}}<v<1500$ $\mathrm{km\,s^{-1}}$ approximately 30 CMEs were randomly selected per bin of $\Delta v=200$ $\mathrm{km\,s^{-1}}$. It is to be noted that the cases when the slower CMEs are likely to be overtaken by faster ones were also taken into consideration, however, these CMEs launched in quick succession were not treated as individual events. CMEs with less than three height-time measurements were discarded, due to uncertainty of the speed estimate. This criterion was relaxed in the case of very fast CMEs ($v>1500$ $\mathrm{km\,s^{-1}}$), where only two height-time measurements are not unusual (see SOHO LASCO CME Catalog, \opencite{yashiro04}, \url{http://cdaw.gsfc.nasa.gov/CME_list/}).

Using plots available on the SOHO-LASCO catalog, which associate the CME height-time measurement and the $Dst$ index, we link the $Dst$ events with CME-flare pairs (see an example shown in Figure~\ref{fig1}). An extrapolation to the distance of 214 solar radii (approximately the distance from the Sun to Earth) was performed using CME "height-time" to derive a proxy time of arrival to the Earth. A $Dst$ event was then sought in a specific time window, chosen to account for possible errors in the SOHO LASCO catalog speed measurements, influence of the drag (acceleration or decceleration by solar wind; see, \emph{e.g.}, \opencite{cargill04}; \opencite{vrsnak04}; \opencite{vrsnak13}) and geometrical effects (ICME hitting with a flank; see, \emph{e.g.}, \opencite{mostl13}).
For CMEs in the speed range $v=400-600$ $\mathrm{km\,s^{-1}}$ the time window starts 24 hours before and ends 36 hours after the proxy of the arrival time. For CMEs with speed $v>600$ $\mathrm{km\,s^{-1}}$ the time window starts six hours before and ends 48 hours after the proxy of the arrival time. In this case a longer time beyond the time of estimated arrival was assumed because of the drag-decceleration effect and possible delayed impact of the flank, both of which depend on the speed of the CME (see \emph{e.g.} \opencite{vrsnak13} and references therein for the drag-decceleration effect and \opencite{mostl13} for the flank-delayed impact). 
Within the time window, the $Dst$ index was measured at the point where it reaches the minimum value ($Dst$ timing). If there was no geomagnetic storm within the time window corresponding to a specific CME, any recognizable variation in the $Dst$ index ($|Dst|\ge 10$ nT) closest to the proxy of arrival time (within the time window) was taken as the associated $Dst$ level. The $Dst$ timing in those cases is not a reliable parameter, therefore, the temporal aspect of geomagnetic storms (\emph{e.g.} duration) is not included in the analysis. If there was no variation in $Dst$ index throughout the time window which could be associated to a specific CME, the value of the $Dst$ index at the proxy of arrival time was taken as the associated $Dst$ level. It should be noted that there are no multiple associations between CME/flare pairs and $Dst$ index values.

\ifpdf
	bla
\else
  \begin{figure}    
   \centerline{\includegraphics[width=0.99\textwidth]{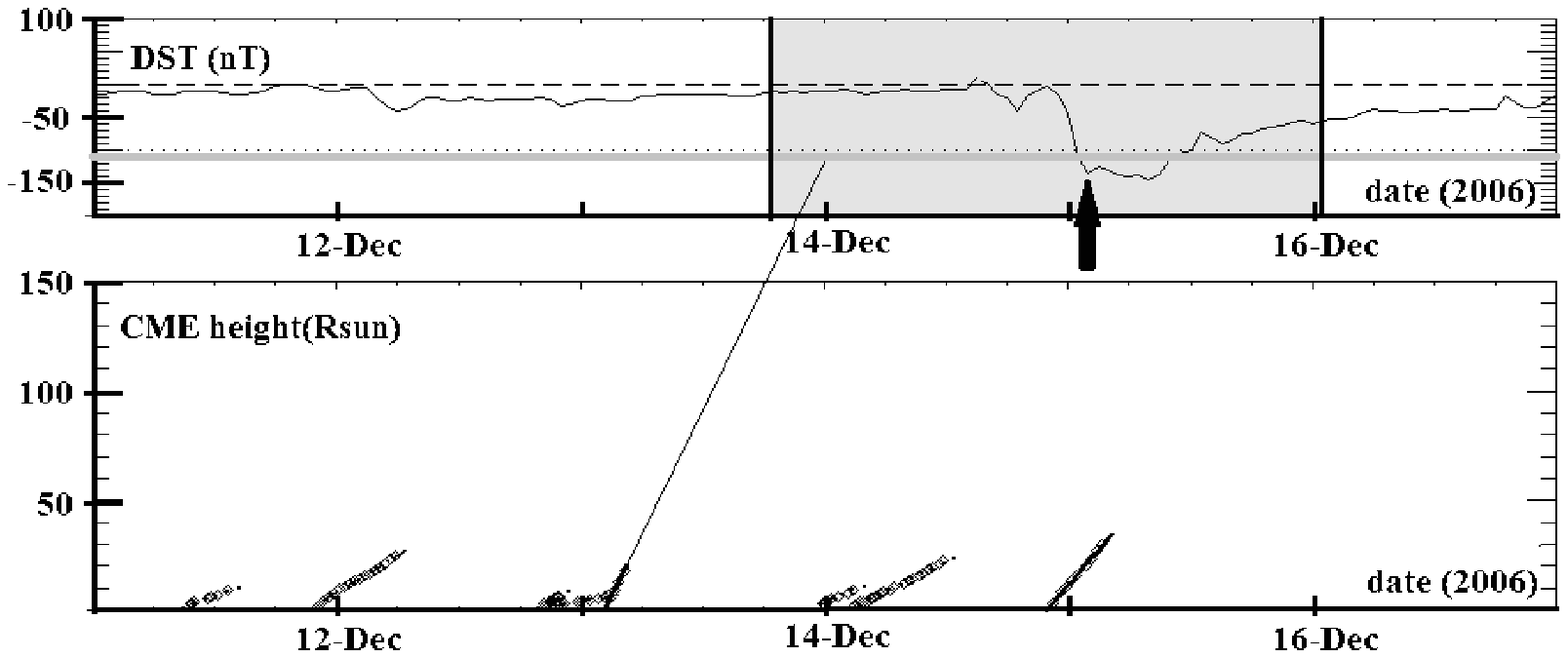}
              }
\caption{Association of a flare-related CME (first LASCO-C2 appearance, 13 December 2006, 02:54 UT) with a $Dst$ event at Earth. The height-time curve (black solid line) is extrapolated to 1 AU (gray solid line). The shaded area represents the time window in which a $Dst$ event was sought (six hours before and 48 hours after the proxy of arrival at Earth). Black arrow denotes the time at which the $Dst$ level is measured.
        }
   \label{fig1}
   \end{figure}
\fi


For each CME in the subsample of 211 events selected for study in the present paper a level of interaction with other CMEs was determined based on the following criteria:
\begin{itemize}
\item \emph{the kinematical criterion} -- interacting CMEs are associated with flares originating from the visible side of the Sun and their extrapolated kinematical curves cross or meet each other;
\item \emph{the timing criterion} -- the liftoff of interacting CMEs is within a reasonable time window ($\approx2$ days);
\item \emph{the source position/width criterion} -- interacting CMEs originating from the same or neighbouring source region, \emph{i.e.}, have close locations (unless halo and partial halo CMEs are involved, in which case this criterion was relaxed due to the fact that they have similar directions, \emph{i.e.} they are presumably Earth-directed).
\end{itemize}
We note that the listed criteria do not mean that CMEs necessarily interacted, they are used only to characterise the CMEs which are likely to interact. The kinematical criterion is based on the linear extrapolation of observed kinematical curves, without considering the drag effect. Furthermore, for simplicity we consider only flare-associated CMEs, for which the source location on the visible side of the Sun is identified. The timing criterion is introduced to prevent the unrealistically long chains of possibly interacting CMEs (\emph{e.g}, a "CME1" kinematically interacts with a "CME2" that was launched a day before, which interacted with a "CME3" that started a day prior to "CME2", \emph{etc.}). Finally, a source position/width criterion resolves cases where, \emph{e.g}, two narrow CMEs from opposite limbs satisfy both kinematical and timing criterion, although they are unlikely to interact due to their different propagation directions. These criteria in many cases do not clearly indicate a possible interaction therefore we introduce the "interaction parameter" by which we specify four levels of "interaction probability":
\begin{itemize}
\item "SINGLE" (S) events - no interaction;
\item "SINGLE?" (S?) - interaction not likely;
\item "TRAIN?" (T?) - probable interaction;
\item "TRAIN" (T) - interaction highly probable.
\end{itemize}
The determination of the interaction parameter is illustrated in Figure \ref{fig2}. The fastest CME (CME1, first appearance in LASCO-C2 15 June 2000, 07:54 UT) was a partial halo CME launched from N16W55; its proxy arrival time is marked with a black dot. It is preceded by three slower flare-related CMEs launched from source positions (chronologically backwards) N23W90 (CME2), N22W74 (CME3), and N21W69 (CME4), within a period of $\approx2$ days prior to the liftoff of the CME1. The extrapolated kinematical curve of CME1 crosses those of CME2 and CME4, but not of CME3. On the other hand, the extrapolated kinematical curves of CME3 and CME4 cross each other, whereas kinematical criterion for CME4 and CME2 is not met. Furthermore, CME2 is a narrow CME with source position at the limb, so we associate CME1 with an interaction level "T?" (interaction likely). The interaction parameter is assigned to each CME in the subsample of 211 events. We note that the whole CME train is then treated as one event that is characterized by solar parameters (\emph{e.g.}, speed, width, flare association, \emph{etc.}) of the fastest CME within a train.

\ifpdf
	bla
\else
  \begin{figure}    
   \centerline{\includegraphics[width=0.99\textwidth]{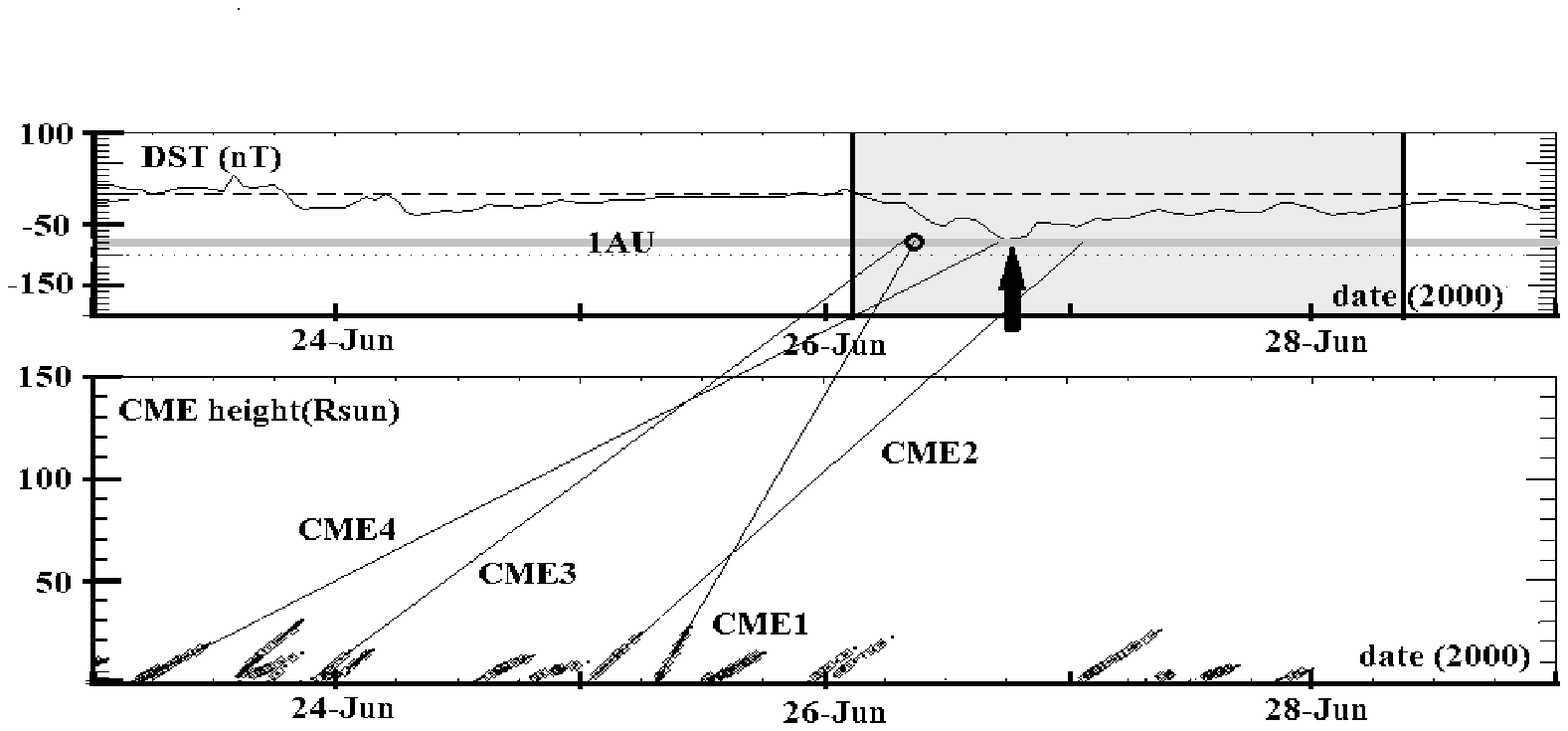}
              }
\caption{Association of a group of flare-related CMEs with a $Dst$ event at Earth. We associate the fastest of the CMEs (CME1) with an interaction parameter "T?" (interaction likely) due to possible interaction with CMEs 2-4 based on the criteria described in Section \ref{data} (for details see the main text). The $Dst$ level is estimated as in Figure \ref{fig1}.
        }
   \label{fig2}
   \end{figure}
\fi

For each event, \emph{in situ} signatures were associated with $Dst$ events. For this purpose we used the ICME list from \cite{richardson10} available at \url{http://www.srl.caltech.edu/ACE/ASC/DATA/level3/icmetable2.htm}, \emph{in situ} data from \emph{Advanced Composition Explorer} satellite (ACE; \opencite{stone98}) \emph{Magnetometer} (MAG; \opencite{smith98}) and \emph{Solar Wind Electron, Proton, and Alpha Monitor} (SWEPAM; \opencite{mccomas98}) instruments (\url{http://www.srl.caltech.edu/ACE/ASC/level2/lvl2DATA_MAG-SWEPAM.html}), and \emph{in situ} data from  \emph{Wind} satellite \emph{Magnetic Field Investigation} (MFI; \opencite{lepping95}) and \emph{Solar Wind Experiment} (SWE; \opencite{ogilvie95}) instruments (\url{http://wind.gsfc.nasa.gov/mfi_swe_plot.php}). In this way we also checked if some of the geomagnetic storms with $|Dst|>100$ nT were caused by a corotating interaction region (CIR) \cite{zhang03,richardson06}. We note that in the following analysis, CMEs associated with $|Dst|<100$ nT are considered as non-relevant events of low geoeffectiveness. They either missed the Earth or did not produce a major storm. We note that some of them are in fact associated with CIRs, but from the prediction point of view, it is only relevant that they did not produce a geomagnetic storm with $|Dst|>100$ nT, which is considered as the threshold for relevant strong geomagnetic activity.

In our sample of 211 CME--flare pairs the majority of the events were associated with ICMEs (57\%), whereas 41\% of events could not be associated with clear ICME signatures, \emph{i.e.}, they were either CIRs, complex ejecta, or there was no \emph{in situ} event at all. For 2\% of events \emph{in situ} data were not available due to measurement gaps. Out of 41\% of events that were not associated with clear ICME signatures, only one had $|Dst|>100$ nT, however we did not discard it because it does not have clear CIR signatures as well.


\section{Statistical analysis method}
				\label{method}
				
The selected sample of 211 CMEs/flares and associated $Dst$ index provides numerous CME and flare parameters, as well as corresponding $Dst$ value. Following the results of previous studies (\emph{e.g.}, \opencite{zhang03}; \opencite{srivastava04}; \opencite{gopalswamy07}; \opencite{zhang07}; \opencite{richardson10}; \opencite{richardson11b}) we focus on specific parameters \emph{viz.} the initial CME speeds and angular width, as well as solar flare soft X-ray class and location. In addition a level of interaction is also defined as a parameter since there are studies that indicate that interaction of CMEs can enhance their geoeffectivness (\emph{e.g.}, \opencite{farrugia04}) and that most intensive storms are associated with trains of successive/multiple CMEs \cite{gopalswamy07,zhang07,mostl12,mishra14}. Distributions are used as a statistical tool for the analysis with the  following bins: $|Dst|<100$\,nT, $100$\,$\mathrm{nT}<|Dst|<200$\,$\mathrm{nT}$, $200$\,$\mathrm{nT}<|Dst|<300$\,$\mathrm{nT}$, and $|Dst|>300$\,nT, where $|Dst|$ represents the magnitude of the $Dst$-index variation.

In order to check how $|Dst|$ distributions change for a specific key parameter, these key parameters were binned as well. For some key parameters the binning was obvious (\emph{e.g.}, interaction parameter) as they are already discrete parameters. For continuous parameters all the bins have approximately the same number of events. The distribution mean, skewness, and kurtosis were calculated as relevant distribution parameters that depict the behavior of the $|Dst|$ distribution with the change in the (discrete) CME/flare parameter. The distribution skewness and kurtosis are coefficients derived from 3rd and 4th order moment of the distribution and represent asymmetry and peakedness/flatness coefficients, respectively.

The statistical significance of results was tested using two-sample t-test (2stt) at the 0.05 level (95\% significance) for assuming dependence (equal variance assumed) and independence (equal variance not assumed) of the test samples. Due to the fact that 2stt is based on the normality assumption, \emph{i.e.}, requires certain sample sizes, nonparametric significance tests were also performed, namely Kolmogorov-Smirnov and Mann-Whitney U-test, but there were no notable differences. In addition, no significant differences were noticed between the dependence and independence assumptions. Therefore we present only 2stt results for unequal variance (\emph{i.e.} assumption of independence).

\ifpdf
	bla
\else
  \begin{figure}    
   \centerline{\includegraphics[width=0.5\textwidth]{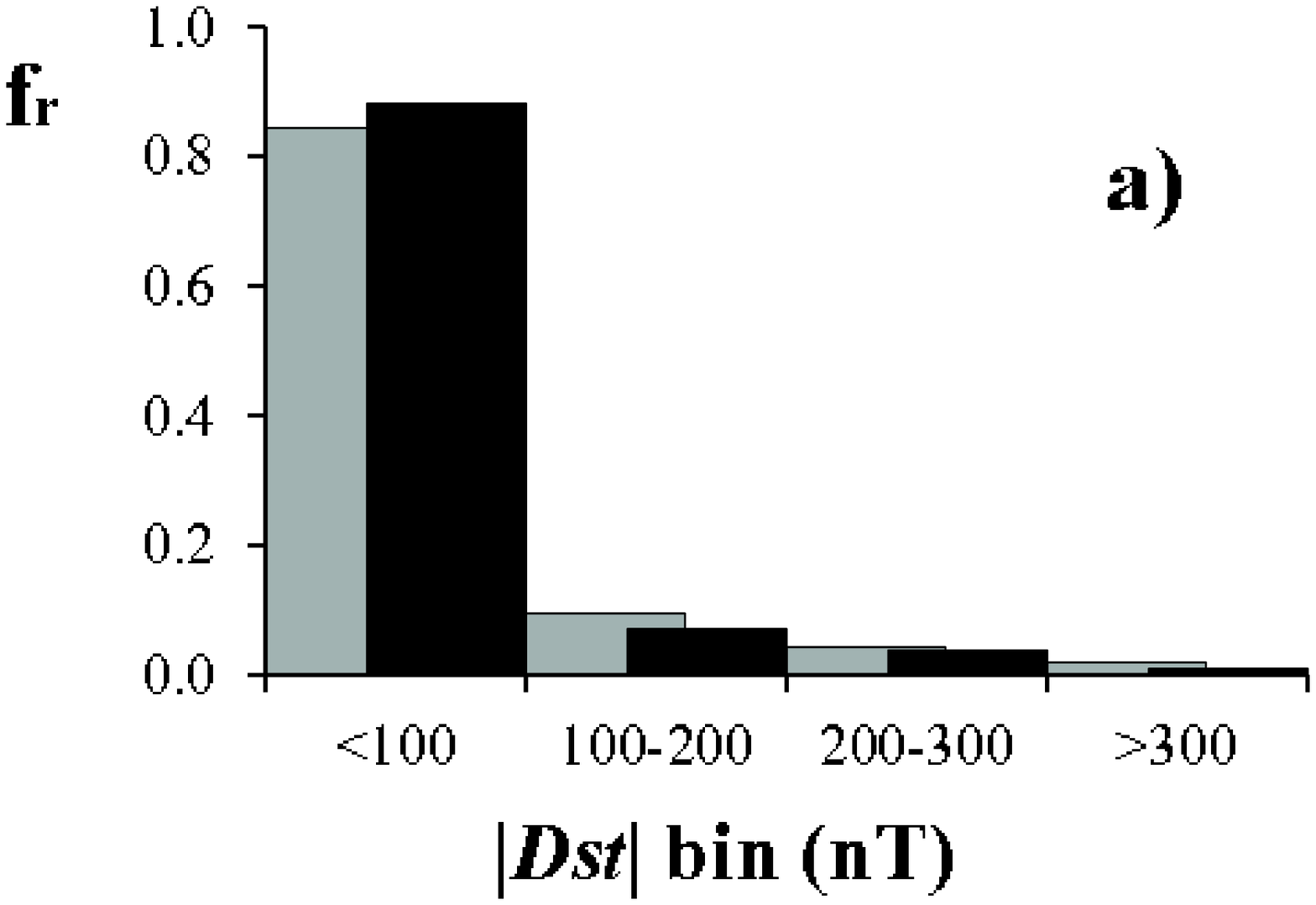}
               \includegraphics[width=0.5\textwidth]{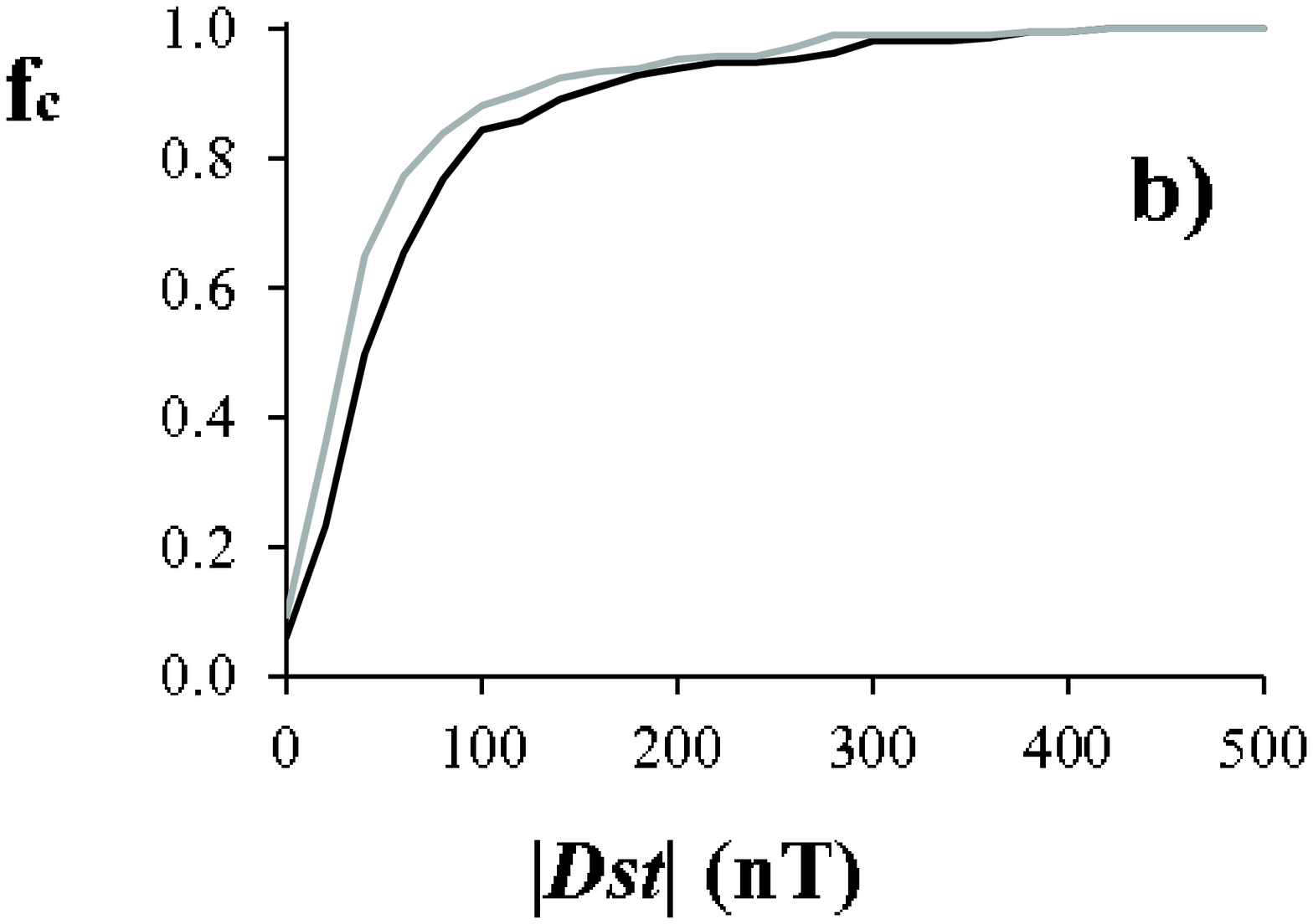}
               }
\caption{Distribution (a) and cumulative distribution (b) of the $Dst$-index variation (gray -- total $|Dst|$; black -- relative $|Dst|$).
        }
   \label{fig3}
   \end{figure}
\fi

First, a general distribution of all measured values of the $Dst$ index was performed, for two different types of measurements. Namely, in the $Dst$\,--\,time plot a geomagnetic storm is seen as a decrease in the $Dst$ index, where the intensity of the storm is given by the magnitude of this depletion. The magnitude of this decrease was measured in two ways: the total magnitude (``total $|Dst|$''), measured from reference value 0, and relative magnitude (``relative'' $|Dst|$), measured from the reference value at the start of the storm, \emph{i.e.}, the amplitude of the $Dst$-index variation. We examine how the measurement process can affect the results, \emph{i.e.}, is there a difference in considering the total or relative magnitudes. The statistical analysis shows that the distributions of total $|Dst|$ and relative $|Dst|$ are somewhat different, with relative $|Dst|$ shifted to lower values (Figure \ref{fig3}). The mean values are 68 nT and 53 nT, respectively, and are found to be significantly different at the 0.05 significance level with a 2stt. Total $|Dst|$ is usually larger than relative $|Dst|$, mostly due to the recovery of the preceding geomagnetic storms. While the total $|Dst|$ includes effects of the preceding storm, they are excluded in relative $|Dst|$, since it is measured from the onset point. Therefore, we focus our study on relative $|Dst|$, as a more realistic measure of storm strength. It should be noted though that the same analysis was repeated for total $|Dst|$ as well and similar results were obtained, therefore we do not present them here (the distributions were systematically shifted towards somewhat larger amplitudes, however, with the similar overall behavior).

In general, the $|Dst|$ distribution is highly asymmetric with over 80\% of non geo-effective events ($|Dst|>100$ nT for only 20 events). On the other hand, there are 110 HALO CMEs (52 \%) and 140 CMEs with $v>$800 km s$^{-1}$ (66 \%), i.e. our sample contains a large number of false alarms. This is to be expected, because the sample was chosen based on the CME observations, where only a subset of CMEs caused geomagnetic storms. Therefore, although false alarms were not studied directly, their influence was taken into account (since they constitute a substantial part of the sample).


\section{Results of the statistical analysis}
				\label{results}

In this section we analyze and discuss the relationship between CME/flare properties derived from remote solar observations and the $|Dst|$ levels at the Earth. In particular, we investigate the relationship between $|Dst|$ levels and the following solar parameters:
\begin{itemize}
\item \emph{CME initial speed} (Section \ref{speed}; Figures \ref{fig4} and \ref{fig5}; Table \ref{table1});
\item \emph{CME/flare source position} (Section \ref{source}; Figures \ref{fig6} and \ref{fig9}a; Table \ref{table2}a);
\item \emph{CME--CME interaction parameter} (Section \ref{interaction}; Figures \ref{fig7} and \ref{fig9}b; Table \ref{table2}b);
\item \emph{CME angular width} (Section \ref{width}; Figures \ref{fig8}a and \ref{fig9}c; Table \ref{table3}a);
\item \emph{solar flare class} (Section \ref{flare}; Figures \ref{fig8}b and \ref{fig9}d; Table \ref{table3}b).
\end{itemize}
In Section \ref{combinations} we investigate the influence of combined solar parameters described in Sections \ref{speed}\,--\,\ref{flare} on the storm intensity ($|Dst|$ levels), whereas in Section \ref{activity} we investigate the overall behavior of the $Dst$ index and compare it with the CME/flare activity to validate the sample and the statistical analysis.


\subsection{CME initial speeds}
				\label{speed}
				
\ifpdf
	bla
\else
\begin{figure}
\centerline{\includegraphics[width=0.5\textwidth]{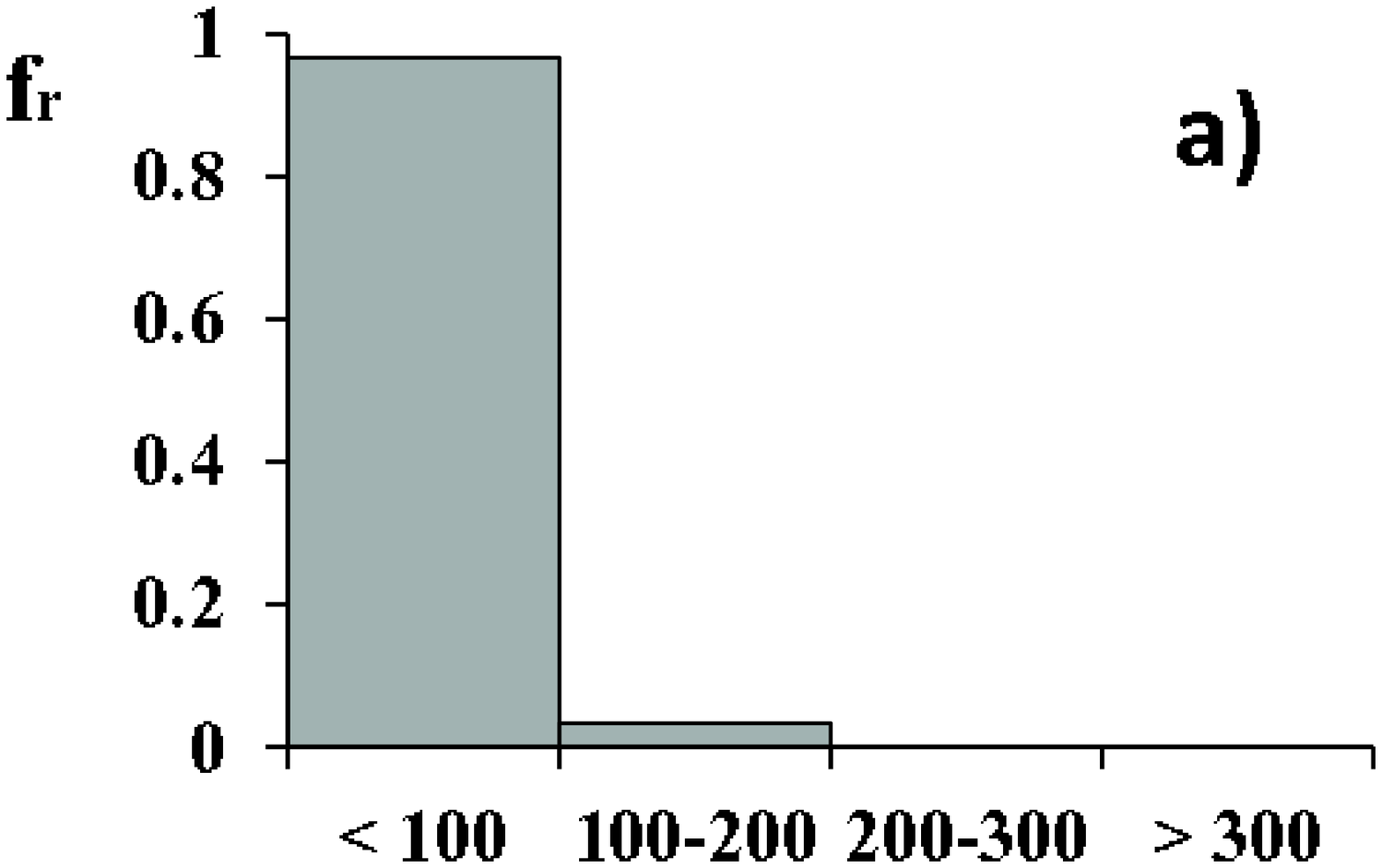}
            \hspace*{0.03\textwidth}
            \includegraphics[width=0.5\textwidth]{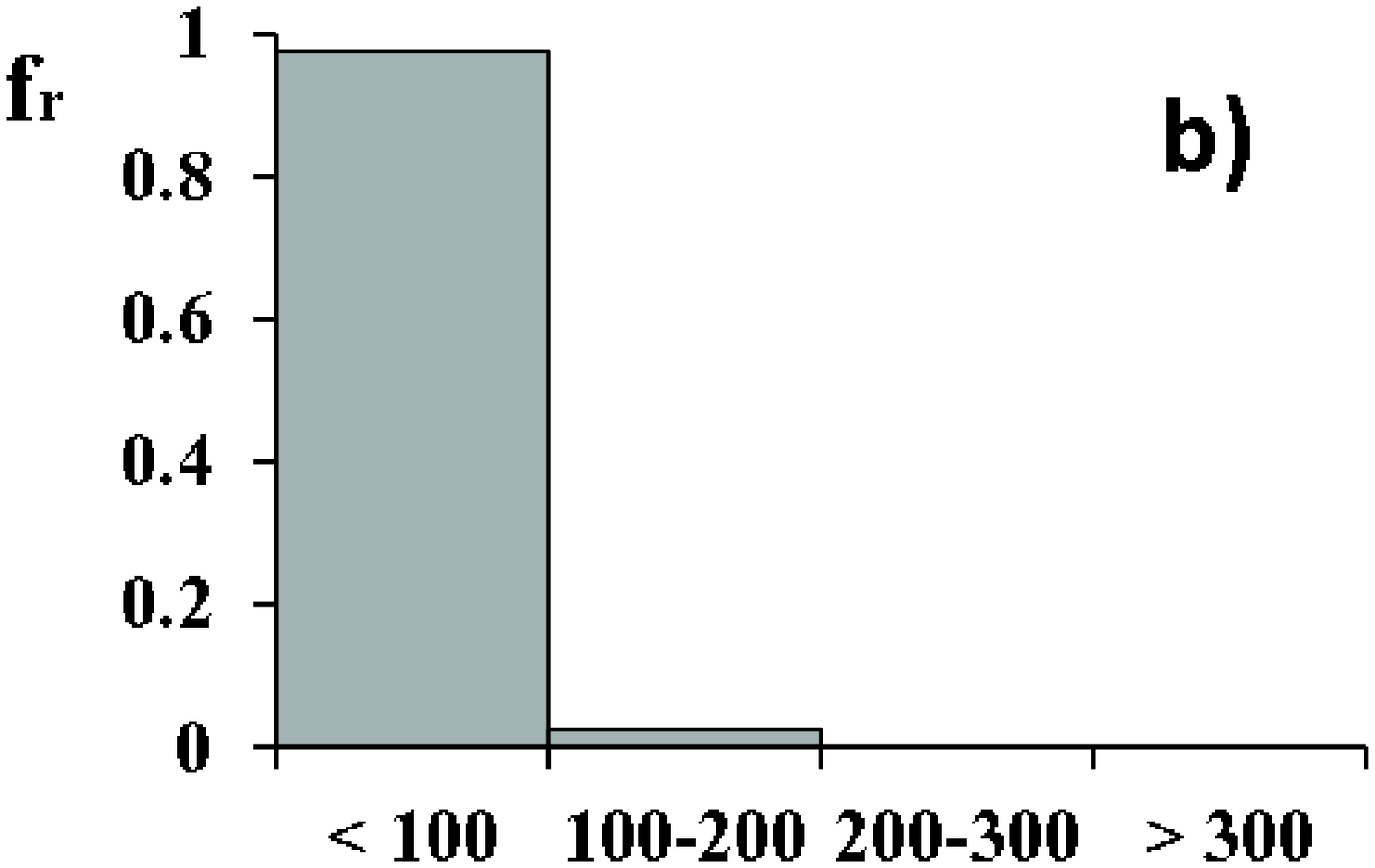}
              }
\vspace{0.03\textwidth}
\centerline{\includegraphics[width=0.5\textwidth]{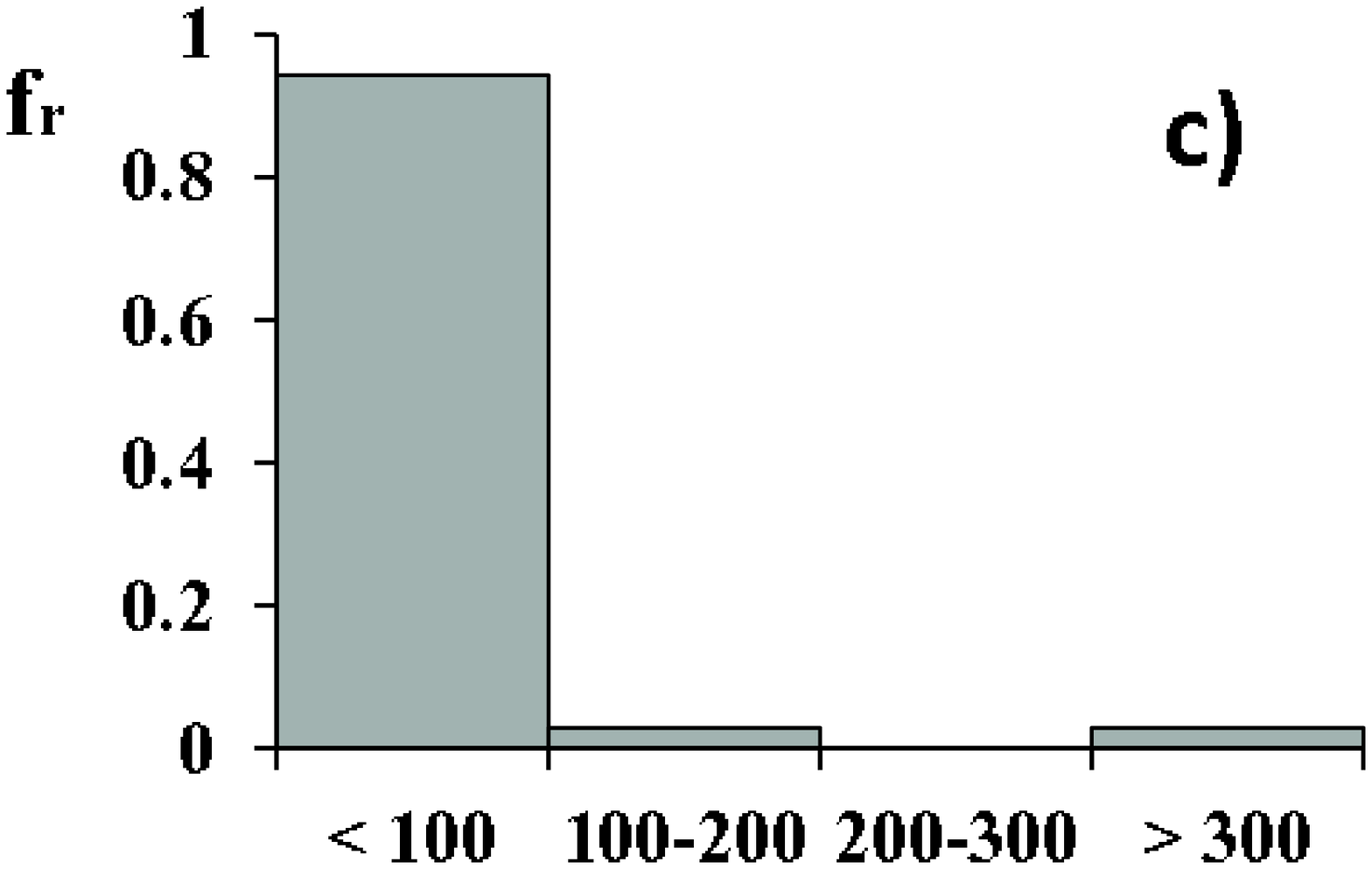}
            \hspace*{0.03\textwidth}
            \includegraphics[width=0.5\textwidth]{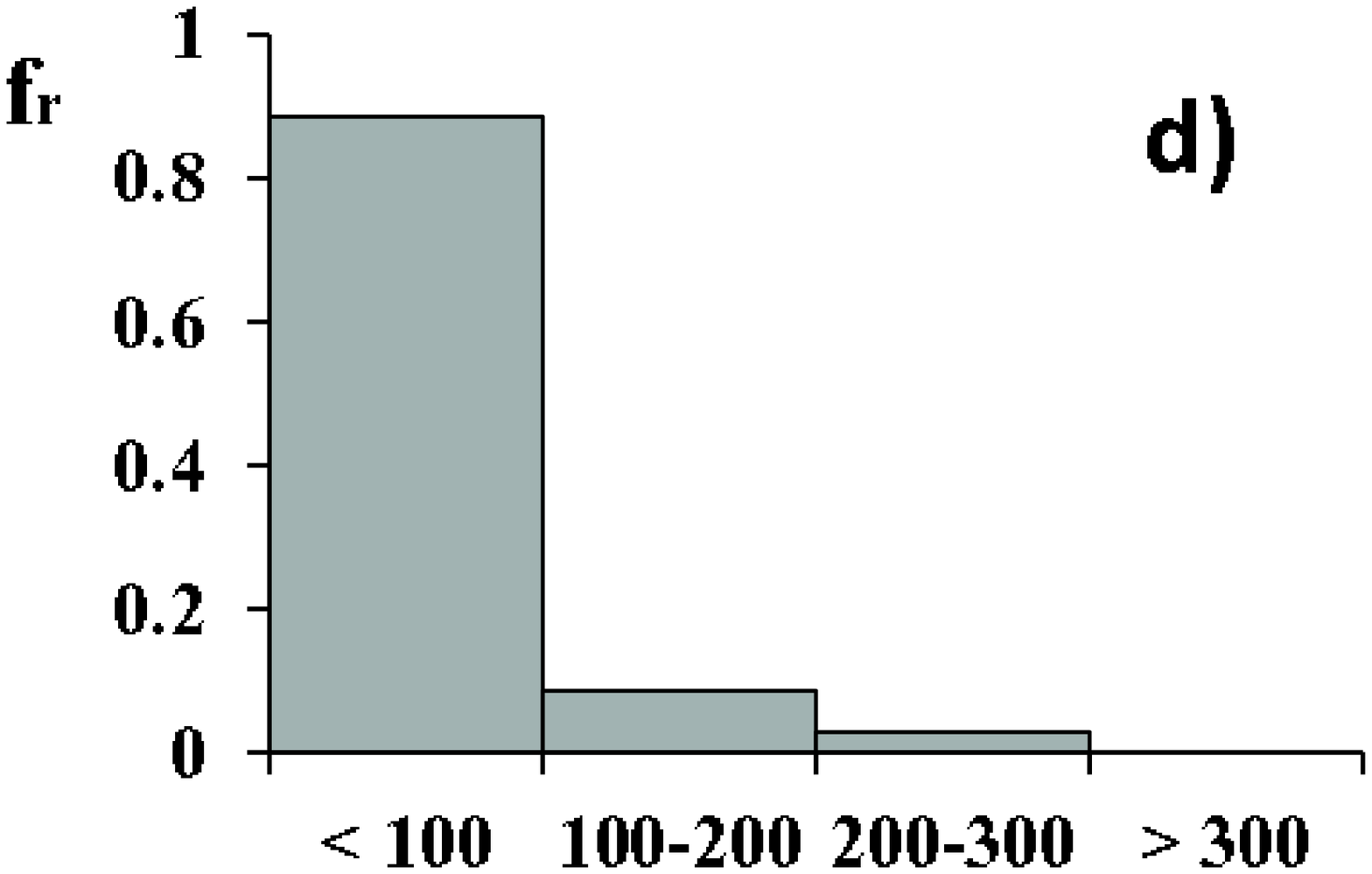}
              }
\vspace{0.03\textwidth}
\centerline{\includegraphics[width=0.5\textwidth]{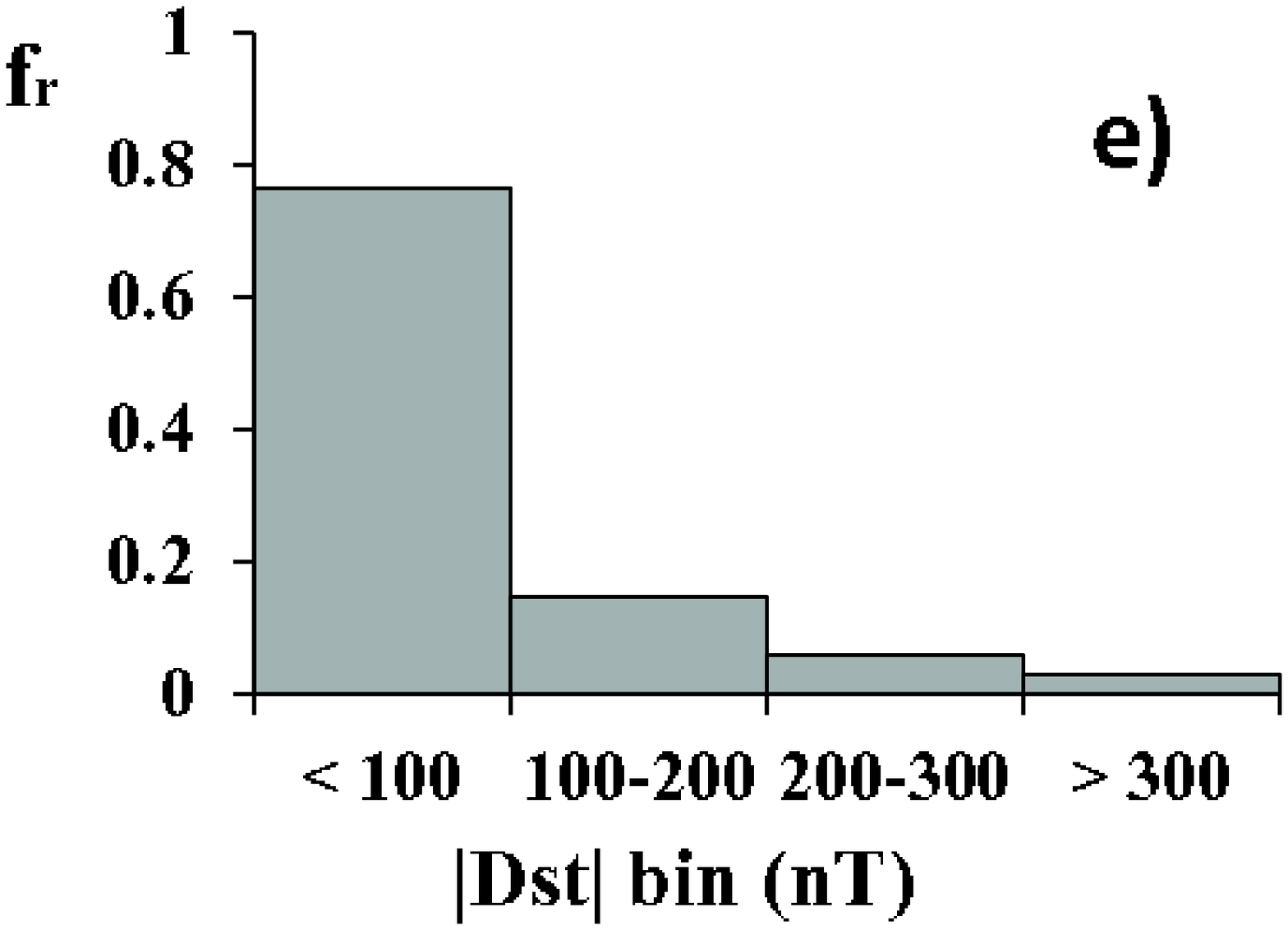}
            \hspace*{0.03\textwidth}
            \includegraphics[width=0.5\textwidth]{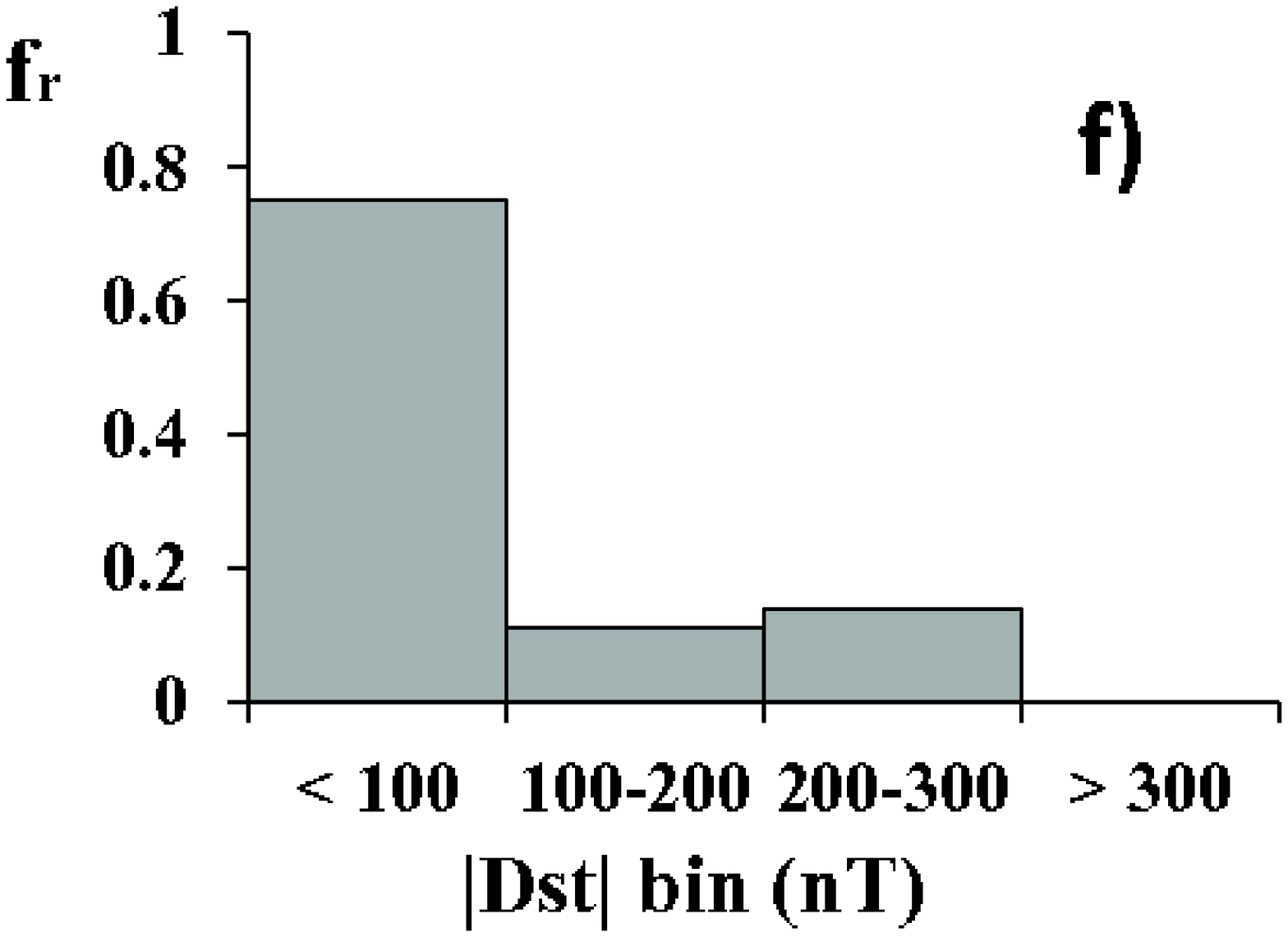}
              }

\caption{$|Dst|$ relative frequencies for different bins of CME speed, $v$: a) 400\,--\,600 $\mathrm{km\,s^{-1}}$; b) 600\,--\,800 $\mathrm{km\,s^{-1}}$; c) 800\,--\,1000 $\mathrm{km\,s^{-1}}$; d) 1000\,--\,1200 $\mathrm{km\,s^{-1}}$; e) 1200\,--\,1700 $\mathrm{km\,s^{-1}}$; f)$>$1700 $\mathrm{km\,s^{-1}}$.
        }
   \label{fig4}
   \end{figure}
\fi

The first CME parameter analyzed is 1st order (linear) CME speed, $v$, derived from LASCO C2 and C3 images. Although this is plane-of-sky speed and subject to projection effects, it can be related to the radial speed of the CME (\emph{e.g.}, \opencite{schwenn05}; \opencite{gopalswamy12}) and can therefore be taken as its proxy. The benefit is that this parameter can be relatively easily and quickly derived using L1 coronagraphic images. The events in our data sets were categorized into six different CME speed bins, with the following CME speed ranges: 400\,--\,600 $\mathrm{km\,s^{-1}}$ (bin 1), 600\,--\,800 $\mathrm{km\,s^{-1}}$ (bin 2), 800\,--\,1000 $\mathrm{km\,s^{-1}}$ (bin 3), 1000\,--\,1200 $\mathrm{km\,s^{-1}}$ (bin 4), 1200\,--\,1700 $\mathrm{km\,s^{-1}}$ (bin 5), and $v > 1700$ $\mathrm{km\,s^{-1}}$ (bin 6). The number of events in each bin is 36, 34, 35, 35, 41, and 30, respectively. In Figure \ref{fig4} the $|Dst|$ distributions are presented for each CME speed bin, using the previously defined bin sizes. The mean value, skewness and kurtosis of the distributions were calculated to quantitatively examine changes in the $|Dst|$ distribution for different CME speed ranges (Figures \ref{fig5}a, \ref{fig5}c, and \ref{fig5}d, respectively). Furthermore, to test the differences between $|Dst|$ distributions (\emph{i.e.}, whether they represent statistically different samples) a two sample t-test between each pair of $|Dst|$ distributions was applied. The results are presented in Table \ref{table1}.


\begin{table}
\caption{The two sample t-test significance levels of the difference between $|Dst|$ mean values in different bins of CME speed, $v$, with unequal variance assumed. Unless marked with an asterisk, the value states that the mean of the two samples are not significantly different; ** denotes that the significance of the difference is $>95$\%; * denotes that the significance of the difference is $>90$\%.}
\label{table1}

\begin{tabular*}{0.99\textwidth}{ccccccc}
\hline
\multicolumn{7}{c}{$v$ bins}\\
\hline
 & & bin1\tabnote{bins 1--6 represent different speed ranges in km s$^{-1}$: 400-600 (bin1), 600-800 (bin2), 800-1000 (bin3), 1000-1200 (bin4), 1200-1700 (bin5), and $>$1700 (bin6)} & bin2 & bin3 & bin4 & bin5 \\
\hline
bin6   & & $2\cdot10^{-4}$** & $9\cdot10^{-4}$** & 0.02**       & 0.01**        & 0.44 \\
bin5   & & 0.02**                     & 0.05**   & 0.22         & 0.10*         & |    \\
bin4   & & 0.30                       & 0.73     & 0.66         & |             & |    \\
bin3   & & 0.16                       & 0.42     & |            & |             & |    \\
bin2   & & 0.33                       & |        & |            & |             & |    \\
\hline
\end{tabular*}
\end{table}

It can be seen in Figures \ref{fig4}a and \ref{fig4}b that for $v<800$ $\mathrm{km\,s^{-1}}$ the distribution is restricted to $|Dst|<200$ nT and is mostly contained within $|Dst|<100$ nT. When compared with distributions in Figures \ref{fig4}c\,--\,\ref{fig4}f, these distributions lack the tail.
This is also seen in the behavior of the distribution parameters (black dots in Figures \ref{fig5}a\,--\,d): the value of the mean, skewness and kurtosis first increase with increasing CME speed, up to the 800\,--\,1000 $\mathrm{km\,s^{-1}}$ bin. Then, as the speed range increases from the intermediate range speed bins (800\,--\,1000 $\mathrm{km\,s^{-1}}$) to the higher range speed bins ($>1700$ $\mathrm{km\,s^{-1}}$) the distribution loses peakedness, as the values of skewness and kurtosis decrease, whereas the value of the distribution mean still increases.
Changing the range of the CME speed leads to a change in the corresponding $|Dst|$ distribution in a way that from a distribution the events are grouped in first two $|Dst|$ bins, the distribution first obtains the tail (mean, skewness and kurtosis increase), and then starts filling the tail (mean increases, whereas skewness and kurtosis decrease). This shift of distribution towards larger $|Dst|$ bins is also evident from the behavior of the distribution mean (Figure \ref{fig5}a), which can be approximated with a linear function (correlation coefficient, cc=0.96), although due to small number of data points this correlation should be taken with caution.

In addition, alternative speed bins were made, to substantiate our results. The alternative CME speed bins cover ranges: 400\,--\,700 $\mathrm{km\,s^{-1}}$ (52 events), 700\,--\,1000 $\mathrm{km\,s^{-1}}$ (54 events), 1000\,--\,1500 $\mathrm{km\,s^{-1}}$ (52 events), and $v>1500$ $\mathrm{km\,s^{-1}}$ (53 events). The distribution parameters for these alternative distributions are shown as gray dots in Figures \ref{fig5}b\,--\,d. Including the distribution parameters of this alternative speed binning does not change the result notably, on the contrary, they follow the same trend.

Consequently, this means that faster CMEs have higher probabilities of producing strong geomagnetic storms, and furthermore, that slow CMEs ($v<600$ $\mathrm{km\,s^{-1}}$) are not likely to produce intense storms ($|Dst|>200$ nT) unless they are involved in a CME--CME interaction with a faster CME. The latter comes from the fact that interacting CMEs in the sample are related to the CME speed of the fastest CME in the train. It should be noted though that one parameter alone (e.g. CME initial speed) does not determine the geoeffectiveness of CMEs (as will be demonstrated in following sections). Therefore, distributions shown in Figure \ref{fig4} are not a suitable measure of CME geoeffectiveness probability. The two sample t-test analysis reveals that there is no significant difference between two neighbouring speed bins (or several, as we go to lower speed bins) indicating that the $v$\,--\,$|Dst|$ relationship should be the most significant for very fast CMEs, whereas for slow CMEs it is unclear.

\ifpdf
	bla
\else
\begin{figure}
\centerline{\includegraphics[width=0.5\textwidth]{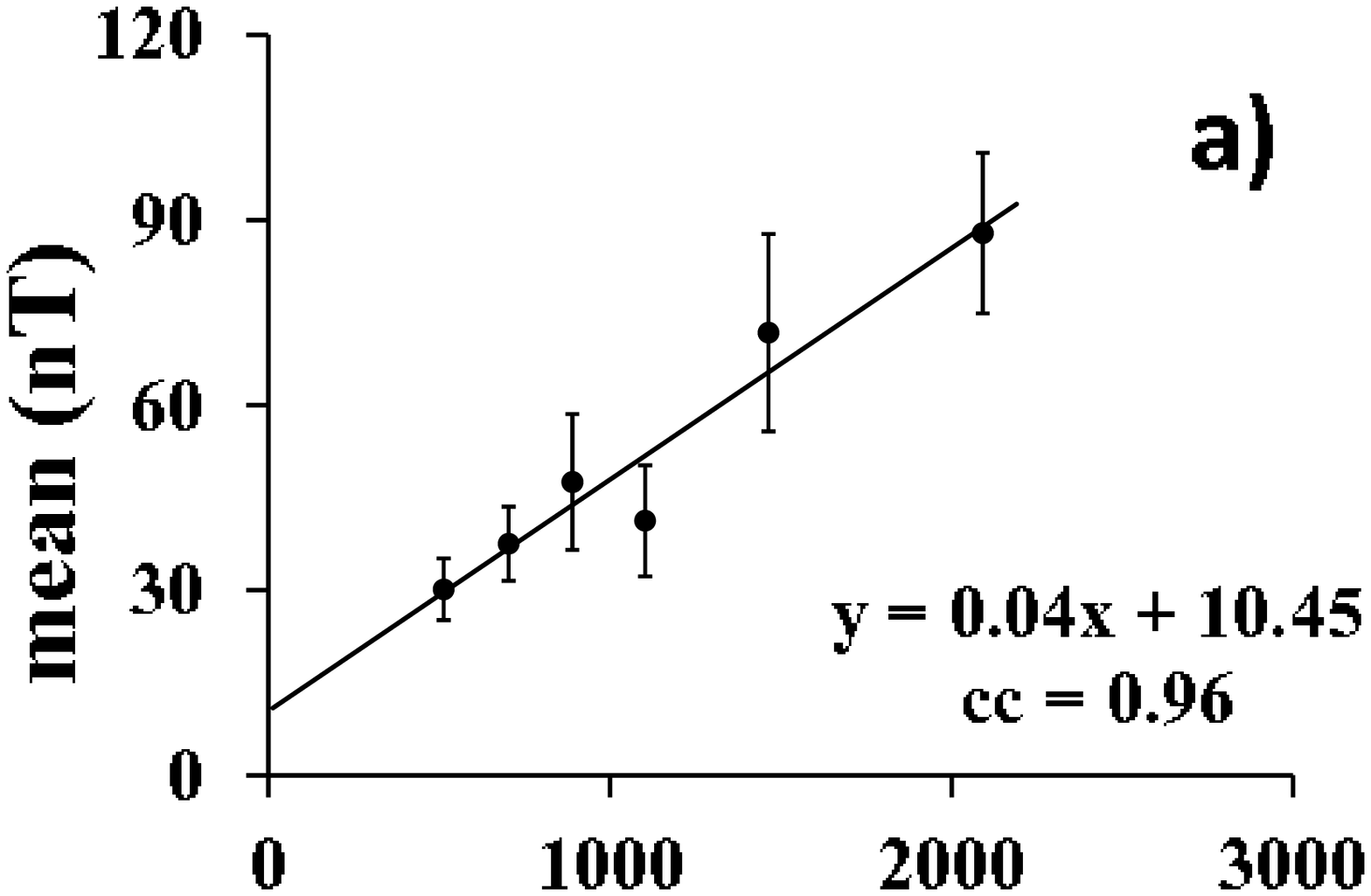}
            \hspace*{0.03\textwidth}
            \includegraphics[width=0.5\textwidth]{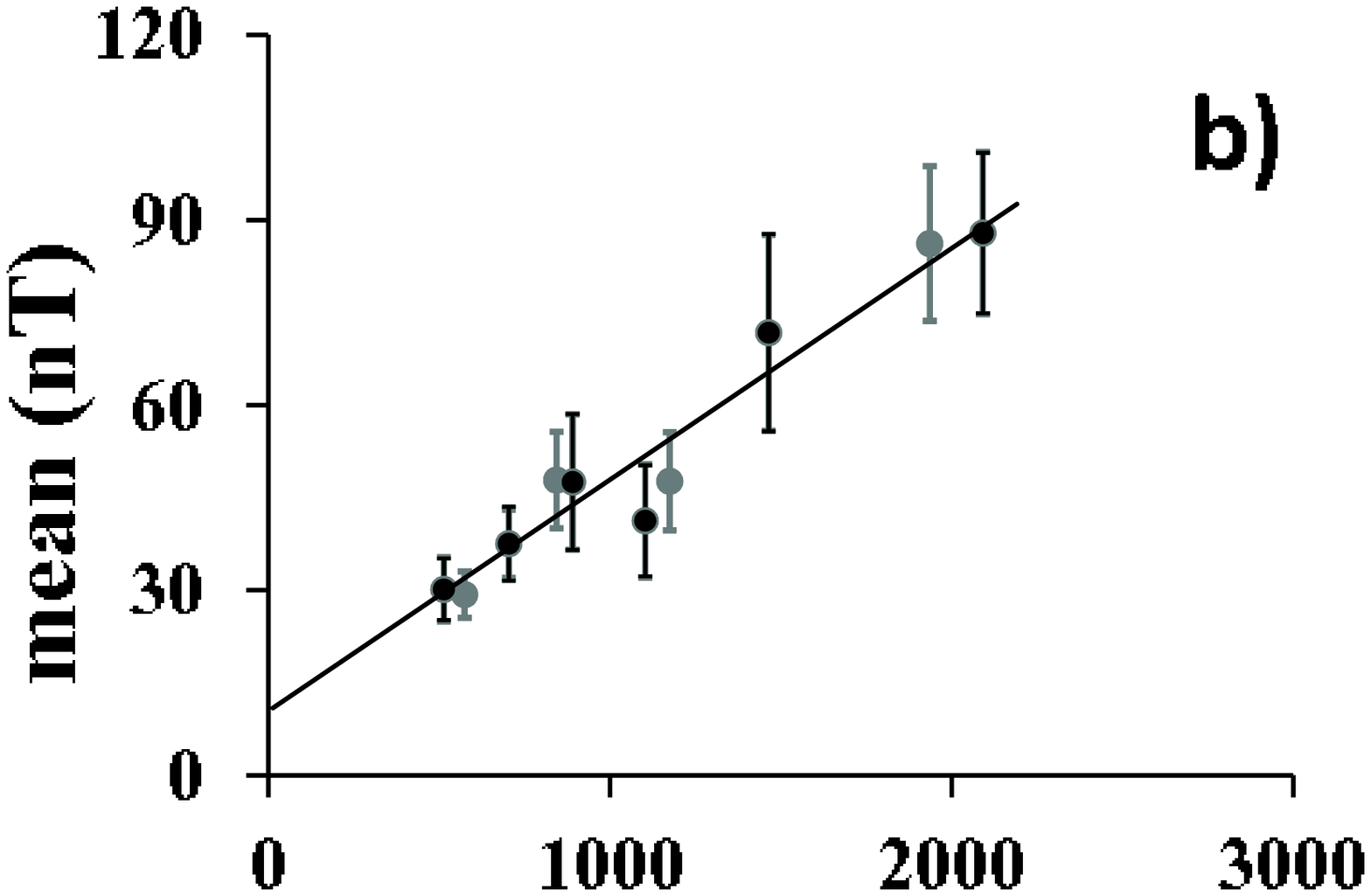}
              }
\vspace{0.03\textwidth}
\centerline{\includegraphics[width=0.5\textwidth]{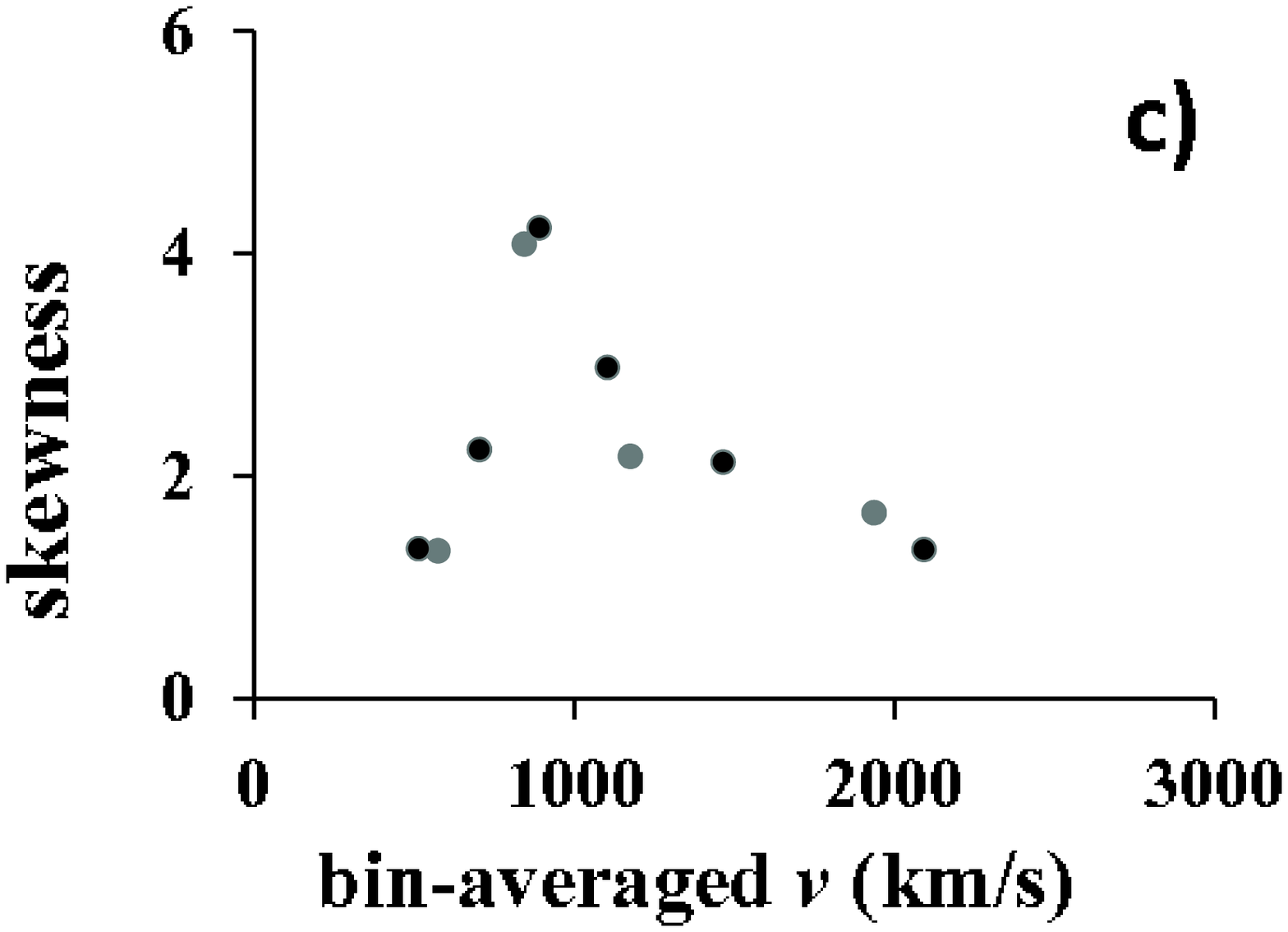}
            \hspace*{0.03\textwidth}
            \includegraphics[width=0.5\textwidth]{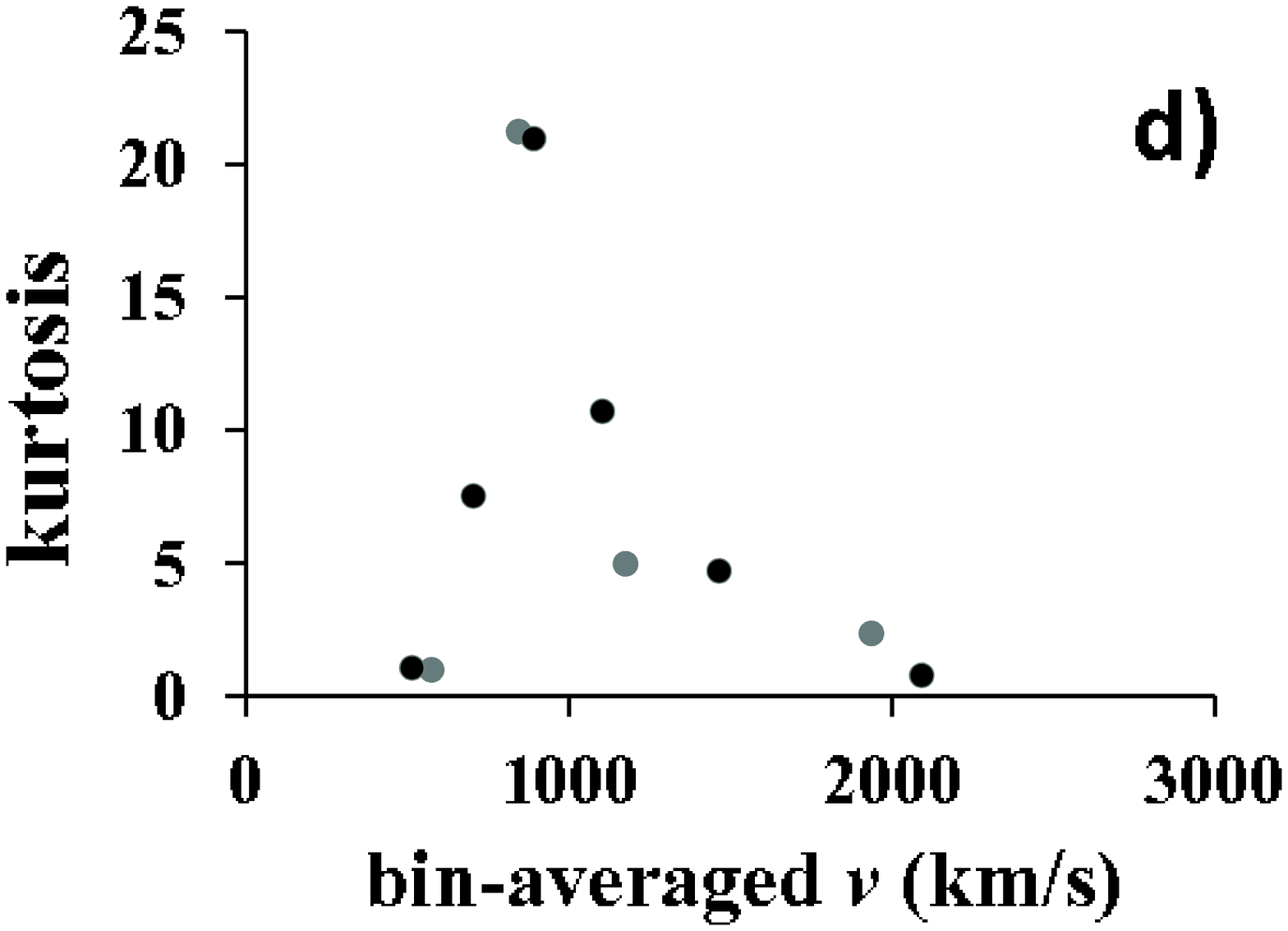}
              }
\caption{$|Dst|$ distribution parameters as a function of the bin-averaged value for the CME speed, $v$, in km s$^{-1}$. Black and gray dots mark the values for two different speed bins. While the black dots correspond to speed bins: 400\,--\,600 $\mathrm{km\,s^{-1}}$, 600\,--\,800 $\mathrm{km\,s^{-1}}$, 800\,--\,1000 $\mathrm{km\,s^{-1}}$, 1000\,--\,1200 $\mathrm{km\,s^{-1}}$, 1200\,--\,1700 $\mathrm{km\,s^{-1}}$, and $v > 1700$ $\mathrm{km\,s^{-1}}$, the gray dots correspond to speed bins: 400\,--\,700 $\mathrm{km\,s^{-1}}$, 700\,--\,1000 $\mathrm{km\,s^{-1}}$, 1000\,--\,1500 $\mathrm{km\,s^{-1}}$, and $v>1500$ $\mathrm{km\,s^{-1}}$. Error bars in a) and b) represent confidence intervals, whereas the straight line shows a linear fit to data marked with black dots.
        }
   \label{fig5}
   \end{figure}
\fi

\subsection{CME/flare source position}
				\label{source}

Several aspects of CME/flare source position were analyzed. First, the events were categorized by the quadrant in which the CME/flare source position is found (northeast, northwest, southeast, and southwest). It was confirmed by a two sample t-test that there is no difference in the samples from different quadrants. Next, it was investigated whether there is an asymmetry regarding the north/south and west/east source position of the CME/flare. Although a small difference in the $|Dst|$ distributions was observed between the west and east hemispheres, the two sample t-test could not confirm the differences.

\ifpdf
	bla
\else
\begin{figure}
\centerline{\includegraphics[width=0.5\textwidth]{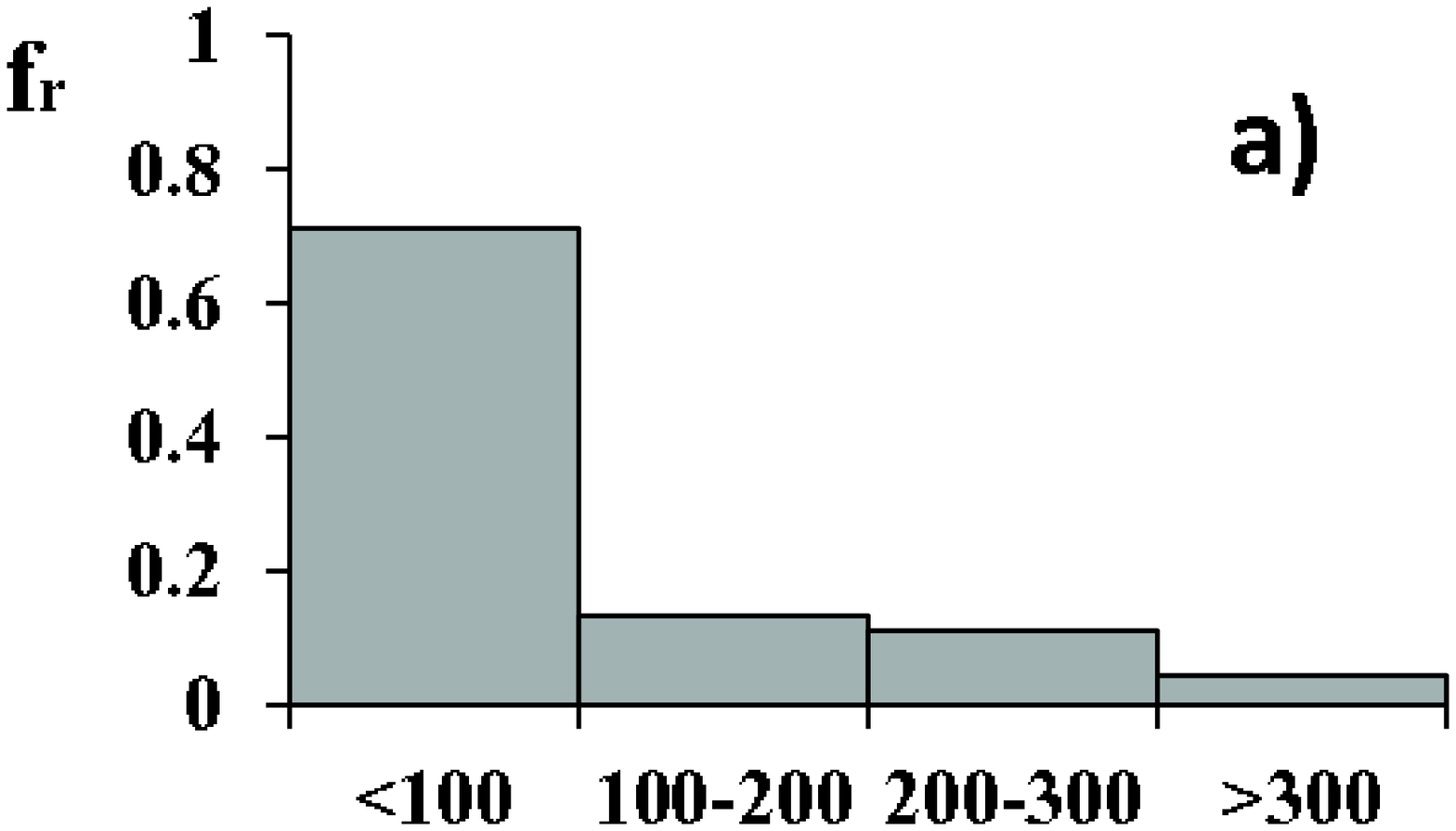}
            \hspace*{0.03\textwidth}
            \includegraphics[width=0.5\textwidth]{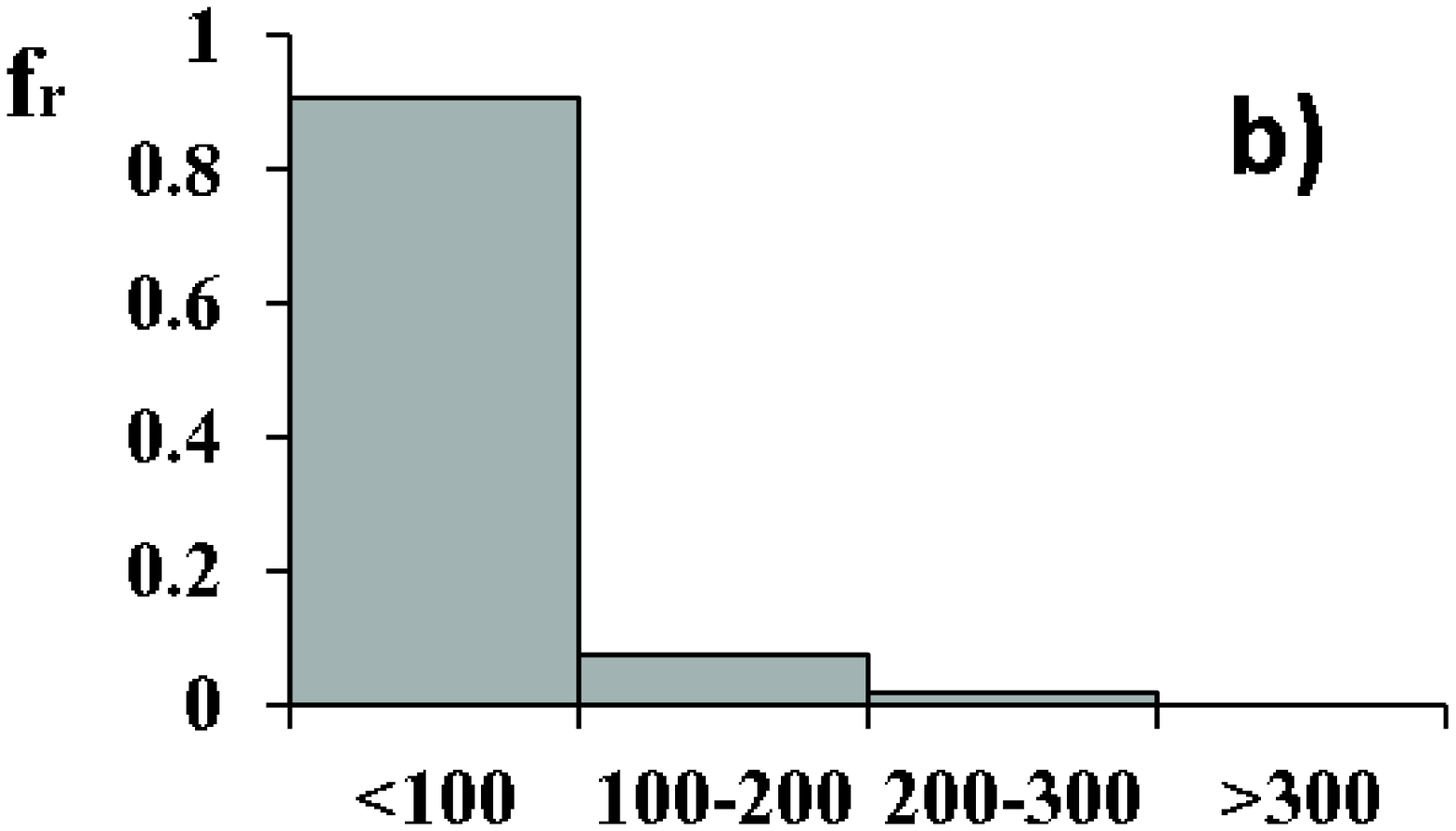}
              }
\vspace{0.03\textwidth}
\centerline{\includegraphics[width=0.5\textwidth]{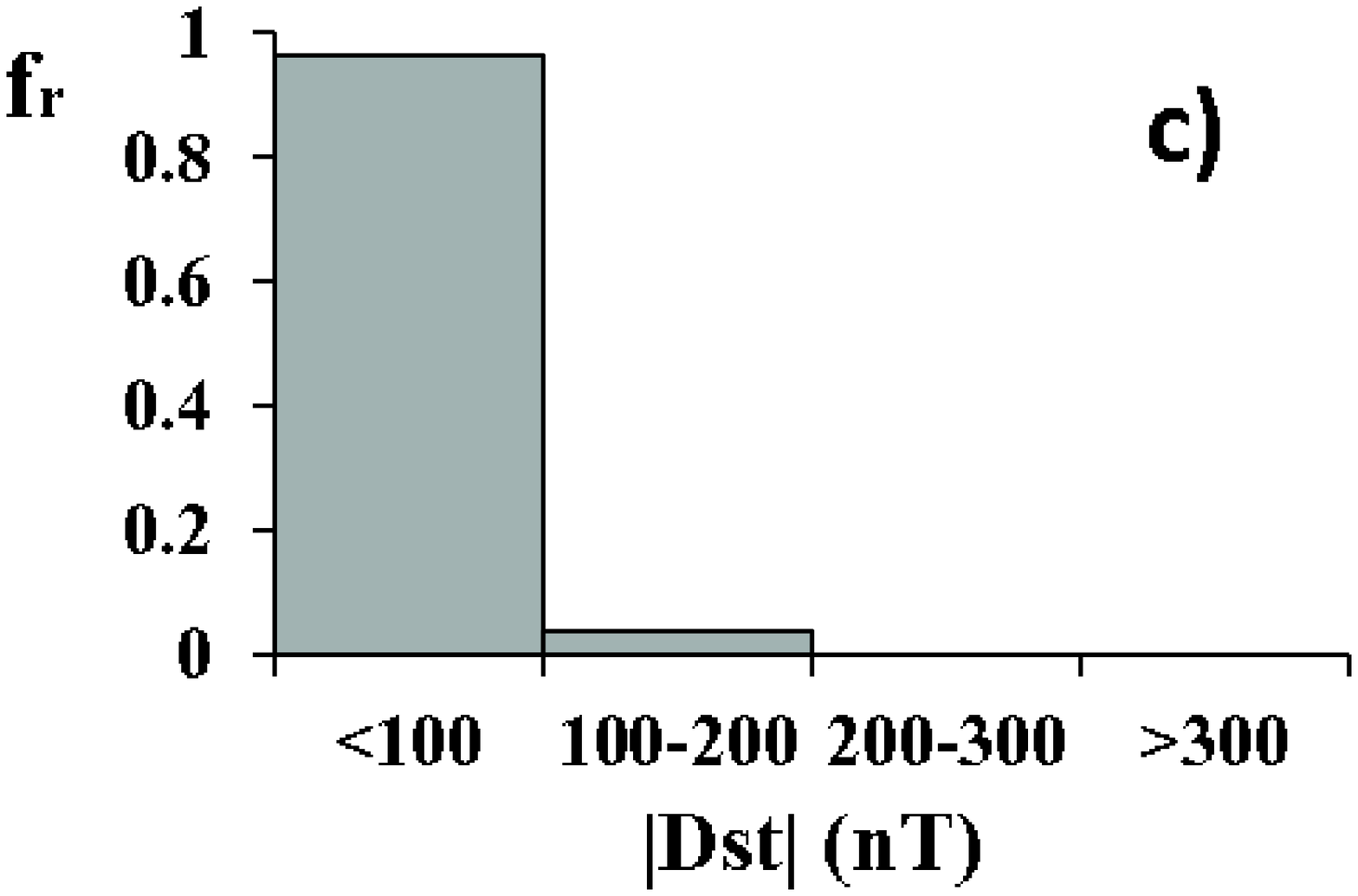}
            \hspace*{0.03\textwidth}
            \includegraphics[width=0.5\textwidth]{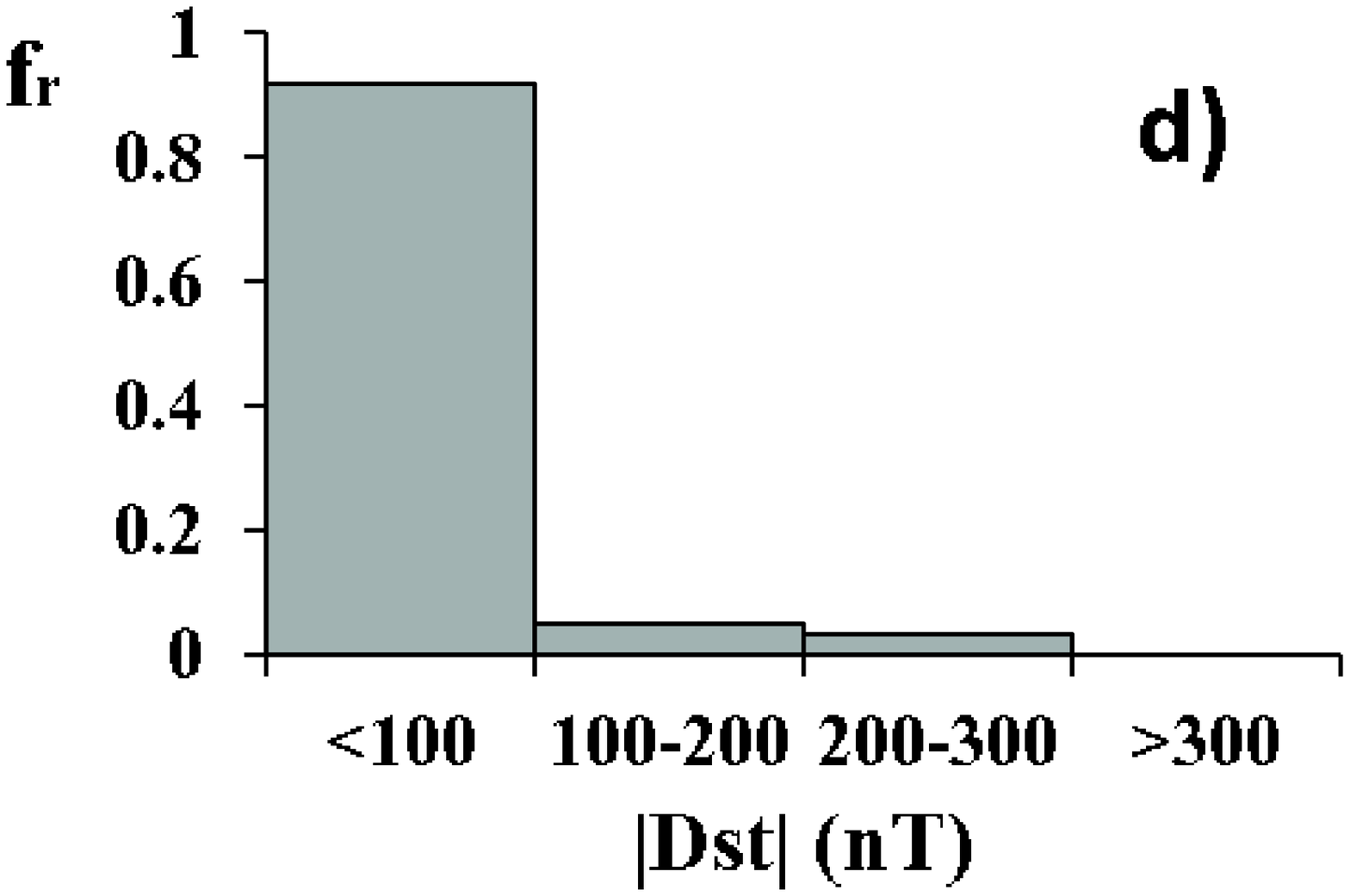}
              }

\caption{$|Dst|$ relative frequencies for different bins of the distance of the CME/flare source from the solar disc centre, $r$, expressed in the units of solar radius: a) $r<0.4$; b) $0.4<r<0.6$; c) $0.6<r<0.8$; d) $r>0.8$
        }
   \label{fig6}
   \end{figure}
\fi

Finally, a source distance from the solar disc centre, $r$, was investigated as a key parameter, ranging from 0 to 1 (in units of solar radii). Similarly as with CME speed, $v$, the events were optimally categorized into four bins:
$r<0.4$ (bin 1), $0.4<r<0.6$ (bin 2), $0.6<r<0.8$ (bin 3), and $r>0.8$ (bin 4).
The number of events in each bin is 45, 53, 53, and 60, respectively. For events involved in (possible) CME--CME interaction the source region of the fastest CME was taken as the relevant one. For each range of $r$, a $|Dst|$ distribution was made, using criteria of $|Dst|$ bins as discussed in Section \ref{speed}. This resulted in four $|Dst|$ distributions for different $r$ ranges of the CME/flare source region (Figure \ref{fig6}). The results of the two sample t-test between each pair of $|Dst|$ distributions are presented in Table \ref{table2}a.

It can be seen from Figure \ref{fig6} that as the distance of the source region of the CME/flare from the disc centre increases the distribution loses the tail and for $0.6<r<0.8$ is restricted to $|Dst|<200$ nT. This is somewhat expected and in agreement with numerous previous studies, where CMEs closer to the centre of the disc were found to be more geoeffective (\emph{e.g.}, \opencite{zhang03}; \opencite{srivastava04}; \opencite{gopalswamy07}; \opencite{richardson10}). However, for the near-limb events the distribution again increases in the tail, showing that limb CMEs can also be highly geoeffective, as pointed out in, \emph{e.g.}, \inlinecite{schwenn05} and \inlinecite{cid12}. The two sample t-test shows significant differences only for the bin around the solar disc centre ($r<0.4$), however, loss of significance for other bins does not seem stochastic, since there is a decrease in significance as we go towards the near-limb source locations (Table \ref{table2}a). It can be seen in Figure \ref{fig9}a how the distribution mean decreases with the increasing distance from the disc centre, following approximately a power law (the curve is given illustratively in Figure \ref{fig9}a to guide the eye).

Due to the fact that the CME--CME trains are considered as one entity (T? and T CMEs, as explained in Section \ref{data}), associated with a source position of the fastest event within the train, the complete analysis was repeated for S and S? samples (a total of 132 CMEs). All of the mentioned categories (quadrant, north/south, east/west and distance from the disc centre) show quite a similar behaviour.

Additionally, we inspected a dependence on the central meridian distance (CMD), \emph{i.e.}, the distance of the CME/flare source position relative to the central meridian on the visible solar disc. The events were categorized as follows:
$-90^{\circ}<$ CMD $<-60^{\circ}$, $-60^{\circ}<$ CMD $<-30^{\circ}$, $-30^{\circ}<$ CMD $<0^{\circ}$, $0^{\circ}<$ CMD $<30^{\circ}$, $30^{\circ}<$ CMD $<60^{\circ}$, and $60^{\circ}<$ CMD $<90^{\circ}$, with number of events per bin 26, 36, 40, 51, 36, and 22, respectively. A small E--W asymmetry can be observed for $30^{\circ}<$ CMD $<60^{\circ}$, due to the fact that out of 36 east events in this bin, none had a $|Dst|>100$ nT. This bin is also significantly different from all other bins. However, there is a lack of significant E--W differences between the rest of the CMD samples. Therefore,  contrary to studies that report E--W asymmetry (\emph{e.g.}, \opencite{zhang03}; \opencite{zhang07}; \opencite{gopalswamy07}), our analysis shows more or less symmetrical longitudinal distribution of geoeffective CMEs in agreement with, \emph{e.g.}, \inlinecite{srivastava04}.


\begin{table}
\caption{The significance results for the two sample t-test with equal variance not assumed, for the $|Dst|$ distribution mean, between different bins of: a) the source-location distance from the solar disc centre, $r$; b) interaction parameter. Unless marked with an asterisk, the value states that the means of the two samples are not significantly different; ** denotes that the significance of the difference is $>95$\%; * denotes that the significance of the difference is $>90$\%. }
\label{table2}
\begin{tabular*}{0.9\textwidth}{cccc}
\hline
\multicolumn{4}{c}{a) $r$ bins}\\
\hline
 & bin1\tabnote{bins 1--4 represent different $r$ ranges in units of solar radii: $<$ 0.4 (bin1), 0.4-0.6 (bin2), 0.6-0.8 (bin3), and $>$ 0.8 (bin4)} &  bin2 &  bin3\\
\hline
bin4 &  0.002** &  0.24 &  0.86 \\
bin3 &  0.001** &  0.10*&  |  \\
bin2 &  0.01**  &  |    &  |  \\
\hline
\multicolumn{4}{c}{b) interaction parameter}\\
\hline
 & S\tabnote{different interaction parameters: no interaction (S), interaction not likely (S?), interaction probable (T?), and interaction highly probable (T)} &  S? &  T? \\
\hline
T  &  0.001** &  0.43 &  0.45 \\
T? &  0.06*   &  0.91 &  | \\
S? &  0.12    &  |    &  | \\
\hline
\end{tabular*}
\end{table}

An alternative binning of the source position distance from the solar disc centre, $r$, was made, with the same purpose as in Section \ref{speed}. The alternative $r$ bins cover ranges: $r<0.35$, $0.35<r<0.5$, $0.5<r<0.65$, $0.65<r<0.78$, $0.78<r<0.92$, and $r>0.92$. The number of events in each bin is 35, 38, 37, 33, 37 and 31, respectively. The distribution mean for these alternative distributions are shown as gray dots in Figure \ref{fig9}a, together with the original binning (black dots). We can see that binning does not change the result notably, as both distributions follow the same trend.


\subsection{CME--CME interaction parameter}
				\label{interaction}

As explained in Section \ref{data} the CME--CME interaction parameter was defined employing four categories: SINGLE (S), SINGLE? (S?), TRAIN? (T?), and TRAIN (T). The number of events in each bin is 98, 34, 28, and 51, respectively. For each interaction parameter, a $|Dst|$ distribution was made, as in Sections \ref{speed} and \ref{source}. This resulted in four $|Dst|$ distributions shown in Figure \ref{fig7}. The results of the two sample t-test are presented in Table \ref{table2}b.

It can be seen in Figure \ref{fig7} that the distribution for ``S-events'' has a long tail, but is very asymmetric and shifted towards lower values of $|Dst|$. As the interaction level shifts from ``S?'' to ``T?'' and ``T'' the distribution ``fills up'' the tail and therefore shifts towards larger values of $|Dst|$. The results of the two sample t-test are somewhat inconclusive, because there is a significant difference only between ``S'' and ``T'' samples. However, we see that the probabilities that the two samples are statistically the same, decreases with the interaction level, and is highest for the neighbouring bins, thus implying that the effect comes from mixing the bins (``S?'' and ``T?'' are actually mixtures of ``S'' and ``T'' events, with ``S?'' presumably dominated with ``S'' events and ``T?'' with ``T'' events, respectively). Therefore, we conclude that indeed CME--CME interaction influences the probability of a certain $|Dst|$ level, where we can associate higher probabilities of intense storms to CME trains. This does not mean that CME trains are more geoeffective due to some physical mechanism, because we also observe ``S-CMEs'' producing extremely intense ($|Dst|>300$ nT) storms. Our results just show that they are less likely to do that.

\ifpdf
	bla
\else
\begin{figure}
\centerline{\includegraphics[width=0.5\textwidth]{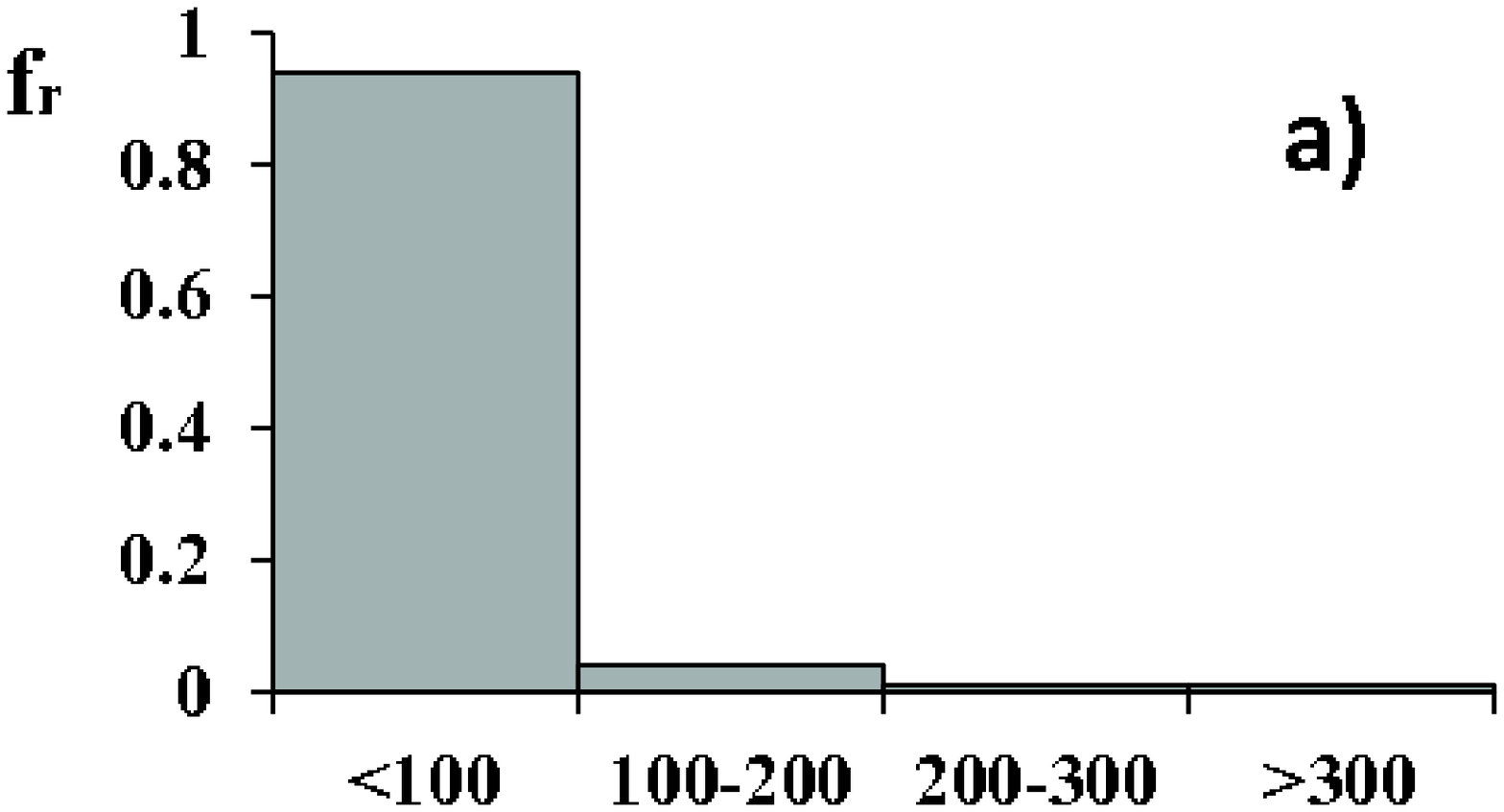}
            \hspace*{0.03\textwidth}
            \includegraphics[width=0.5\textwidth]{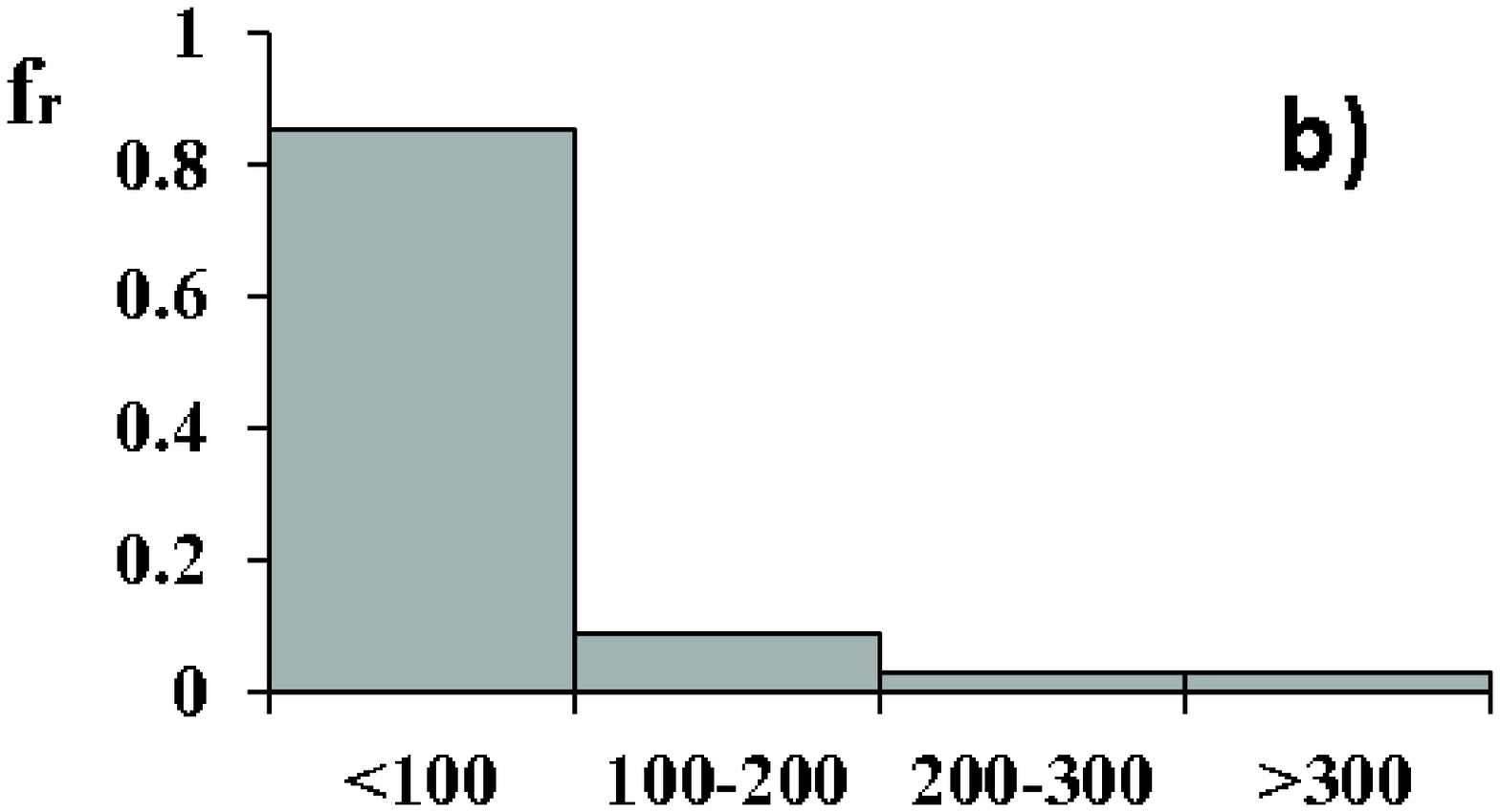}
              }
\vspace{0.03\textwidth}
\centerline{\includegraphics[width=0.5\textwidth]{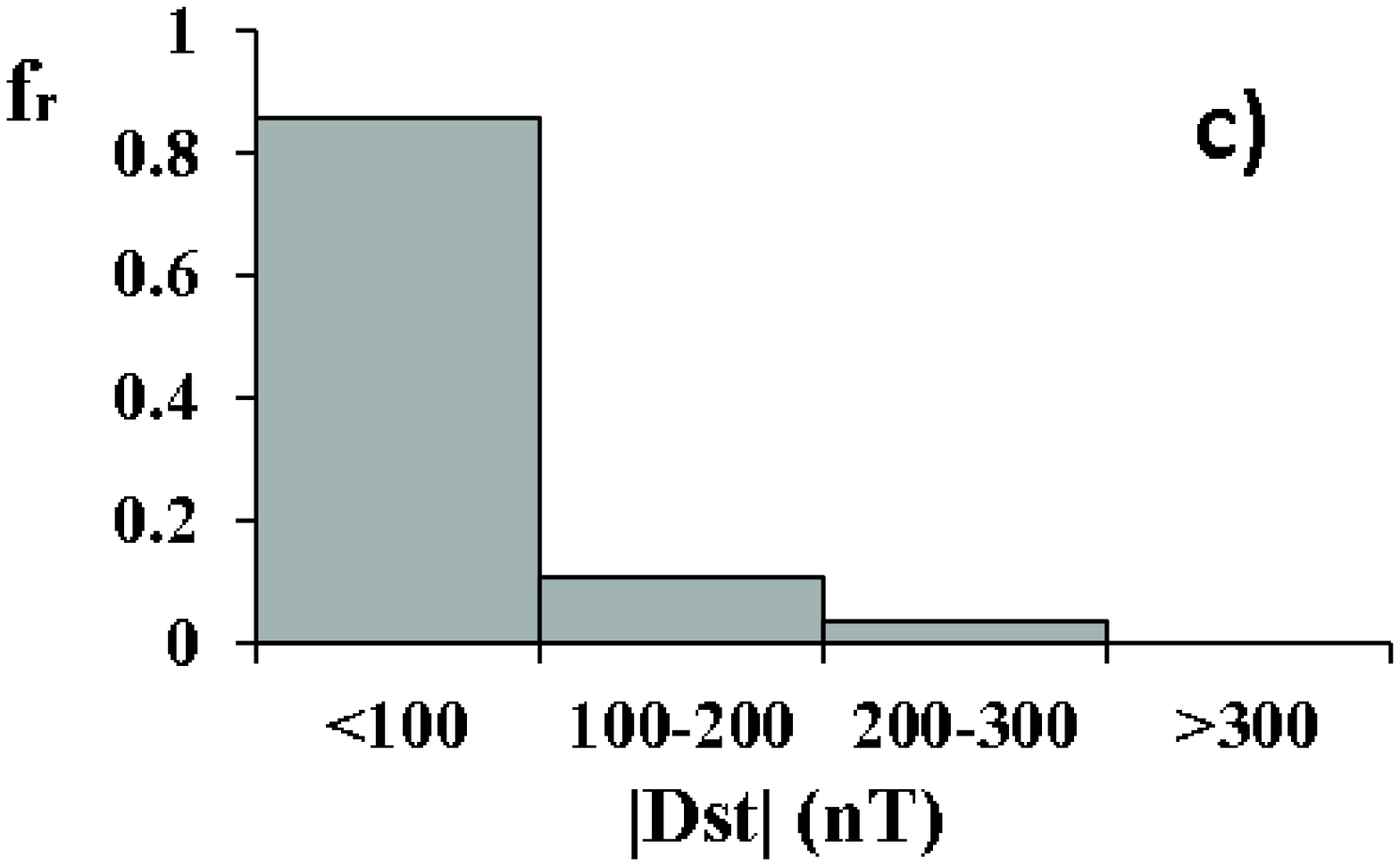}
            \hspace*{0.03\textwidth}
            \includegraphics[width=0.5\textwidth]{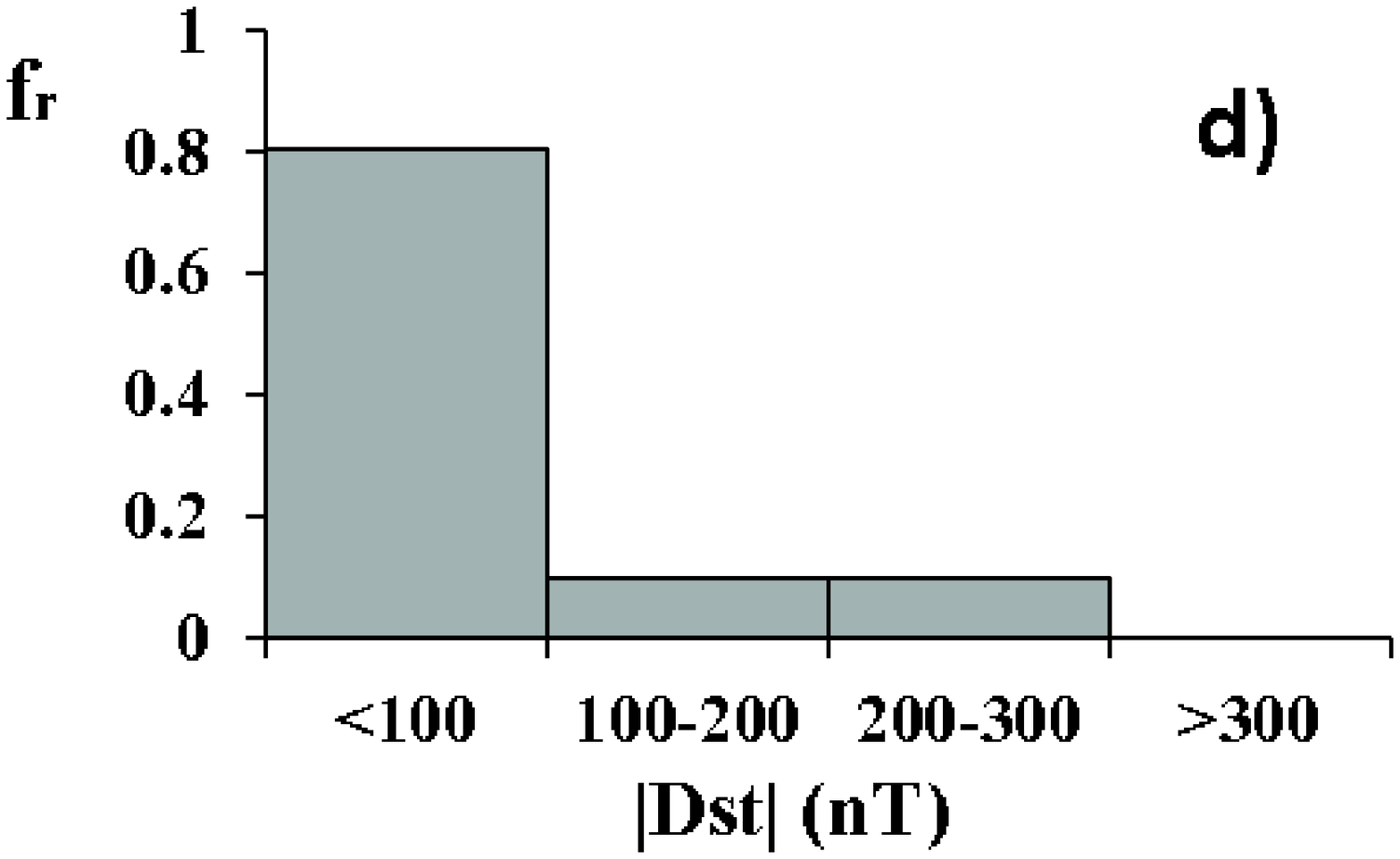}
              }
\caption{$|Dst|$ relative frequencies for different interaction levels:
a) S events - no interaction;
b) S? events - interaction not likely;
c) T? events - interaction probable;
d) T events - interaction highly probable.
        }
   \label{fig7}
   \end{figure}
\fi

However, due to the fact that CMEs in a train are regarded as one entity, with the CME parameters defined by that of the fastest CME in the train, the relationship between the interaction parameter and geoeffectiveness could simply be a byproduct of the relationship between CME speed and geoeffectiveness. Indeed, the speed distribution of ``T'' CMEs is shifted to larger speeds as opposed to ``S'' CMEs, \emph{i.e.}, ``T'' CMEs are generally associated with larger $1^{\mathrm{st}}$ order (linear) CME speed than ``S'' CMEs. On the other hand, when we exclude the fastest CMEs from the sample ($v>$1700 km s$^{-1}$) the difference in the speed distribution between ``T'' and ``S'' CMEs is lost, whereas the relationship between the interaction parameter and $|Dst|$ does not change notably. Therefore, although there is a relationship between the CME speed and the CME interaction parameter, it seems it is not the source of the relationship between the interaction parameter and geoeffectiveness.

To substantiate our results, we mixed the neighbouring bins (S with S?, S? with T?, and T? with T) and thus obtained three additional distributions. The distribution mean for the original interaction bins (black dots) and mixed interaction bins (gray dots) is plotted in Figure \ref{fig9}b, where numerical values were attributed to different interaction levels for quantification reasons (``S''=1; ``S?''=2; ``T?''=3; ``T''=4). It can be seen that the mixed bins follow the same trend as original bins. A power-law function is fitted to the four levels of interaction to illustrate this trend.


\subsection{CME angular width}
				\label{width}

\ifpdf
	bla
\else
\begin{figure}
\centerline{\includegraphics[width=0.5\textwidth]{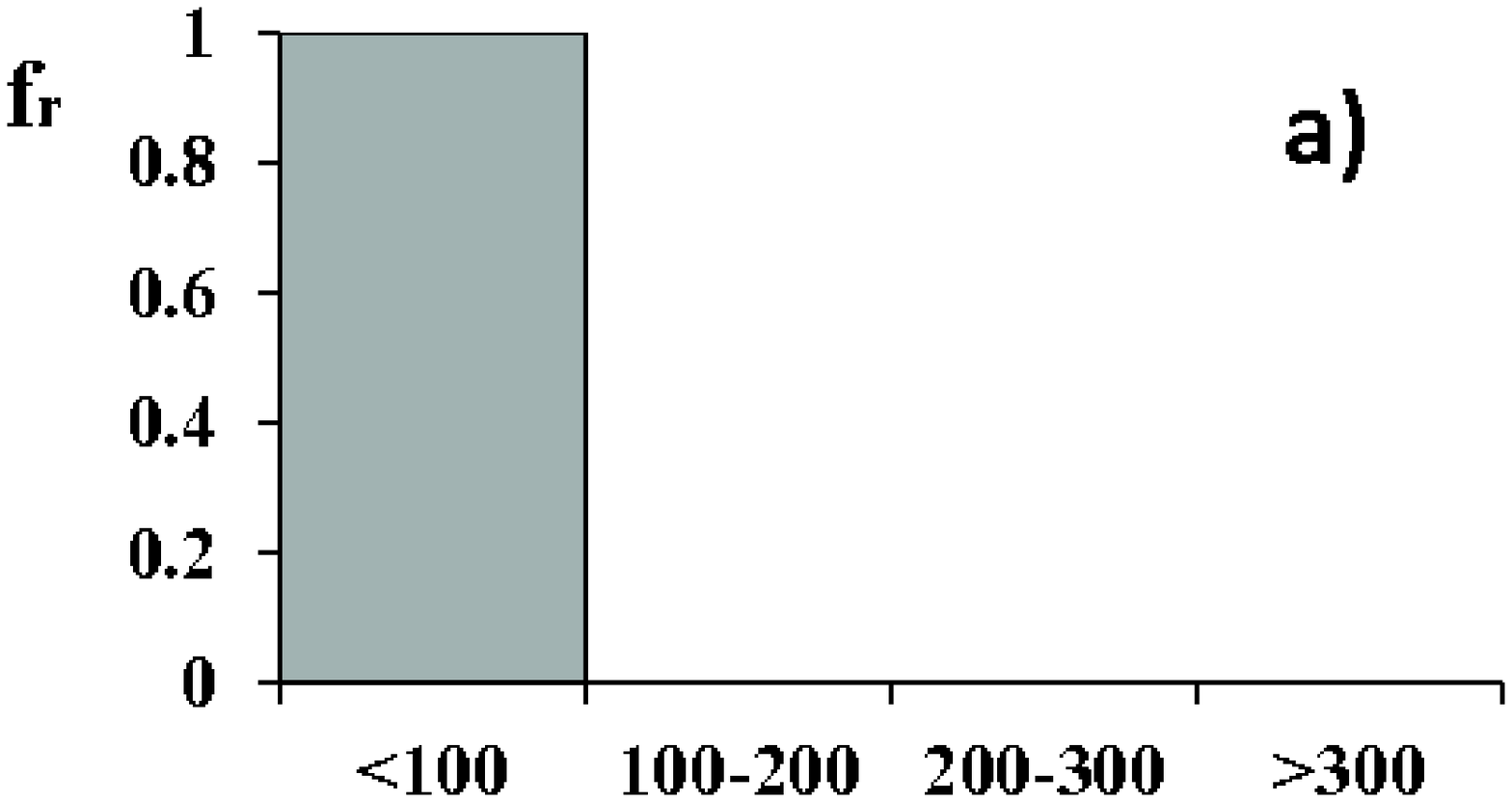}
            \hspace*{0.03\textwidth}
            \includegraphics[width=0.5\textwidth]{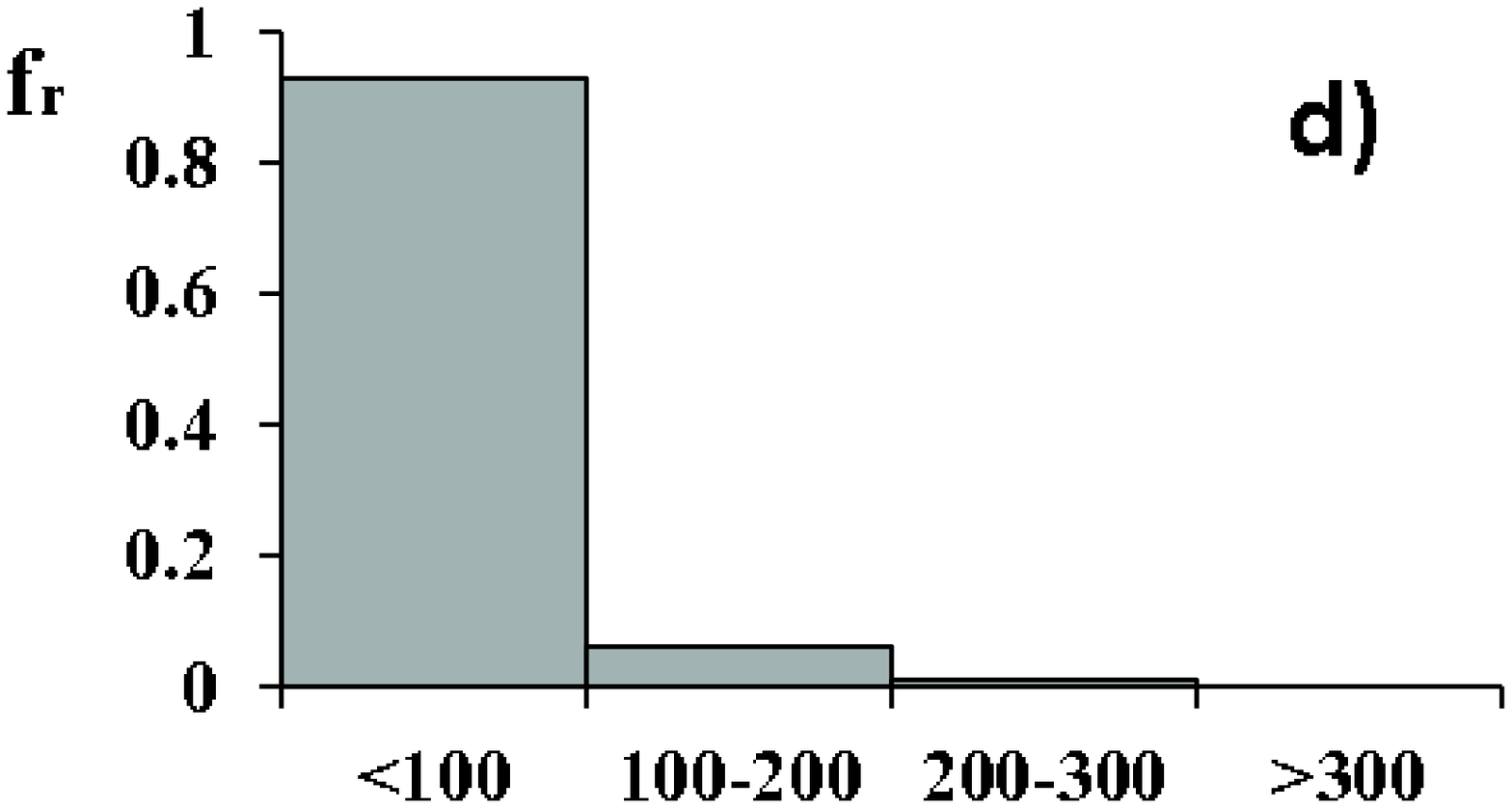}
              }
\vspace{0.03\textwidth}
\centerline{\includegraphics[width=0.5\textwidth]{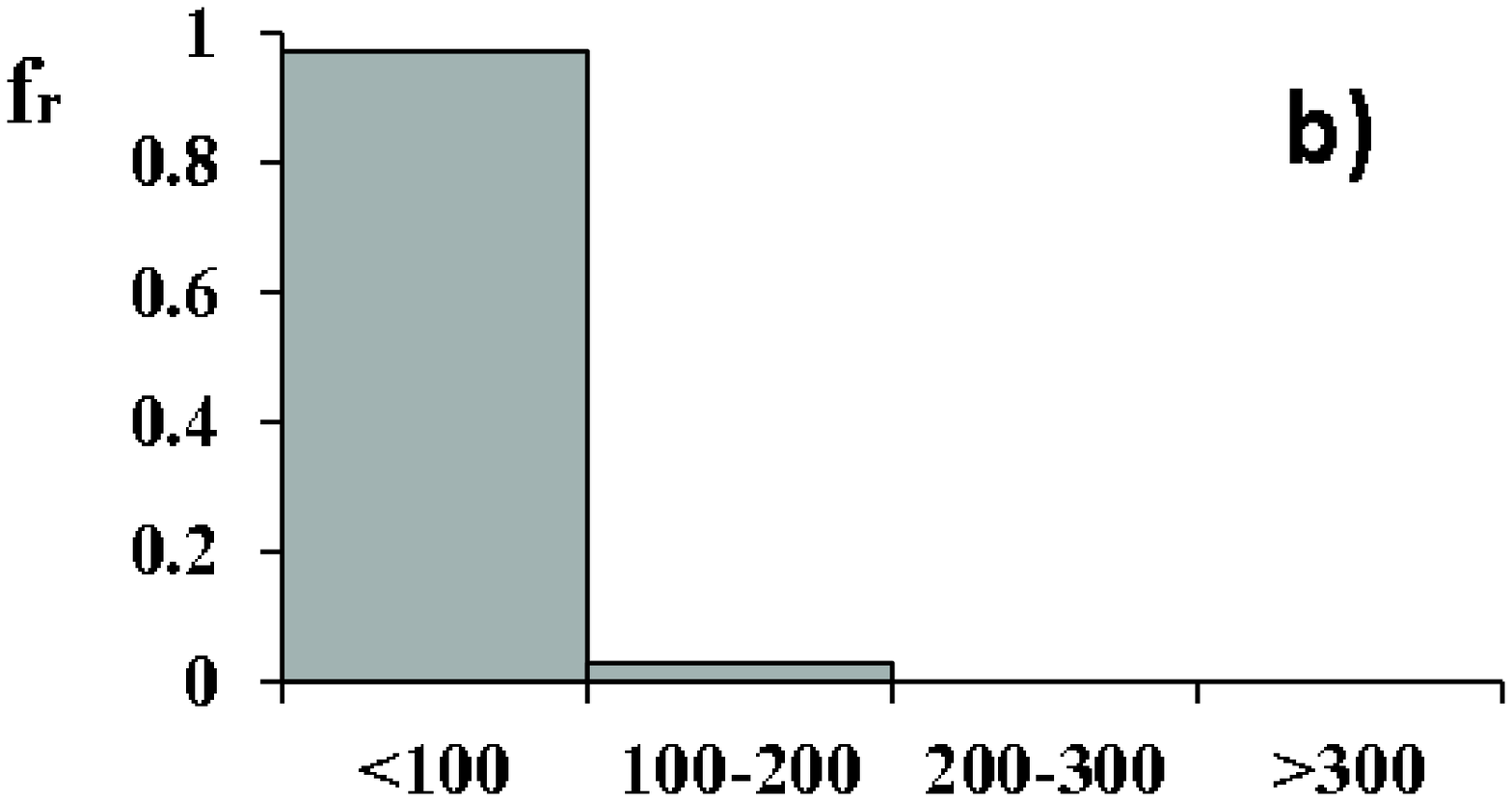}
            \hspace*{0.03\textwidth}
            \includegraphics[width=0.5\textwidth]{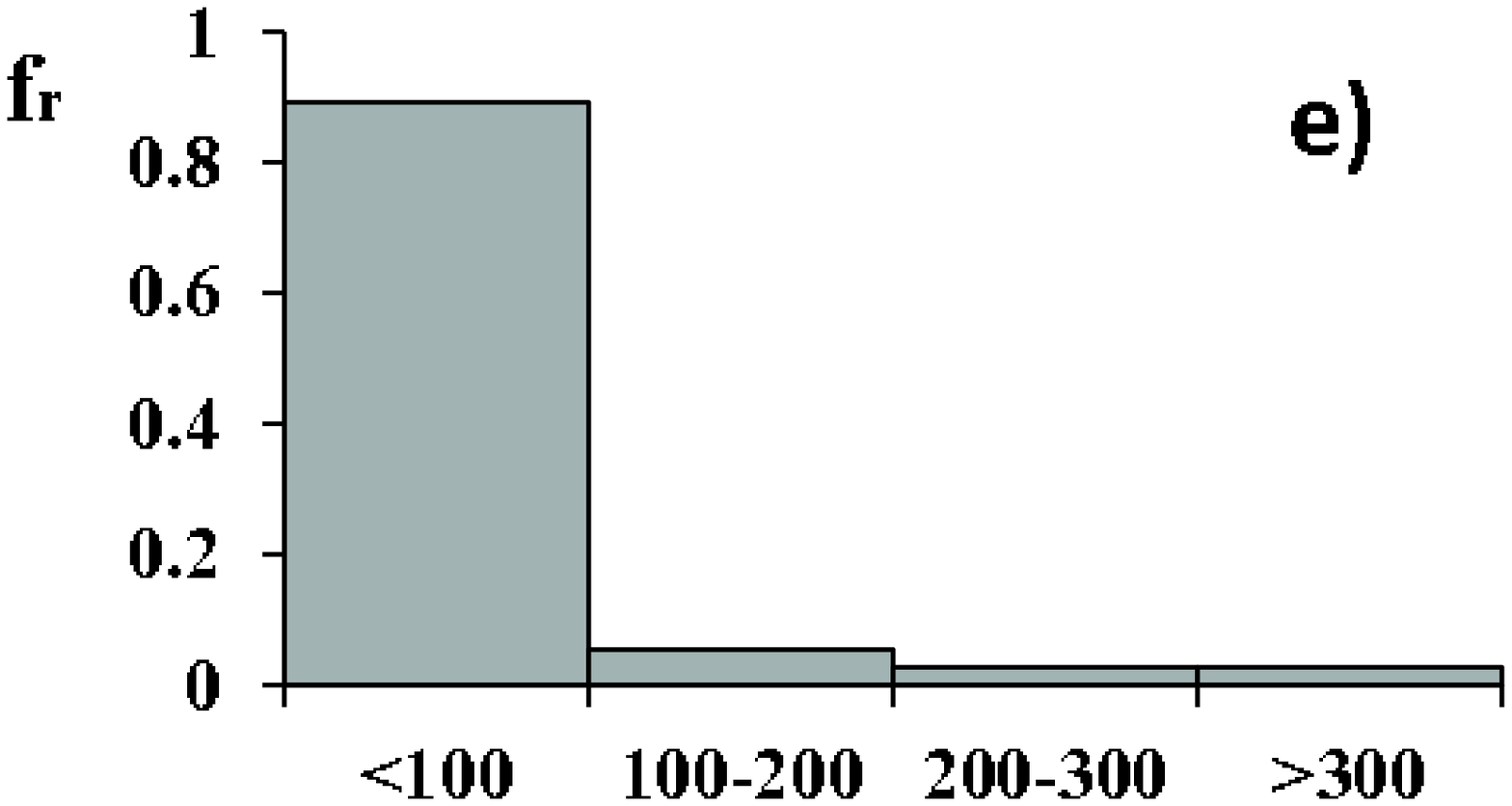}
              }
\vspace{0.03\textwidth}
\centerline{\includegraphics[width=0.5\textwidth]{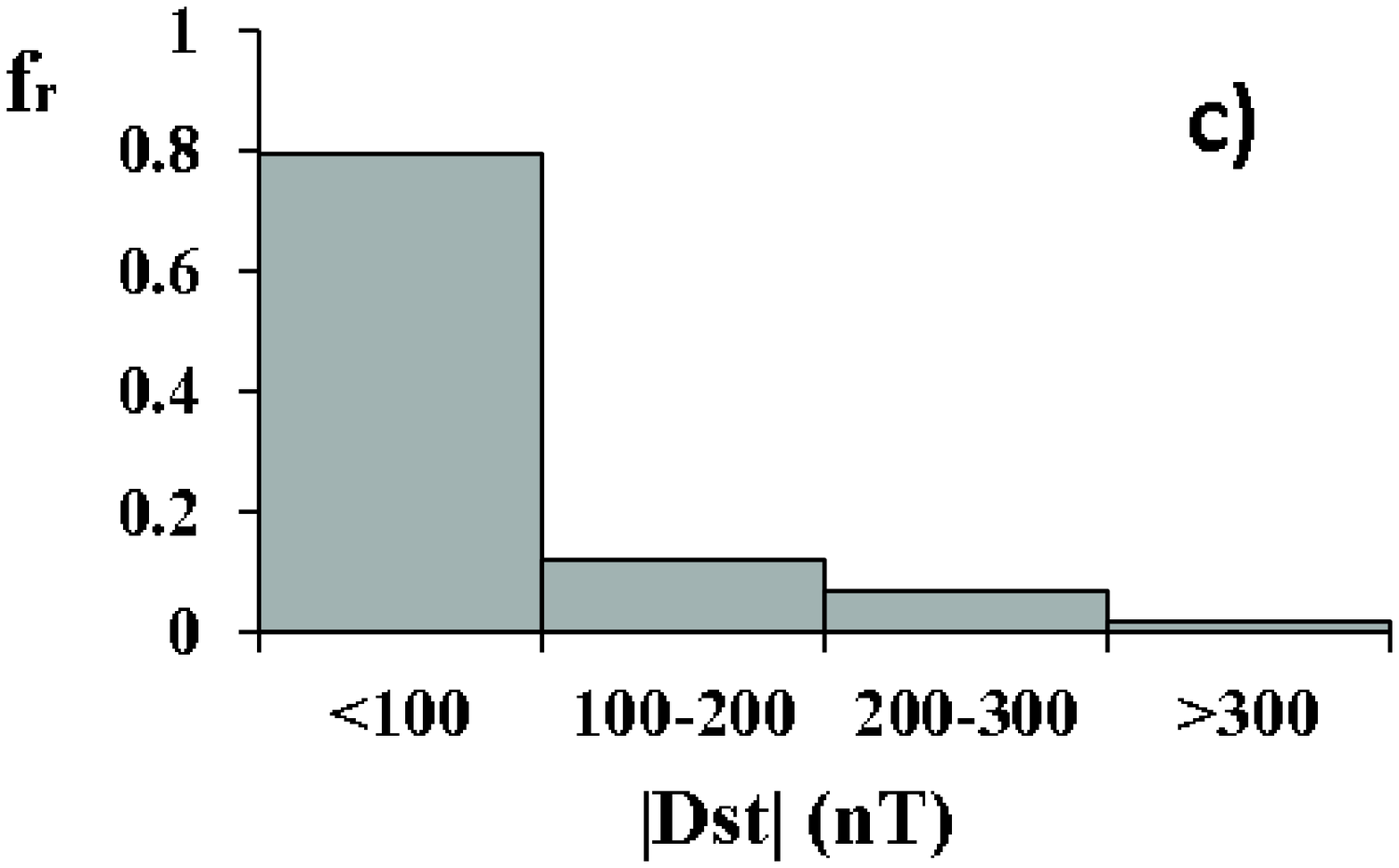}
            \hspace*{0.03\textwidth}
            \includegraphics[width=0.5\textwidth]{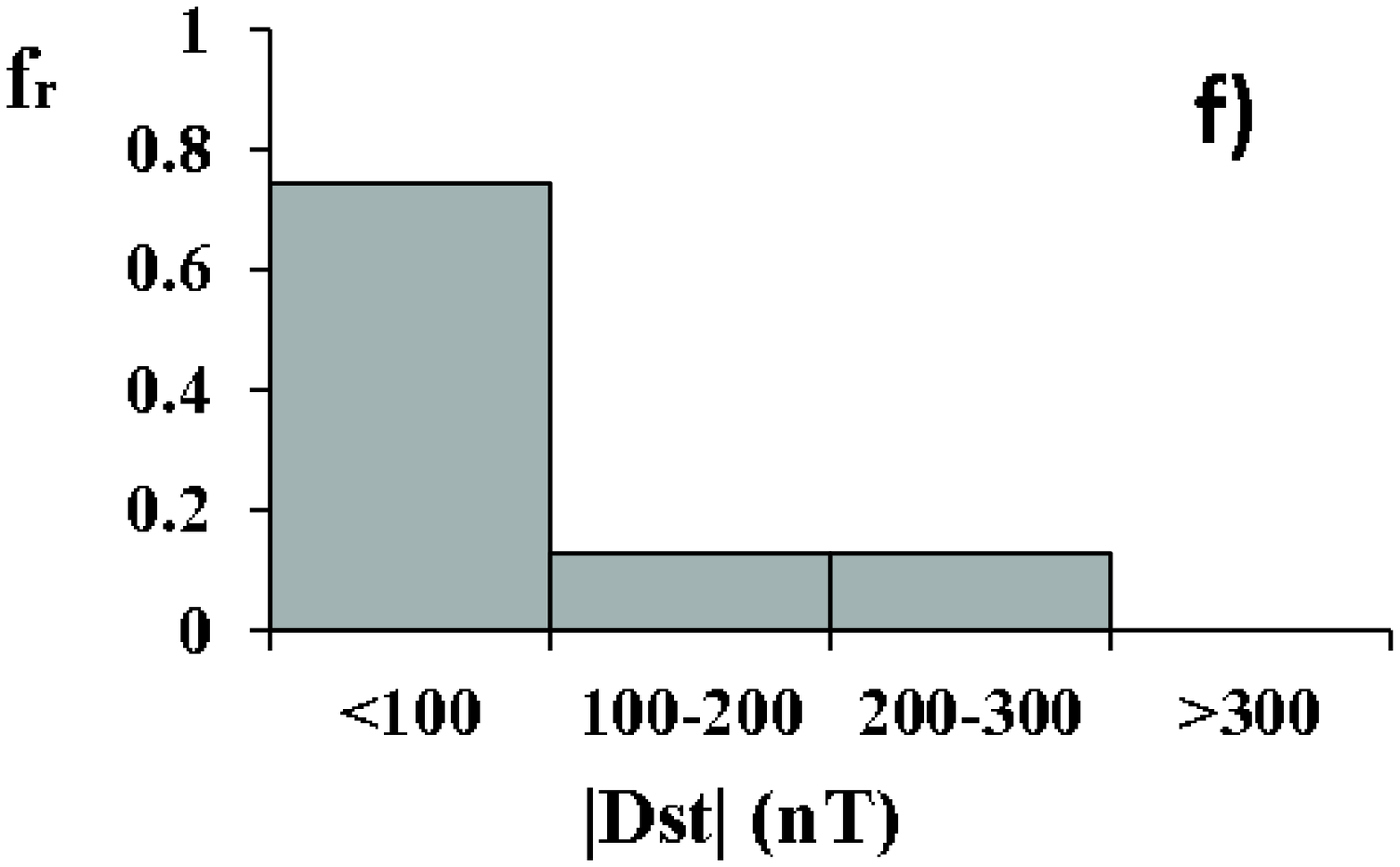}
              }

\caption{$|Dst|$ relative frequencies for different CME-width bins (a\,--\,c) and flare-class bins (d\,--\,f):
a) non halo CMEs;
b) partial halo CMEs;
c) halo CMEs;
d) B\&C class flares;
e) M class flares;
f) X class flares;
        }
   \label{fig8}
   \end{figure}
\fi

The binning for CME (apparent) width, $w$, follows the categorization from the SOHO LASCO CME catalog into non-halo ($w<120^{\circ}$), partial halo ($120^{\circ}< w <360^{\circ}$) and halo CMEs ($w=360^{\circ}$). Due to the fact that interacting CMEs are regarded as one entity (see Section \ref{data}), "T" and "T?" events were associated with the width of the widest CME within a train (\emph{i.e.}, halo or partial halo, if present). The number of events within a certain width bin for non-halo, partial halo, and halo CMEs are 59, 35, and 117, respectively. Using $|Dst|$ binning explained previously, three $|Dst|$ distributions were made (Figures \ref{fig8}a--c). The results of the two sample t-test are presented in Table \ref{table3}a.).

In Figure \ref{fig8} we see an obvious progression in the $|Dst|$ distribution towards larger $|Dst|$ as the apparent width of the CME increases. For non-halos we find one-bin distribution within $|Dst|<100$ nT, for partial halos the distribution gains a small tail, whereas for halos a long tail is observed. The distribution mean has an obvious increasing trend with larger widths (black dots in Figure \ref{fig9}c), which can be fitted by a quadratic function. These results are confirmed with the two sample t-test, showing that non-halo, partial halo and halo CME associated $|Dst|$ distributions are significantly different (Table \ref{table3}a.).

The analysis was repeated separately for ``S'' and ``S?'' CMEs, with the same results and minor loss in significance (due to smaller number of events). Furthermore, we associate the width of the fastest CME in a train (as opposed to the previous association of the widest CME in a train) to ``T'' and ``T?'' events and repeat the analysis. Similar results are obtained. Both the distribution for non-halo and partial halo CMEs are restricted to $|Dst|<200$ nT, but the distribution mean for partial halos is somewhat larger (although not statistically significant).


\begin{table}
\caption{Two sample t-test significance levels for the $|Dst|$ distribution mean with equal variance not assumed for: a) different CME-width bins; b) different flare-class bins. Unless marked with an asterisk, the value states that the means of the two samples are not significantly different; ** denotes that the significance of the result is $>95$\%; * denotes that the significance of the result is $>90$\%}
\label{table3}

\begin{tabular}{cccccccc}
\hline
\multicolumn{3}{c}{a) Width bins}
															
 & & & \multicolumn{3}{c}{b) Flare-class bins}\\
\hline
NH \emph{vs} PH\tabnote{NH = non-halo CMEs ($w<120^{\circ}$); PH = partial halo CMEs ($120^{\circ}<w<360^{\circ}$); H = halo CMEs ($w=360^{\circ}$)} & NH \emph{vs} H & PH \emph{vs} H & & & \verb+B&C+ \emph{vs} M\tabnote{B, C, M, X = B, C, M, X class flares} & \verb+B&C+ \emph{vs} X & M \emph{vs} X\\
\hline
0.06* & $9\cdot10^{\mathrm{-10}}$** & $9\cdot10^{\mathrm{-6}}$** & & & 0.17 & $10^{\mathrm{-3}}$** & 0.03**\\
\hline
\end{tabular}

\end{table}

These results confirm a widely accepted view that halo CMEs are more geoeffective (\emph{e.g.}, \opencite{zhang03}; \opencite{srivastava04}; \opencite{gopalswamy07}), as they clearly show that halo CMEs have larger probabilities to cause intense storms. In addition, we can conclude that non-halo CMEs are not likely to produce major storms ($|Dst|>100$ nT) unless they are involved in a CME--CME interaction with a wider CME.

Finally, an alternative width binning was applied, as in previous sections (see Sections \ref{speed}--\ref{interaction}), using SOHO LASCO CME catalog table values for the apparent width. Again, ``T'' and ``T?'' events were associated with the width of the widest CME within a train. The alternative width bins are: $w<70 ^{\circ}$ (29 events), $70^{\circ}<w<130^{\circ}$ (32 events), $130^{\circ}<w<360^{\circ}$ (33 events), and $w=360^{\circ}$ (halo CMEs, 117 events). The distribution mean for the original width bins (black dots) and alternative width bins (gray dots) is plotted in Figure \ref{fig9}c. The numbers were associated to different width bins for quantitative reasons (non-halo CME=1; partial halo CME=2, halo CME=3). It can be seen that the alternative width bins follow the same trend as the original bins. A quadratic function is fitted to the original width bin data (non-halo, partial halo and halo CMEs) to illustrate this trend.


\subsection{Flare X-ray class}
				\label{flare}

\ifpdf
	bla
\else
\begin{figure}
\centerline{\includegraphics[width=0.5\textwidth]{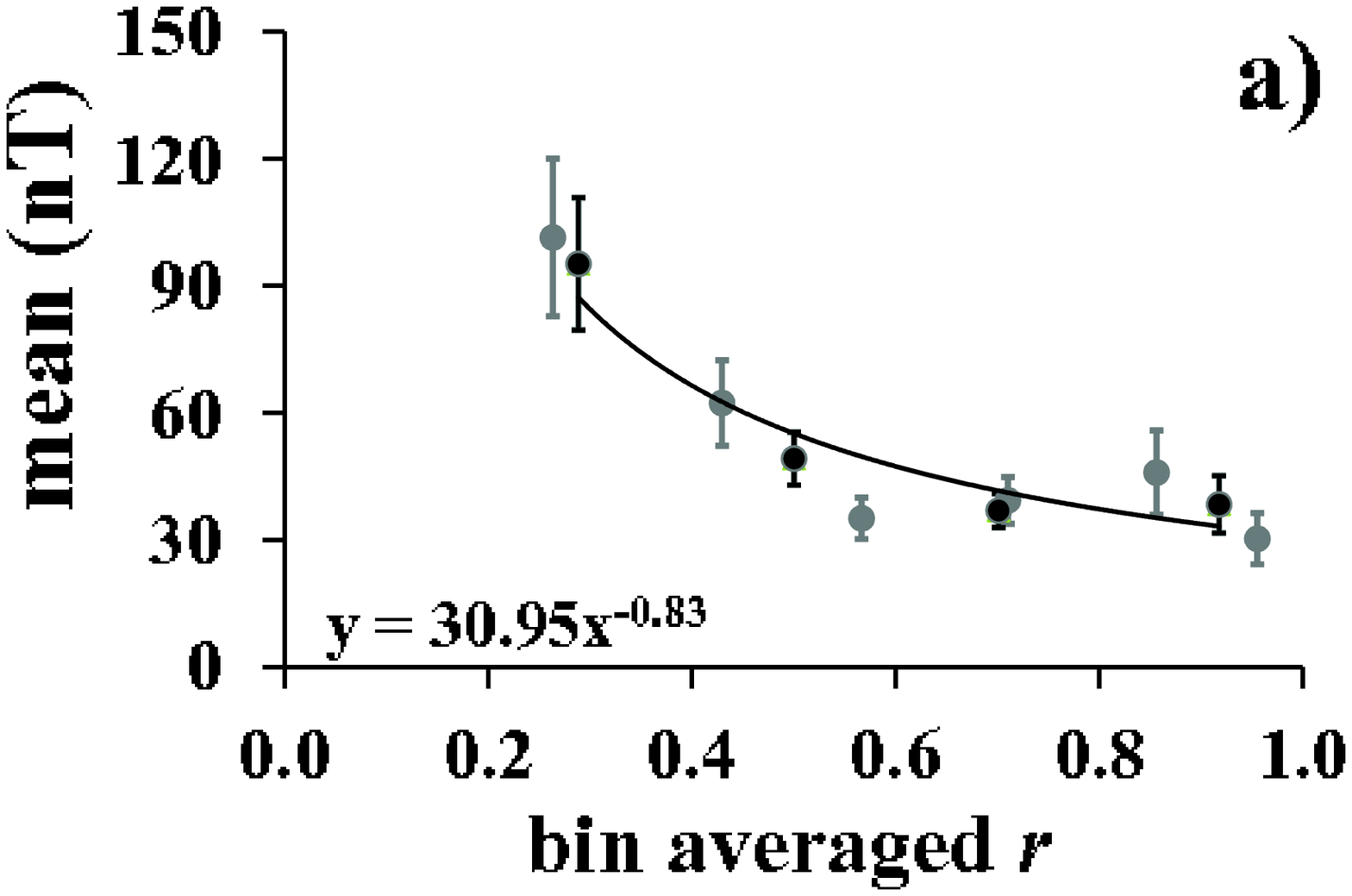}
            \hspace*{0.03\textwidth}
            \includegraphics[width=0.5\textwidth]{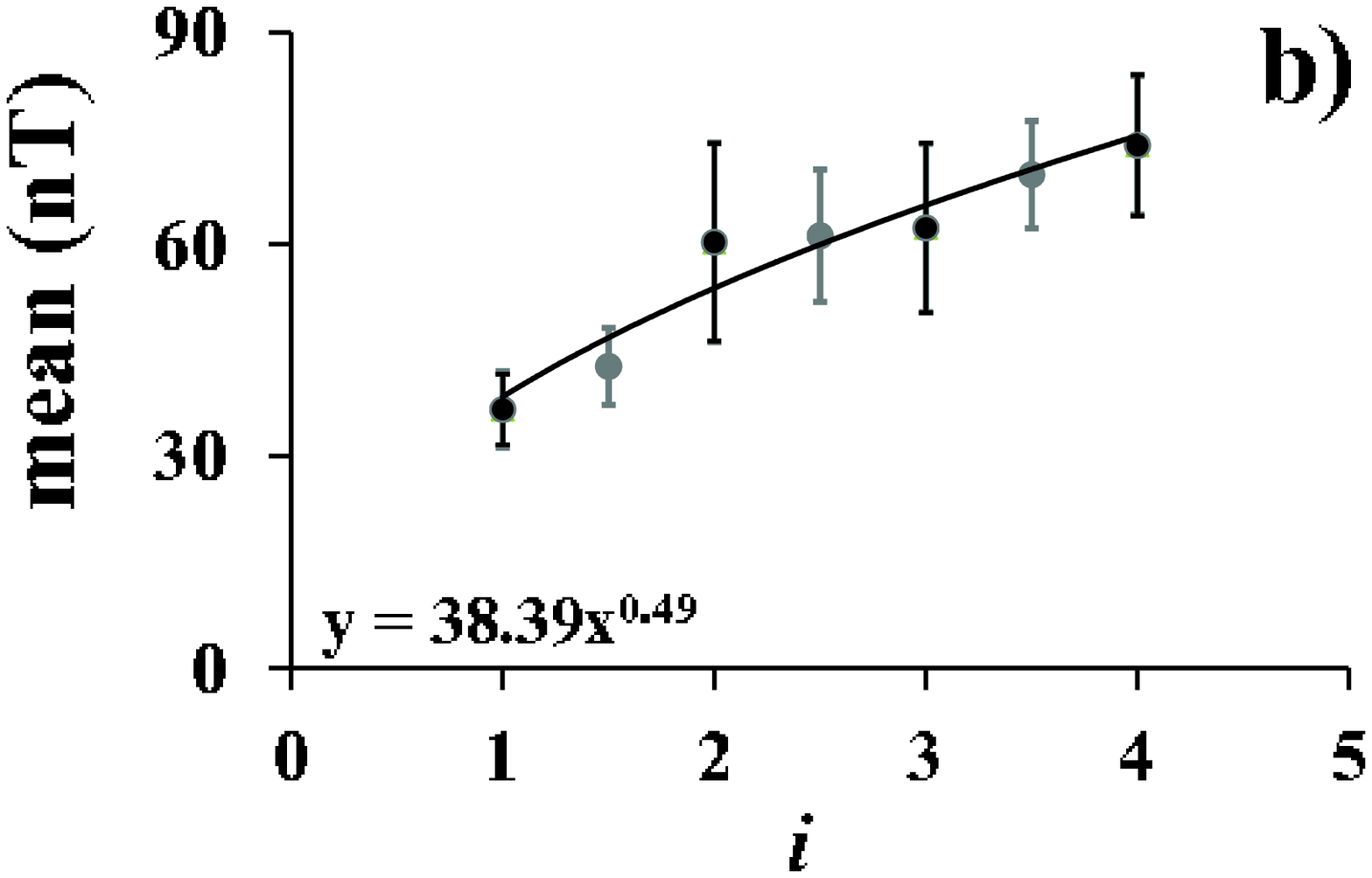}
              }
\vspace{0.03\textwidth}
\centerline{\includegraphics[width=0.5\textwidth]{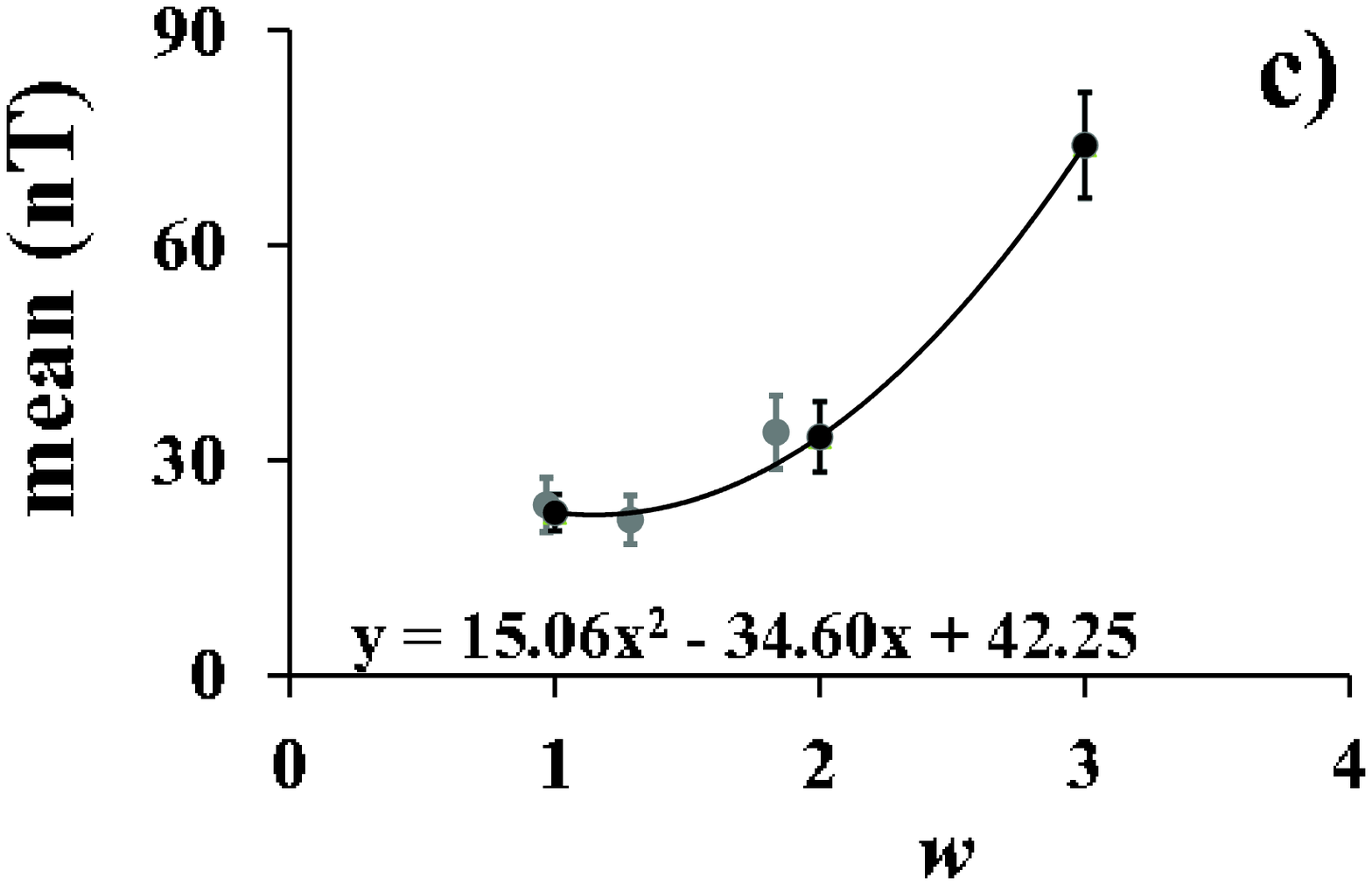}
            \hspace*{0.03\textwidth}
            \includegraphics[width=0.5\textwidth]{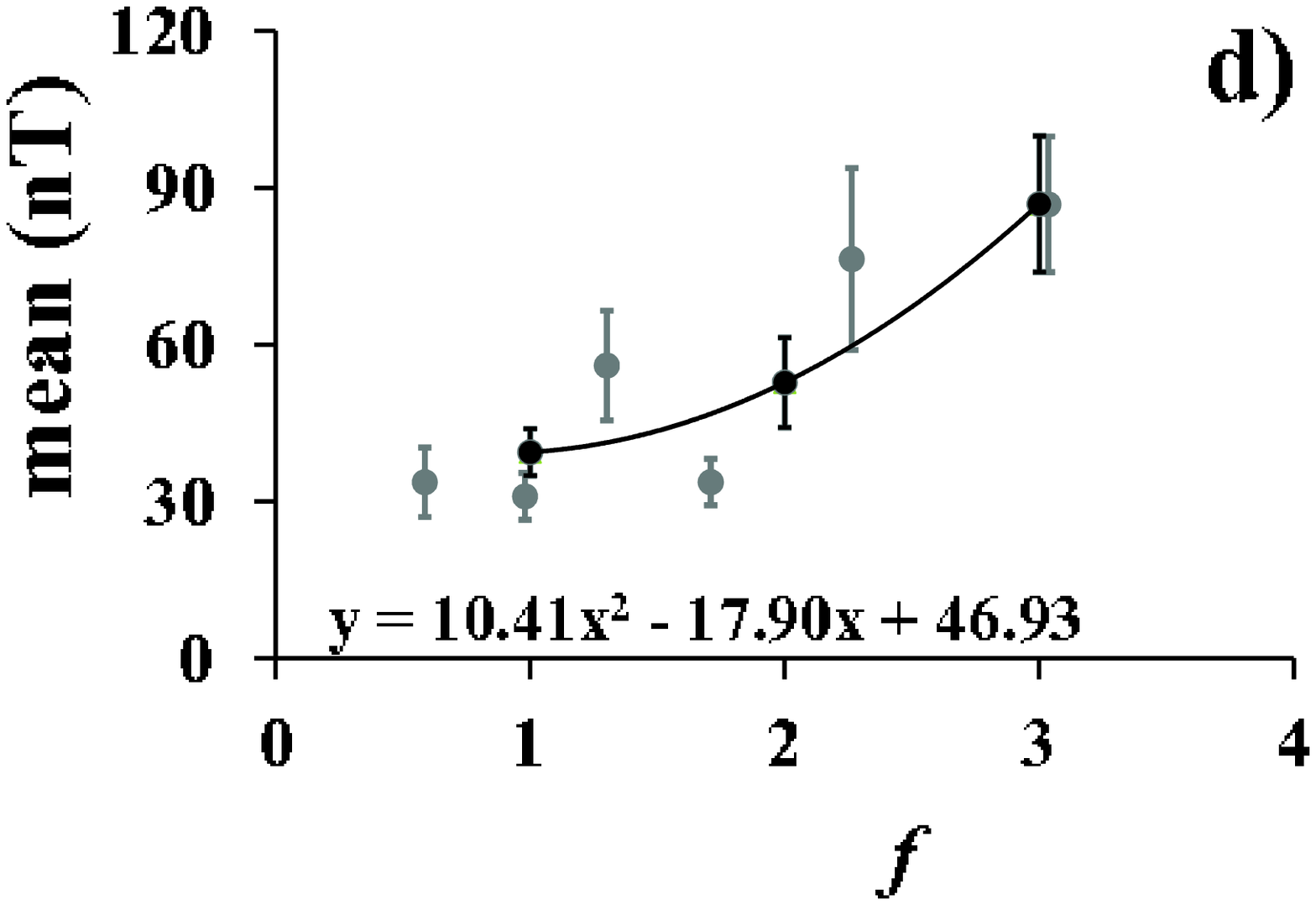}
              }
\caption{$|Dst|$ distribution mean as a function of: a) average value of the source position distance from the solar disc centre, $r$, within a specific bin; b) interaction parameter, $i$ (``S''=1, ``S?''=2, ``T?''=3; ``T''=4); c) width, $w$ (non-halo=1, partial halo=2, halo=3); d) flare class, $f$ (B\&C=1, M=2, X=3). Black and gray dots mark different types of binning (for detailed explanation see Sections \ref{source}-\ref{flare}). Error bars represent confidence intervals, whereas the line shows the fitting curve (fitted trough black dots).
        }
   \label{fig9}
\end{figure}
\fi

The binning of the solar flare class follows the categorization of soft X-ray flares according to their soft X-ray flux peak value ($\mathrm{F_{max}}$ in units $\mathrm{Wm^{-2}}$): $\mathrm{F_{max}}<10^{-6}$ (B class flare), $10^{-6}\le \mathrm{F_{max}}<10^{-5}$ (C class flare), $10^{-5}\le \mathrm{F_{max}}<10^{-4}$ (M class flare), and $\mathrm{F_{max}}\ge 10^{-4}$ (X class flare). Due to lack of events associated with a B class flare, they were put in the same bin with C class flares. Therefore, binning into three flare categories was applied, namely B\&C class flares, M flares, and X flares. The number of events are 98, 74, and 39, respectively. Three $|Dst|$ distributions were made (Figures \ref{fig8}d--f). We can observe minor differences in $|Dst|$ distribution between B\&C class flares and M class flares, whereas there is a clear difference compared to X class flares distribution, which contains substantially larger fraction of events in its tail than other two distributions. This is also reflected in the two sample t-test, showing that although B\&C class flares and M flares are not significantly different samples, they are both significantly different from X flares (Table \ref{table3}).

The distribution mean has an increasing trend with flare class, which can be illustrated by a quadratic function, similar to that shown in Section \ref{width} (Figure \ref{fig9}d). An alternative binning was again applied (gray dots) showing similar trend, however with a large scatter. For alternative binning, we used the peak value of the X-ray flux ($\mathrm{F_{max}}$ in units $\mathrm{Wm^{-2}}$) in the following ranges: $\mathrm{F_{max}}<2.5\cdot10^{-6}$ (31 event), $2.5\cdot10^{-6} \le\mathrm{F_{max}}<5 \cdot10^{-6}$ (40 events), $5 \cdot10^{-6}\le\mathrm{F_{max}}<1.2 \cdot10^{-5}$ (34 events), $1.2 \cdot10^{-5}\le\mathrm{F_{max}}<3 \cdot10^{-5} $ (35 events), $3 \cdot10^{-5}\le\mathrm{F_{max}}< 10^{-4}$ (36 events), $\mathrm{F_{max}}\ge10^{-4}$ (35 events). The numbers were associated to different flare bins for quantitative reasons, similarly as in Sections \ref{interaction} and \ref{width} (B\&C class=1, M class=2, X class=3).

The analysis was repeated for S and S? CMEs, with similar results, but with a loss in significance (only 19 X class flare events). Nevertheless, we can conclude that geoeffective CMEs are associated with stronger flares, in agreement with previous studies (e.g. \opencite{srivastava04}; \opencite{zhang07}).


\begin{table}
\caption{Minimum/maximum/median of the $|Dst|$ distribution for the parameter combinations:
a) CME speed, $v$, and apparent width, $w$;
b) CME speed, $v$, and source position distance from the solar disc centre, $r$;
c) CME speed, $v$, and interaction parameter;
d) apparent width, $w$, and source position distance from the solar disc centre, $r$.
Values are not displayed whenever there are five or less events included.}
\label{table4}

\begin{tabular}{cccccccccc}
\hline
\multicolumn{10}{c}{a) $v$ \emph{versus} $w$}\\
\hline
 & & & $v$ (km s$^{-1}$) & NH\tabnote{NH = non-halo CMEs ($w<120^{\circ}$); PH = partial halo CMEs ($120^{\circ}<w<360^{\circ}$); H = halo CMEs ($w=360^{\circ}$)} & PH & H & & & \\
\hline
 & & & 400-600 & 0/80/20 & | & 10/60/40 & & & \\
 & & & 600-800 & 0/90/20 & 0/80/30 & 10/190/45 & & & \\
 & & & 800-1000 & 0/60/20 & 0/55/35 & 0/380/50 & & & \\
 & & & 1000-1200 & 0/30/10 & 0/40/30 & 0/280/40 & & & \\
 & & & 1200-1700 & 20/190/30 & 0/140/40 & 0/410/35 & & & \\
 & & & $>$1700 & | & | & 0/280/70 & & & \\
\end{tabular}

\begin{tabular}{ccccc}
\hline\hline
\multicolumn{5}{c}{b) $v$ \emph{versus} $r$}\\
\hline
$v$ (km s$^{-1}$) & $r<0.4$ & $0.4<r<0.6$ & $0.6<r<0.8$ & $r>0.8$ \\
\hline
400-600 & 0/110/50 & | & 0/75/20 & 10/30/10 \\
600-800 & 0/190/33 & 0/80/30 & 15/90/40 & 0/50/20 \\
800-1000 & 0/380/38 & 10/100/40 & 10/70/40 & 0/40/10 \\
1000-1200 & | & 0/140/40 & 0/30/10 & 0/280/30 \\
1200-1700 & 0/410/215 & 0/150/35 & 0/140/33 & 0/190/30 \\
$>$1700 & 40/280/130 & 30/250/85 & 0/90/45 & 10/210/50 \\
\hline\hline
\multicolumn{5}{c}{c) $v$ \emph{versus} interaction parameter}\\
\hline
$v$ (km s$^{-1}$) & S\tabnote{for detailed explanation see Section 2.} & S? & T? & T \\
\hline
400-600 & 0/90/20 & | & | & | \\
600-800 & 0/190/20 & | & 10/90/30 & 15/90/45 \\
800-1000 & 0/380/30 & 10/100/55 & | & 0/150/33 \\
1000-1200 & 0/140/30 & 0/280/15 & | & 20/130/30 \\
1200-1700 & 0/140/30 & 0/280/20 & 20/270/150 & 10/250/40 \\
$>$1700 & 0/270/30 & 20/140/85 & | & 60/280/90 \\
\hline\hline
\multicolumn{5}{c}{d) $w$ \emph{versus} $r$}\\
\hline
width & $r<0.4$ & $0.4<r<0.6$ & $0.6<r<0.8$ & $r>0.8$ \\
\hline
NH & 0/30/10 & 0/80/30 & 0/90/20 & 0/190/20 \\
PH & 30/110/45 & 25/30/28 & 0/140/45 & 0/75/15 \\
H & 0/410/65 & 0/250/40 & 0/110/35 & 0/280/55 \\
\hline
\end{tabular}
\end{table}



\subsection{Combinations of solar parameters}
				\label{combinations}

To investigate the influence of combined solar parameters on the $|Dst|$ level, bivariate $|Dst|$ distributions for different combinations of two solar parameters were estimated. The same $|Dst|$ bins were used as in Sections \ref{speed}--\ref{flare}. Table \ref{table4} shows the range and median values of $|Dst|$ distributions for different combinations of two solar parameters.

It can be seen in Table \ref{table4}a that the median of the $|Dst|$ distribution has highest values for speed bins where $v>1200$ $\mathrm{km\,s^{-1}}$ in case of non-halo and partial halo CMEs, whereas for halo CMEs this is the case when $v>1700$ $\mathrm{km\,s^{-1}}$. Therefore, the combination of a larger apparent width and larger speed of CMEs increases the probability of a larger $|Dst|$ level (\emph{i.e.}, stronger geomagnetic storm). It should be noted that the median value for halo CMEs in almost all of the speed bins is larger compared to non-halo and partial-halo CMEs. This again indicates the importance of the apparent width as a relevant solar parameter regarding geoeffectiveness, in agreement with previous studies (\emph{e.g.}, \opencite{zhang03}; \opencite{srivastava04}; \opencite{gopalswamy07}).

In Table \ref{table4}b the combination of CME speed, $v$, and source distance from the solar disc centre, $r$, is investigated. It can be seen that the central source positions are associated with higher values of the $|Dst|$ distribution median, as well as larger speeds. Furthermore, combination of a very fast CME ($v>1200$ $\mathrm{km\,s^{-1}}$) and a source position close to the disc centre have significantly higher median, \emph{i.e.}, highly increases the chance of a strong geomagnetic storm. For the source positions closer to the limb we also observe that the CME speed plays a role in causing higher $|Dst|$ levels. We also observe a change in the median value of the $|Dst|$ distribution for the combination of very fast CMEs and very high interaction level (Table \ref{table4}c). However, this is not so pronounced as in the case of the speed/source-position combination. Finally, the width/source-position combination of solar parameters leads to the highest $|Dst|$ levels for halo CME from the disc centre (Table \ref{table4}d) although halo CMEs closer to the limb can also cause higher $|Dst|$ values.

From the presented analysis we conclude that the combination of favorable solar parameters can result in enhanced geoeffectiveness. These model parameters individually do not have the same impact. The latter is also visible from the fitted curves in Figures \ref{fig5} and \ref{fig9}: the fitted functions for different solar parameters have different growth/descend rates. A combination of solar parameters for investigating geoeffectiveness has been attempted by \inlinecite{srivastava05} and \inlinecite{srivastava06}. However, her conclusion was that without the interplanetary parameters, the forecast of CME geoeffectiveness is insufficiently precise.


\subsection{Monthly CME/flare and $Dst$ activity}
				\label{activity}
			
Finally, we investigate the monthly CME/flare activity in the SOHO era. For that purpose we define monthly CME/flare activity parameter in a given month, $A_{\mathrm{i}}$ (i=1,2,...,12), based on the solar parameters that influence the geoeffectiveness, as derived throughout Sections \ref{speed}--\ref{flare}:

\begin{itemize}
\item number of CMEs in month ``i'' of the year, $N_{\mathrm{CME,i}}$
\item average CME speed in month ``i'', $v_{\mathrm{avg,i}}$
\item number of CMEs with speed $>1000$ $\mathrm{km\,s^{-1}}$ in month ``i'', $N_{v,\mathrm{i}}$
\item number of X-class flares in month ``i'', $N_{\mathrm{X,i}}$
\item number of HALO CMEs in month ``i'', $N_{\mathrm{HALO,i}}$
\item average source position distance from the solar disc centre of CMEs/flares in month ``i'', $r_{\mathrm{avg,i}}$ (in units of solar radii)
\item SOHO downtime in hours in month ``i'', $t_{\mathrm{SOHO,i}}$
\end{itemize}

These parameters were chosen in a way that they not only represent the ``quality'' of events (\emph{i.e.}, connection to geoeffectiveness) but also the occurrence rate of events. We also include the SOHO downtime as the parameter to account for possible lack of important CME data. Each of these parameters was ranked, \emph{i.e.}, associated with an ordinal 1\,--\,12, depending on the value it obtained for each month (see Table 5). The ranks were associated so that the highest rank (\emph{i.e.} lowest ordinal) roughly corresponds to highest geoeffectiveness.


\begin{table}
\caption{Values and ranks of solar parameters in a specific month for 1392 CME--flare pairs in the SOHO era. Definitions of solar parameters are given in the main text.
}
\label{table5}

\begin{tabular*}{0.99\textwidth}{cccccccc}
\hline
 & \multicolumn{7}{c}{Solar parameter value}\\
 Month, i & $N_{\mathrm{CME,i}}$ & $v_{\mathrm{avg,i}}$(km s$^{-1}$) & $N_{v,\mathrm{i}}$ & $N_{\mathrm{X,i}}$ & $N_{\mathrm{HALO,i}}$ & $r_{\mathrm{avg,i}}$ & $t_{\mathrm{SOHO,i}}(\mathrm{h})$\\
\hline
1	& 90	& 621	& 13	& 4	 & 18	& 0.6828	& 1456\\
2	& 69	& 543	& 6	  & 2	 & 11	& 0.7170	& 1740\\
3	& 99	& 552	& 9	  & 3	 & 11	& 0.7334	& 1232\\
4	& 137	& 659	& 21	& 9	 & 20	& 0.7058	& 999 \\
5	& 133	& 593	& 17	& 5	 & 24	& 0.7028	& 808 \\
6	& 130	& 604	& 15	& 4	 & 18	& 0.7610	& 3665\\
7	& 142	& 585	& 19	& 14 & 23	& 0.7583	& 825 \\
8	& 133	& 559	& 15	& 10 & 13	& 0.7276	& 839 \\
9	& 90	& 672	& 17	& 4	 & 14	& 0.6460	& 1081\\
10&	131	& 586	& 18	& 10 & 16	& 0.6648	& 500 \\
11&	139	& 682	& 27	& 16 & 36	& 0.6441	& 1680\\
12&	99	& 573	& 11	& 5	 & 21	& 0.7249	& 1745\\
\hline
\hline
 & \multicolumn{7}{c}{Solar parameter rank}\\
  Month, i & $N_{\mathrm{CME,i}}$ & $v_{\mathrm{avg,i}}$ & $N_{v,\mathrm{i}}$ & $N_{\mathrm{X,i}}$ & $N_{\mathrm{HALO,i}}$ & $r_{\mathrm{avg,i}}$ & $t_{\mathrm{SOHO,i}}$\\
\hline
1	& 2	& 9	& 4	& 3	& 6	& 9	& 8\\
2	& 1	& 1	& 1	& 1	& 1	& 6	& 10\\
3	& 4	& 2	& 2	& 2	& 2	& 3	& 7\\
4	& 10&	10&	11&	8	& 8	& 7	& 5\\
5	& 8	& 7	& 7	& 6	& 11&	8	& 2\\
6	& 6	& 8	& 5	& 4	& 7	& 1	& 12\\
7	& 12&	5	& 10&	11&	10&	2	& 3\\
8	& 9	& 3	& 6	& 9	& 3	& 4	& 4\\
9	& 3	& 11& 8	& 5	& 4	& 11&	6\\
10&	7	& 6	& 9	&10	& 5	& 10&	1\\
11&	11&	12&	12&	12&	12&	12&	9\\
12&	5	& 4	& 3	& 7	& 9	& 5	& 11\\
\hline
\end{tabular*}

\end{table}


The monthly CME--flare activity parameter in month ``i'' (i=1,2,...,12), $A_{\mathrm{i}}$, is defined as follows:


\begin{equation}
A_\mathrm{i}=N_{{\mathrm{CME}},i}+v_{{\mathrm{avg}},i}+N_{{v},i}+N_{{\mathrm{X}},i}+N_{{\mathrm{HALO}},i}+r_{{\mathrm{avg}},i}+t_{{\mathrm{SOHO}},i}.
\label{eq1}
\end{equation}


The monthly CME/flare activity parameter was normalized to obtain the relative monthly CME/flare activity parameter (in the month i=1,2,...12), $A_{\mathrm{rel,i}}$:


\begin{equation}
A_{\mathrm{rel,i}}=\frac{A_{{\mathrm{i}}}}{\sum_{i=1}^{12} A_{{\mathrm{i}}}}  .
\label{eq2}
\end{equation}

\begin{figure}
   \centerline{\includegraphics[width=0.5\textwidth]{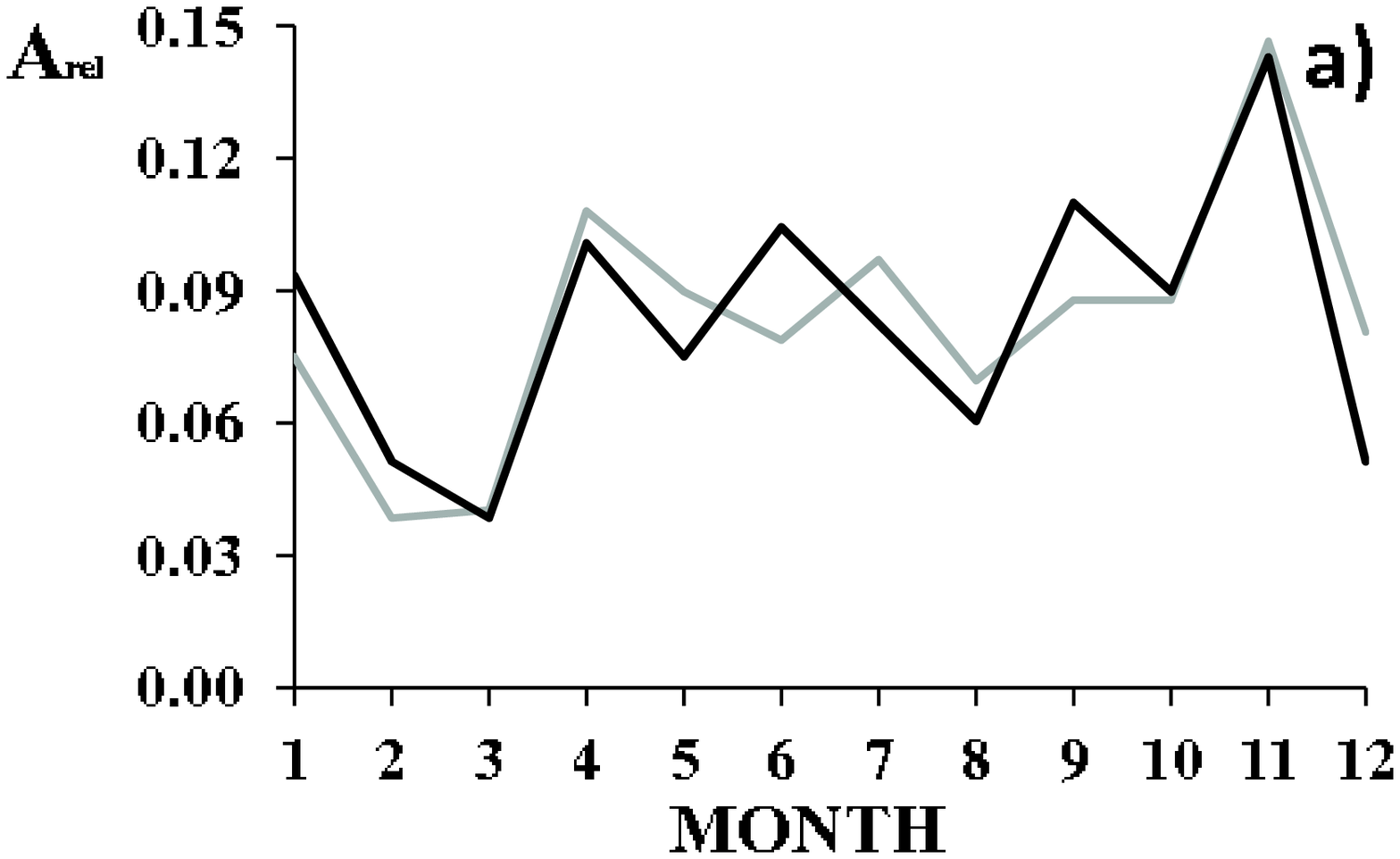}
               \includegraphics[width=0.5\textwidth]{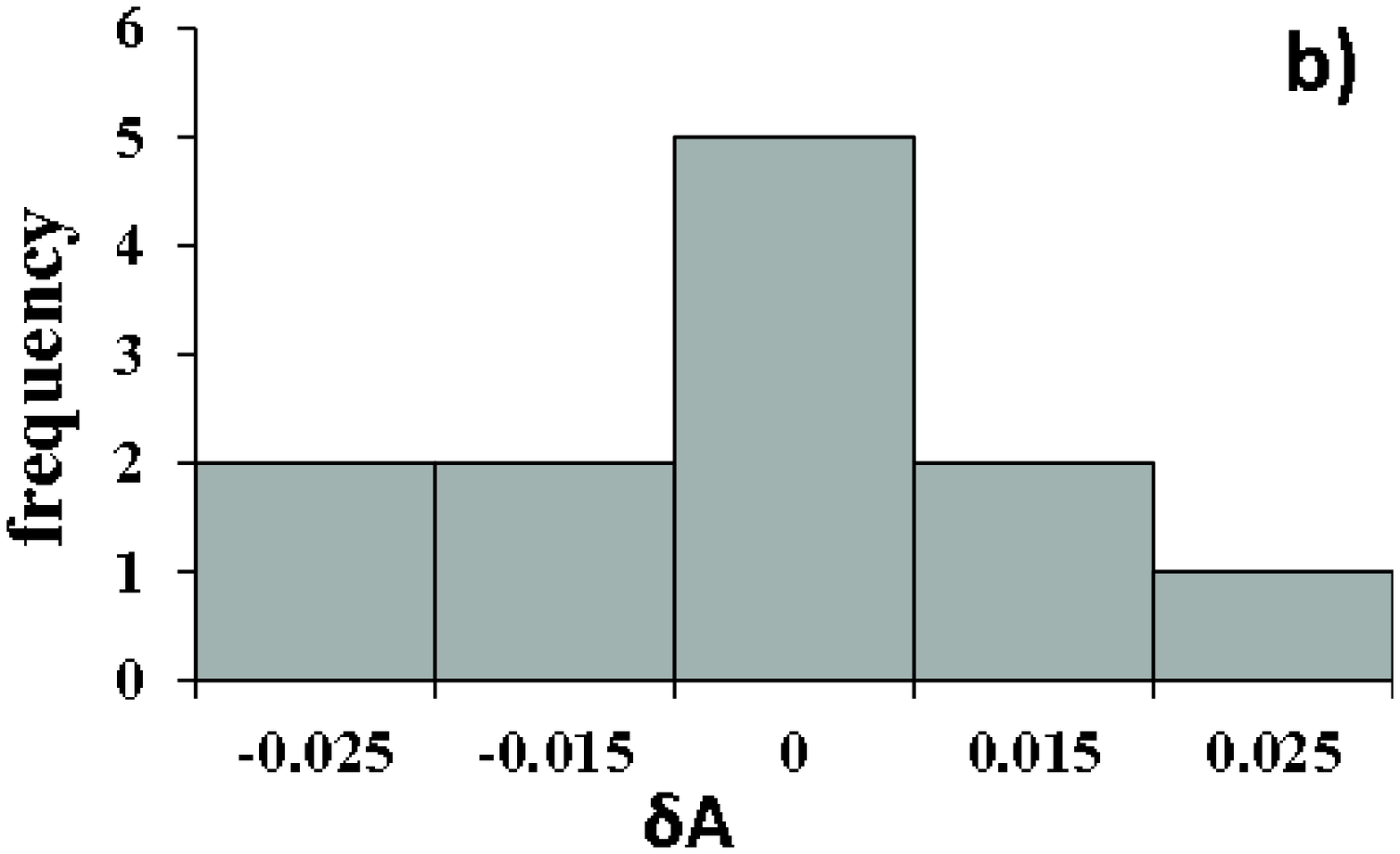}
               }
   \vspace{0.03\textwidth}
   \centerline{\includegraphics[width=0.5\textwidth]{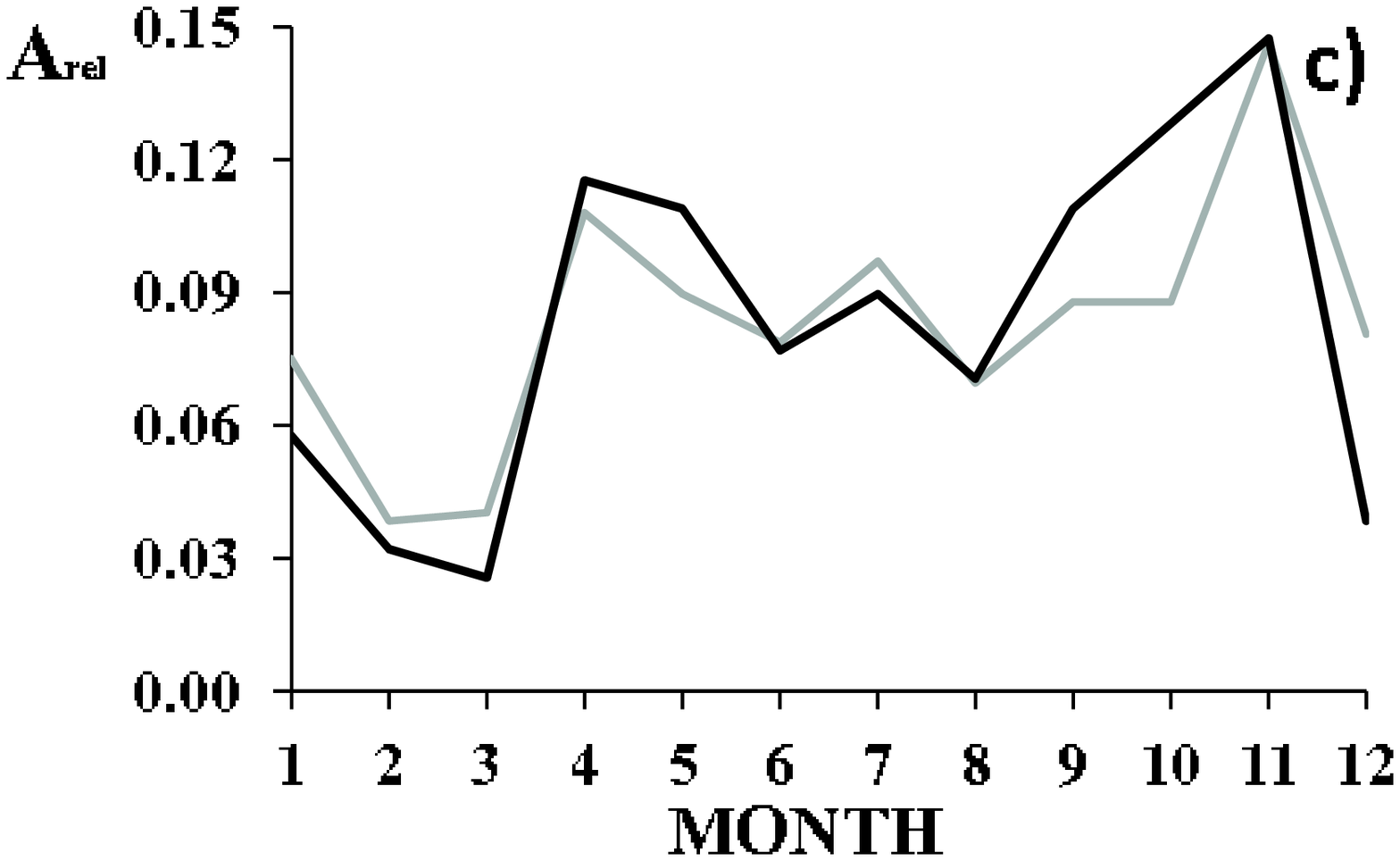}
               \includegraphics[width=0.5\textwidth]{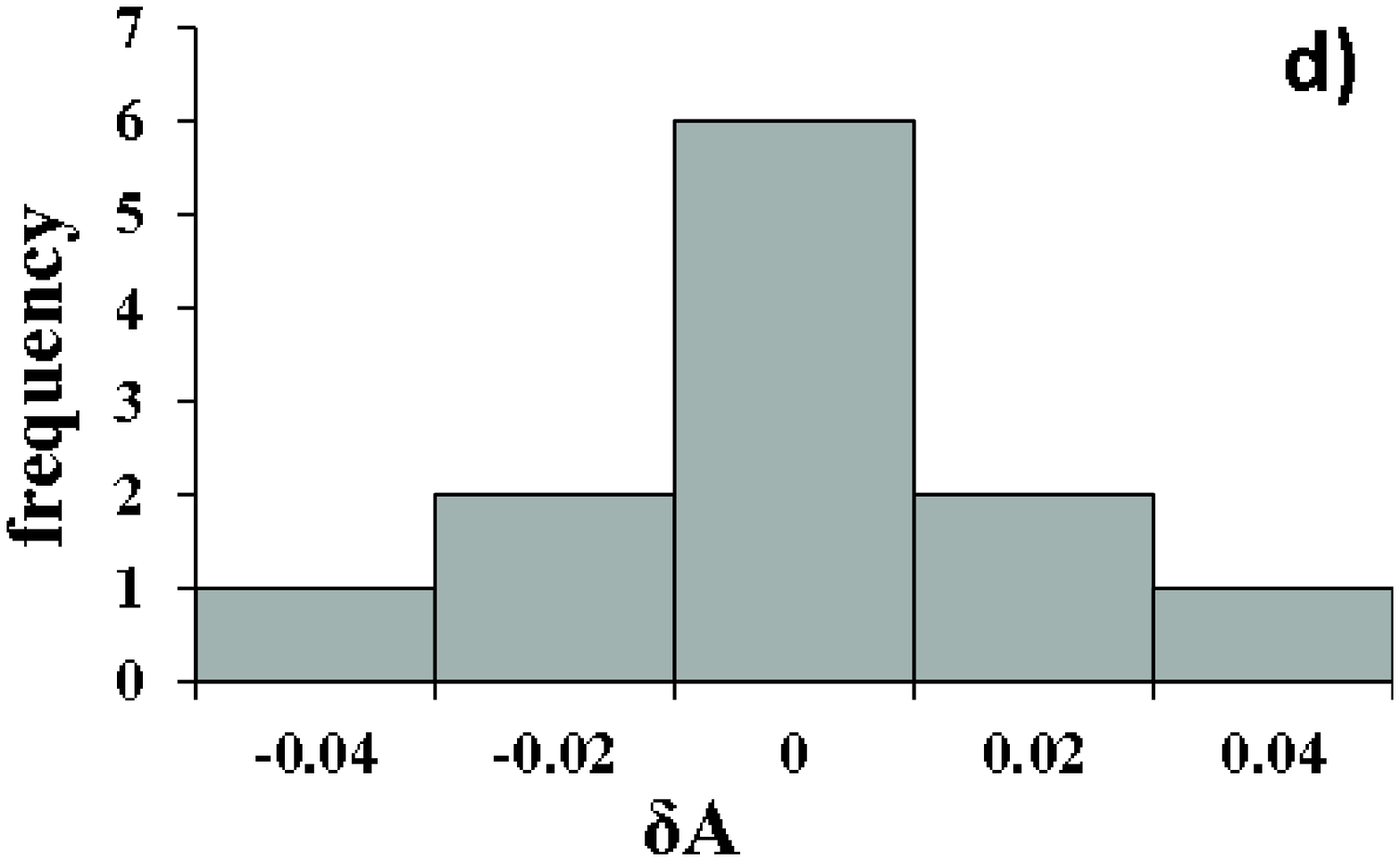}
               }
	 \caption{
	a) Relative monthly CME/flare activity parameter ($A_{\mathrm{rel}}$) derived from the 1392 CME--flare pairs in the SOHO era (gray) and from the 211 $Dst$-associated CME--flare pairs (black); b) distribution of variations, $\delta A$, between curves displayed in Figure 10a. The mean value is $3.5\cdot 10^{-18}$ and standard deviation is $0.017$; 	c) relative monthly CME/flare activity parameter ($A_{\mathrm{rel}}$) derived from the 1392 CME--flare pairs in the SOHO era (gray) and relative monthly $Dst$ activity parameter obtained from the 211 $Dst$-associated CME--flare pairs (black); d) distribution of variations, $\delta A$,  between curves shown in Figure 10c. Mean value is $2.3\cdot 10^{-18}$ and standard deviation is $0.021$.
        }
   \label{fig10}
\end{figure}

The relative monthly CME/flare activity parameter was obtained both for the 1392 CME--flare pairs in the SOHO era and the 211 $Dst$-associated CME--flare pairs used for the statistical analysis throughout Sections \ref{speed}--\ref{combinations}. The two are compared in Figure \ref{fig10}a, where it can be seen that they have a very similar trend. Furthermore, we examined their variations, $\delta A$, \emph{i.e.}, the residuals when the two are subtracted (Figure \ref{fig10}b). Variations of the two curves, $\delta A$, are distributed in a normal-like distribution centred around $\approx 0$, therefore, we conclude that the two curves indeed have the same trend. This means that the CME/flare activity of our sample (211 $Dst$-associated CMEs) is a good representative of the population (1392 CME--flare pairs in the SOHO era).


Then, using 211 $Dst$-associated CME--flare pairs, monthly $Dst$ activity parameter was obtained using the following parameters:
\begin{itemize}
\item number of events in month ``i'' of the year, $N_{\mathrm{i}}$
\item average $|Dst|$ values for month ``i'', $|Dst|_{\mathrm{avg,i}}$
\end{itemize}
Again, both qualitative and quantitative aspects were taken into account. The number of events, $N_{\mathrm{i}}$, was ranked by the value in each month from 1 to 12, where 1 was associated to the month where there was the smallest number of events, and 12 was associated to the month where there was the largest number of events. Similarly, the average values, $|Dst|_{\mathrm{avg,i}}$, were also ranked, where 1 was associated to the month where $|Dst|_{\mathrm{avg,i}}$ assumes the lowest value and 12 was associated to the month where $|Dst|_{\mathrm{avg,i}}$ assumes the highest value. Using $Dst$ parameter ranks, monthly $Dst$ activity parameter, $A_\mathrm{i}$, for specific month i (i=1,2,...,12) was obtained:


\begin{equation}
A_\mathrm{i}=N_{\mathrm{i}}+|Dst|_{\mathrm{avg,i}}.
\label{eq3}
\end{equation}


The monthly $Dst$ activity parameter was normalized to obtain the relative monthly $Dst$ activity parameter (in the month i=1,2,...12) using Equation (\ref{eq2}). We compare it to the relative monthly CME/flare activity parameter in Figure \ref{fig10}c, and again we find that the two have a very similar trend. The variations, $\delta A$, of the two curves (Figure \ref{fig10}d) are distributed in a normal-like distribution centred around $\approx 0$, similarly as in Figure \ref{fig10}b. Therefore, we reach the same conclusion, that the two curves have the same trend. This means that the monthly $Dst$ activity, derived from our sample reflects the monthly CME/flare activity in the SOHO era.


\section{Empirical statistical model for predicting geomagnetic storm levels}
				\label{model}

In Sections \ref{speed}--\ref{flare} the key solar parameters were examined and were found to be related to the $|Dst|$ levels. Namely, the  distribution of $|Dst|$ amplitudes was found to change with CME speed, $v$, CME/flare source region location (distance from the centre of the solar disc, $r$), CME apparent width, $w$, flare class, $f$, and CME--CME interaction level, $i$. These relationships are quantified and can be used to predict the probability for the $|Dst|$ levels based on the remote solar observations. We use the results of the statistical analysis to construct $|Dst|$ distributions, depending on the specific solar parameter. For this purpose we use the geometric distribution:


\begin{equation}
P(X=k)= p\cdot(1-p)^{k-1}\,,
\label{eq4}
\end{equation}


\noindent
where $P(X=k)$ is the probability that the $k^{\mathrm{th}}$ trial is a first success and $p$ is the probability of the success in each trial ($k=1,2,3,...$ is the number of trials). The geometric distribution is suitable for several reasons. It is a rapidly descending discrete distribution, like the $|Dst|$ distribution we observe, and therefore is restricted to a small number of bins. In addition, it can be simply mathematically reconstructed based on the distribution mean ($p=m^{-1}$, where $m$ is the distribution mean). The association between the number of trials and the $|Dst|$ bins was made in the following way:
\begin{itemize}
\item $k=1 \longleftrightarrow |Dst|<100\,\mathrm{nT}$;
\item $k=2 \longleftrightarrow 100\,\mathrm{nT}<|Dst|<200\,\mathrm{nT}$;
\item $k=3 \longleftrightarrow 200\,\mathrm{nT}<|Dst|<300\,\mathrm{nT}$;
\item $k=4 \longleftrightarrow |Dst|>300\,\mathrm{nT}$.
\end{itemize}
In this way, the conversion of the $|Dst|$ distribution mean, $m_{DST}$, into the geometric distribution mean, $m_{GD}$, can be done in a simple way ($m_{GD}=1+m_{DST}\,\mathrm{[nT]}/100$, for details see Appendix).

It was shown in Figures \ref{fig5} and \ref{fig9} that the trend of the change in the $|Dst|$ distribution mean, $m_{DST}$, with a specific solar parameter can be fitted by a corresponding function. Namely, $m_{DST}(v)$ was fitted with a linear function, $m_{DST}(r)$ and $m_{DST}(i)$ with a power-law function, and $m_{DST}(w)$ and $m_{DST}(f)$ with a quadratic function. Therefore, based on the $|Dst|$-solar parameter relationships found, a corresponding geometric distribution can be obtained. We note that the CME speed, $v$, and the CME source distance from the centre of the solar disc, $r$, are regarded as continuous parameters in the ranges of $v\ge400$ $\mathrm{km\,s^{-1}}$ and $0<r\le1$, respectively. The range of $v$ is determined based on the limitations of the sample, whereas the range of $r$ is restricted by the mathematical singularity of the power-law function ($r=0$) and the physical boundary ($r=1$, \emph{i.e.}, the solar limb). The other three solar parameters, the apparent width, $w$, the associated flare class, $f$, and the level of interaction, $i$, are considered as discrete parameters associated with integers 1--3 and 1--4, respectively (1 meaning least significant, \emph{i.e.}, the lowest interaction parameter, width, and flare class).


  \begin{figure}
   \centerline{\includegraphics[width=0.43\textwidth]{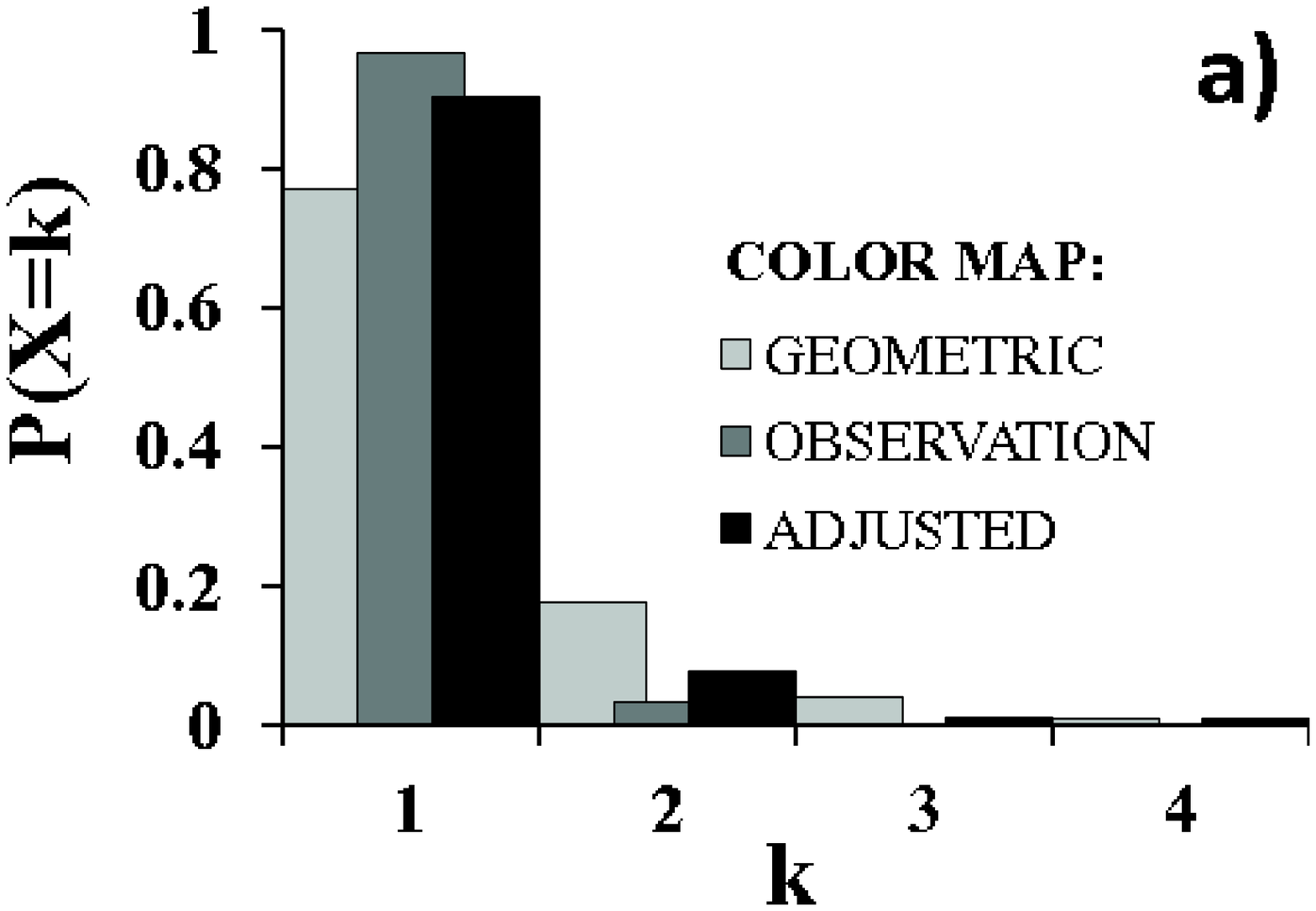}
               \includegraphics[width=0.43\textwidth]{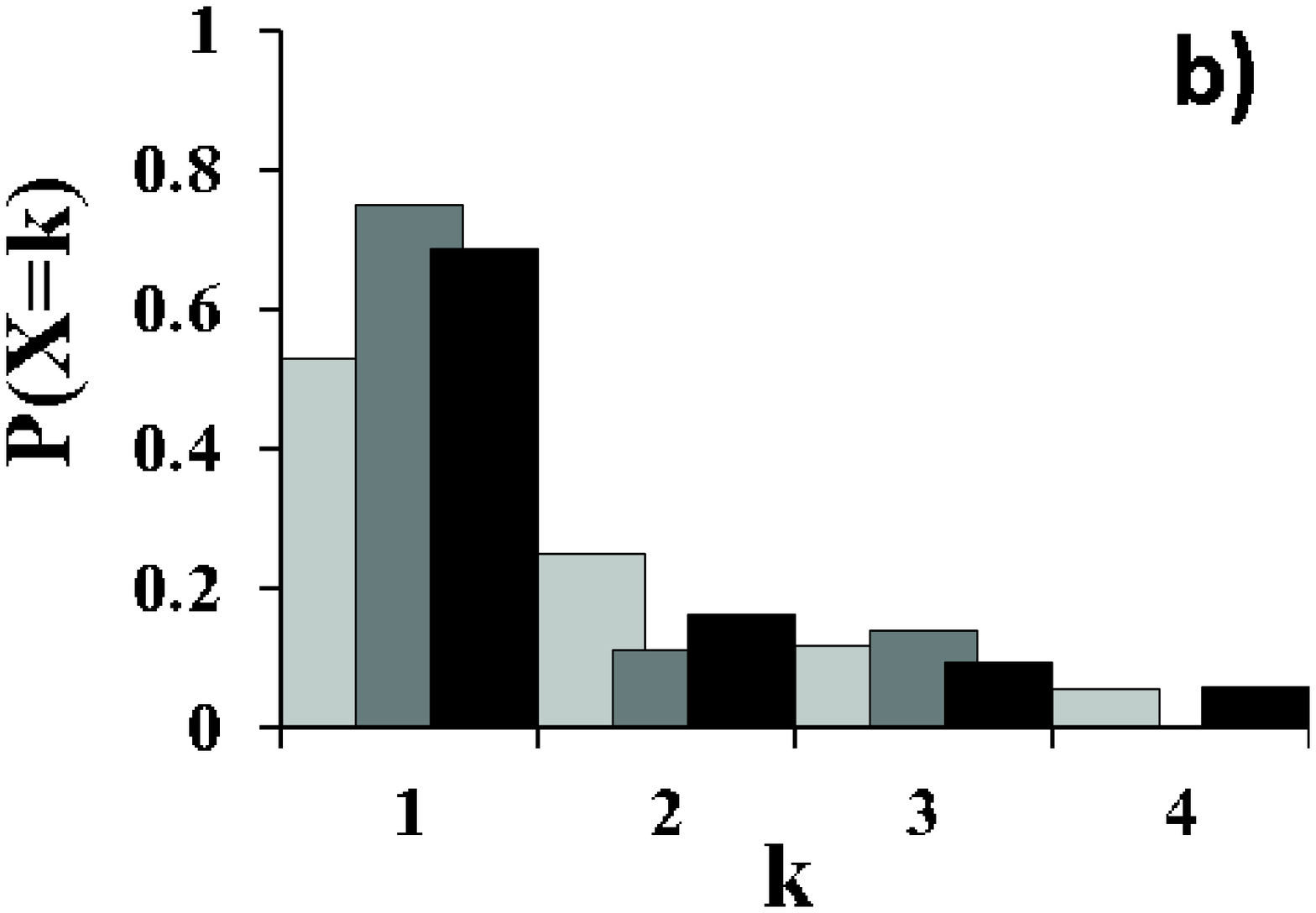}
               }
   \vspace{0.02\textwidth}
   \centerline{\includegraphics[width=0.43\textwidth]{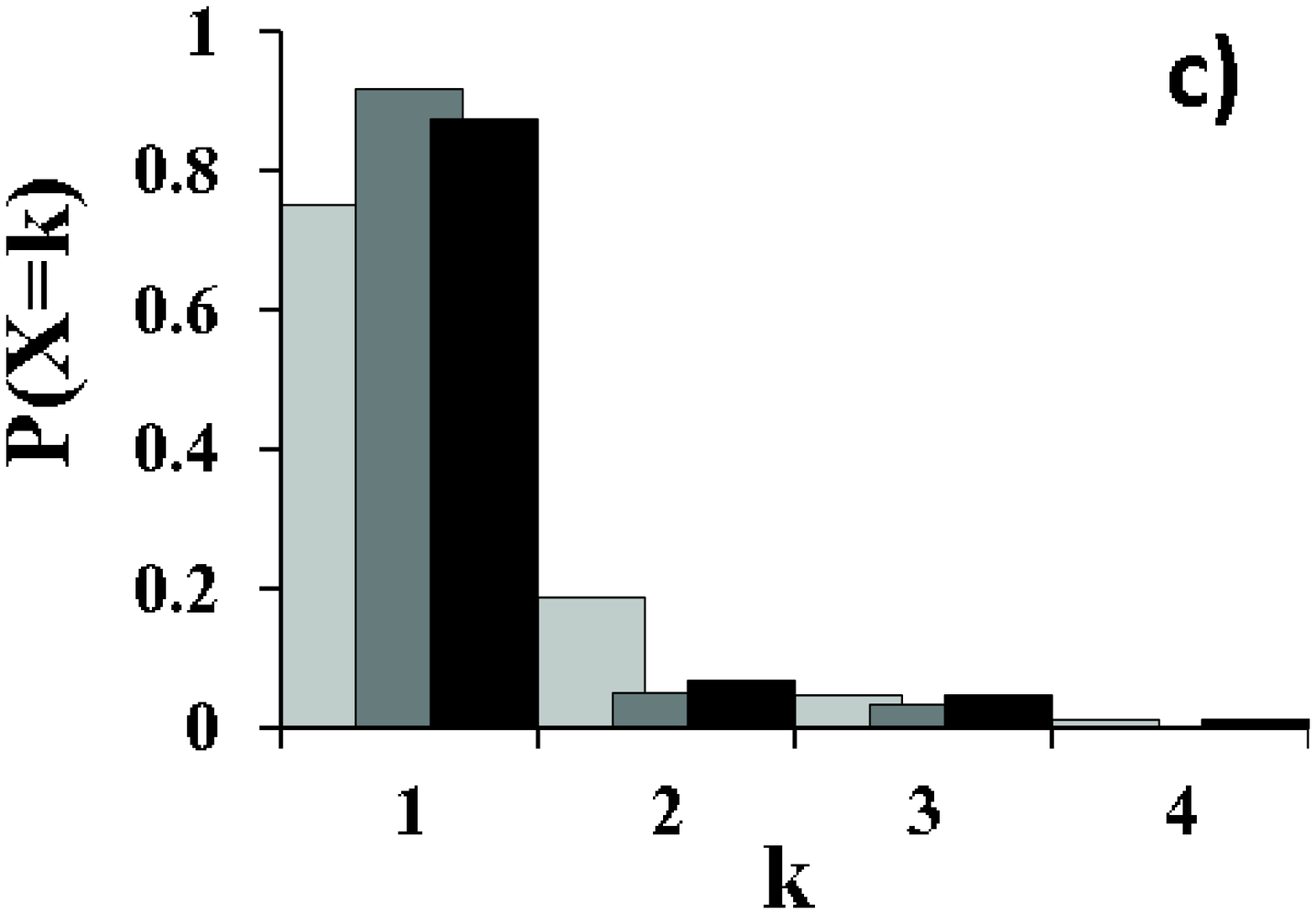}
               \includegraphics[width=0.43\textwidth]{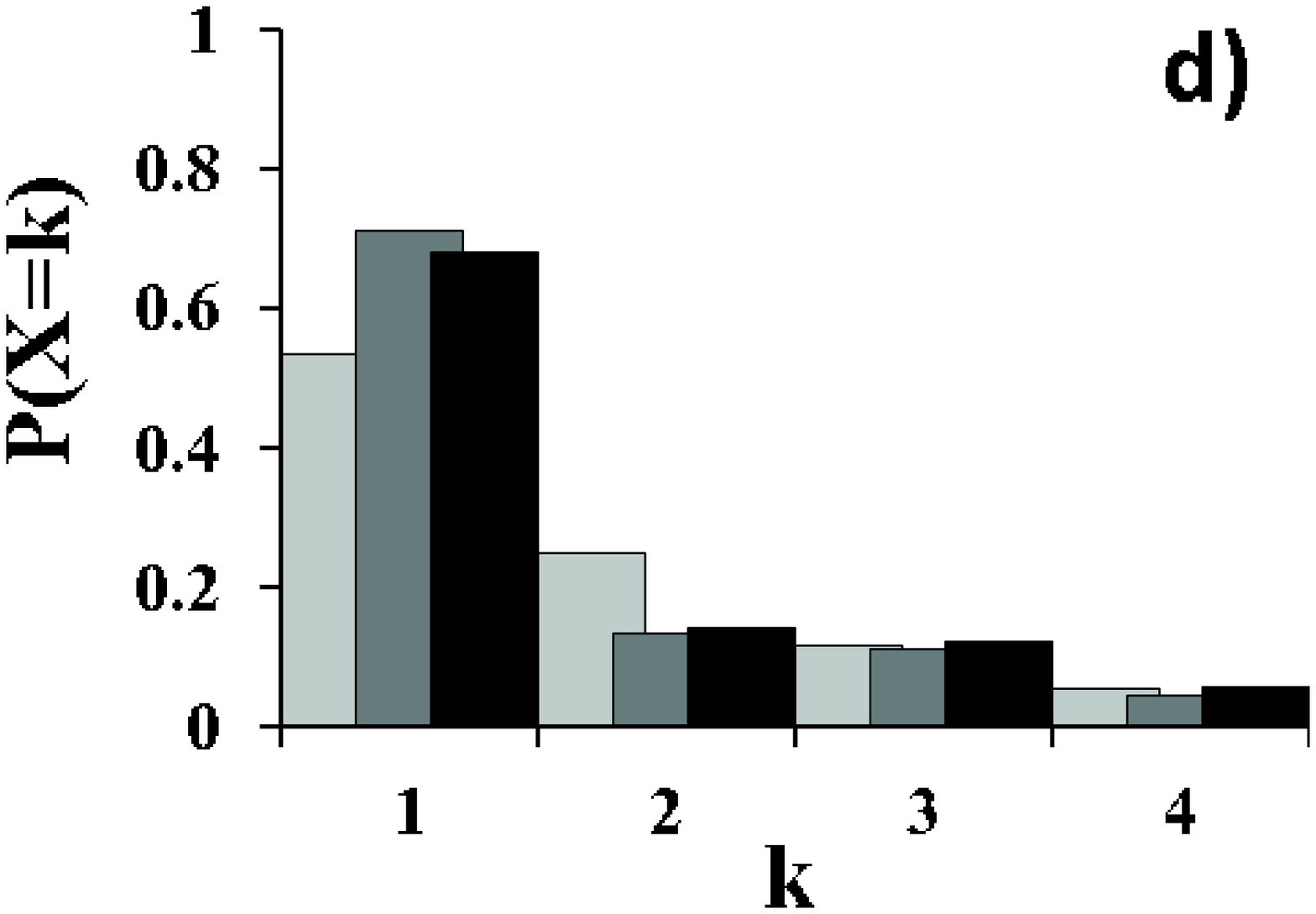}
               }
   \vspace{0.02\textwidth}
   \centerline{\includegraphics[width=0.43\textwidth]{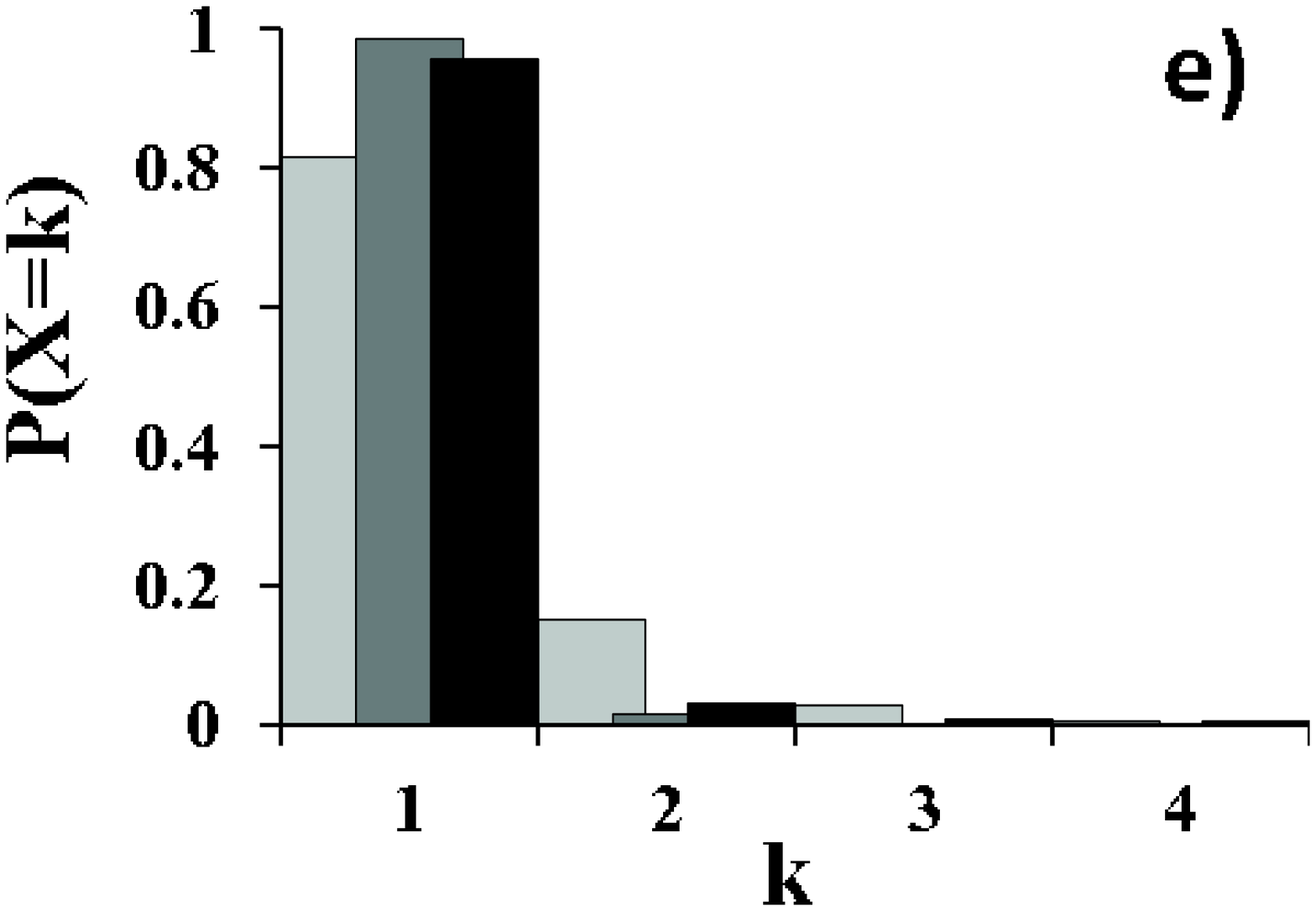}
               \includegraphics[width=0.43\textwidth]{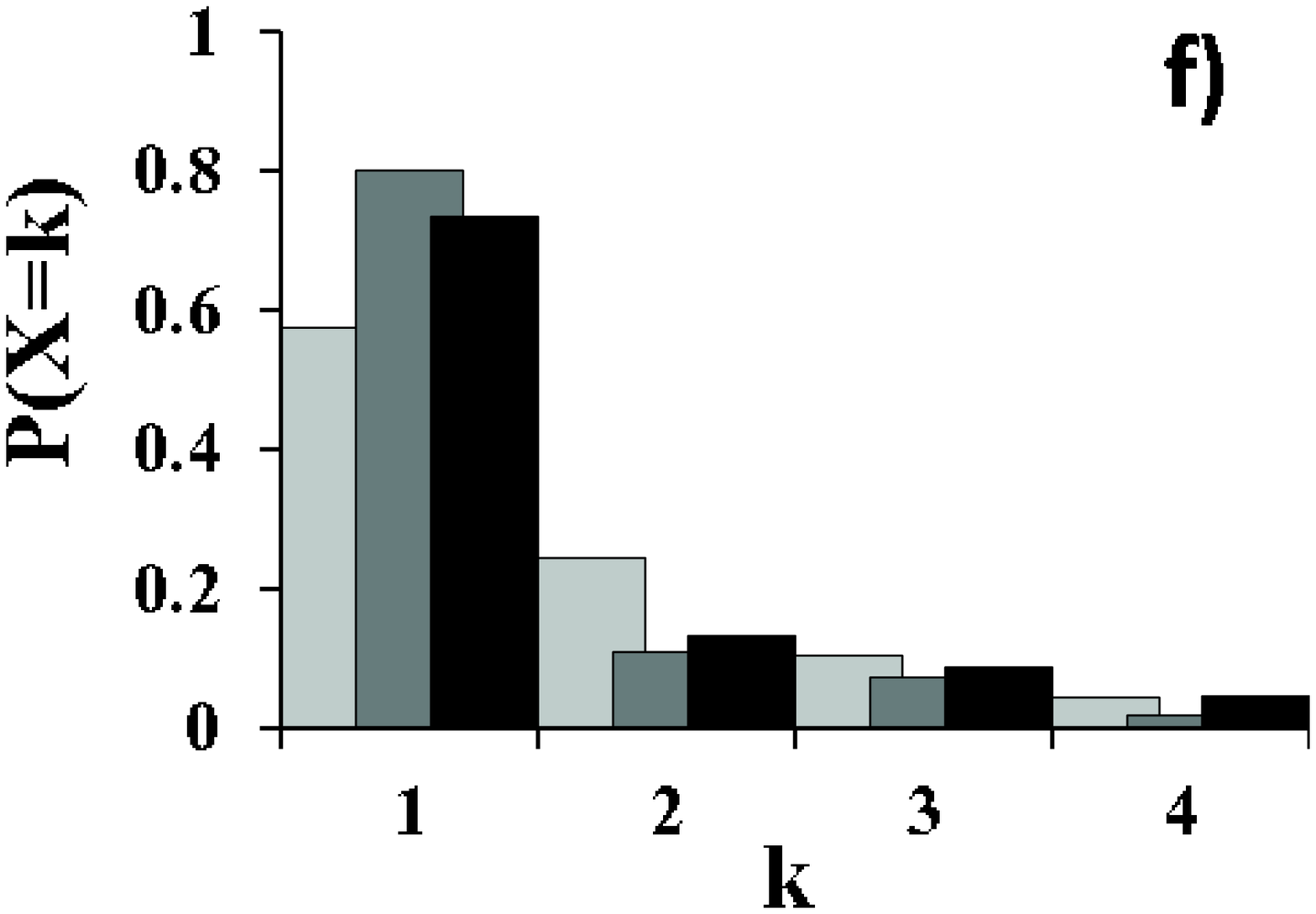}
               }
   \vspace{0.02\textwidth}
   \centerline{\includegraphics[width=0.43\textwidth]{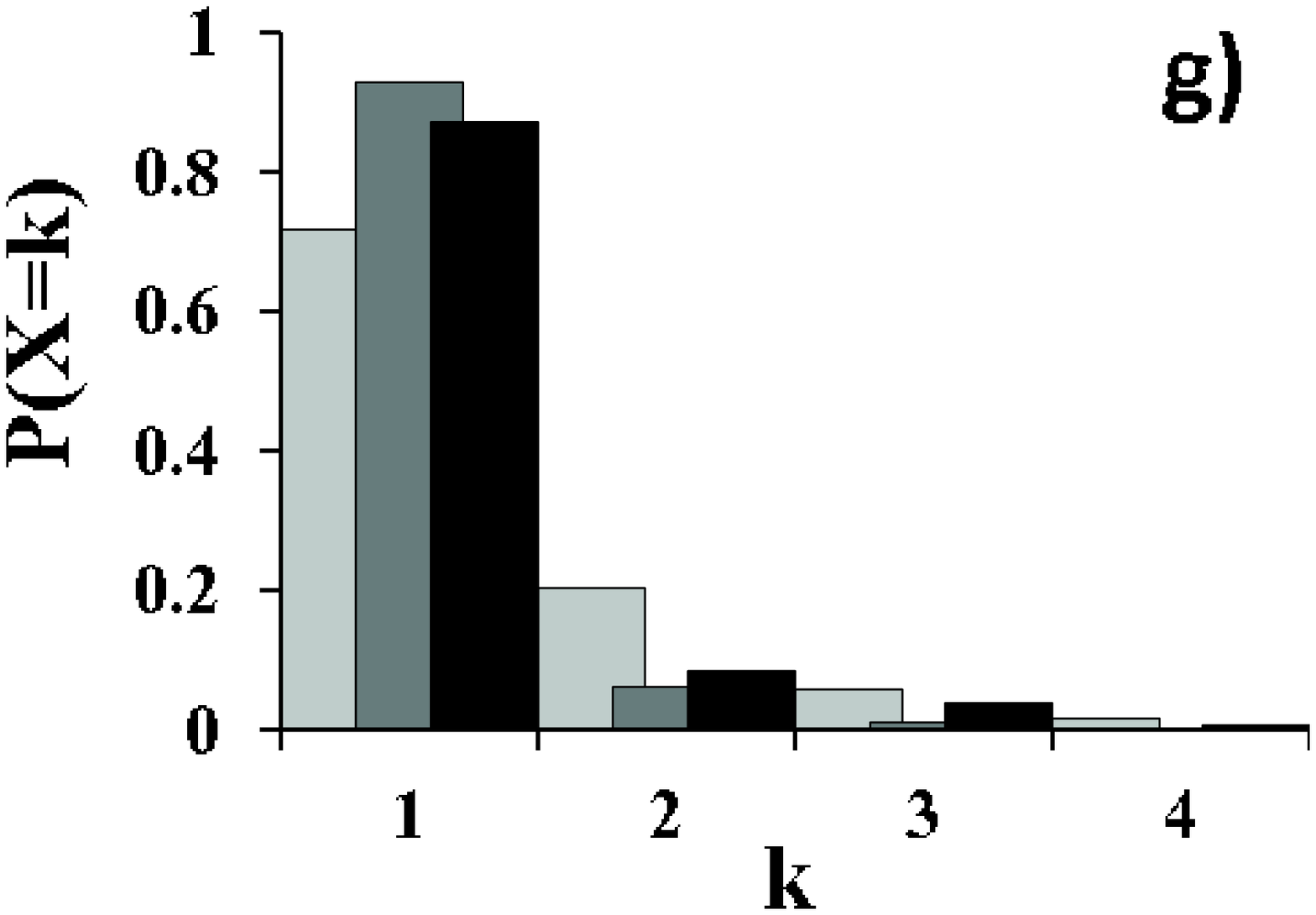}
               \includegraphics[width=0.43\textwidth]{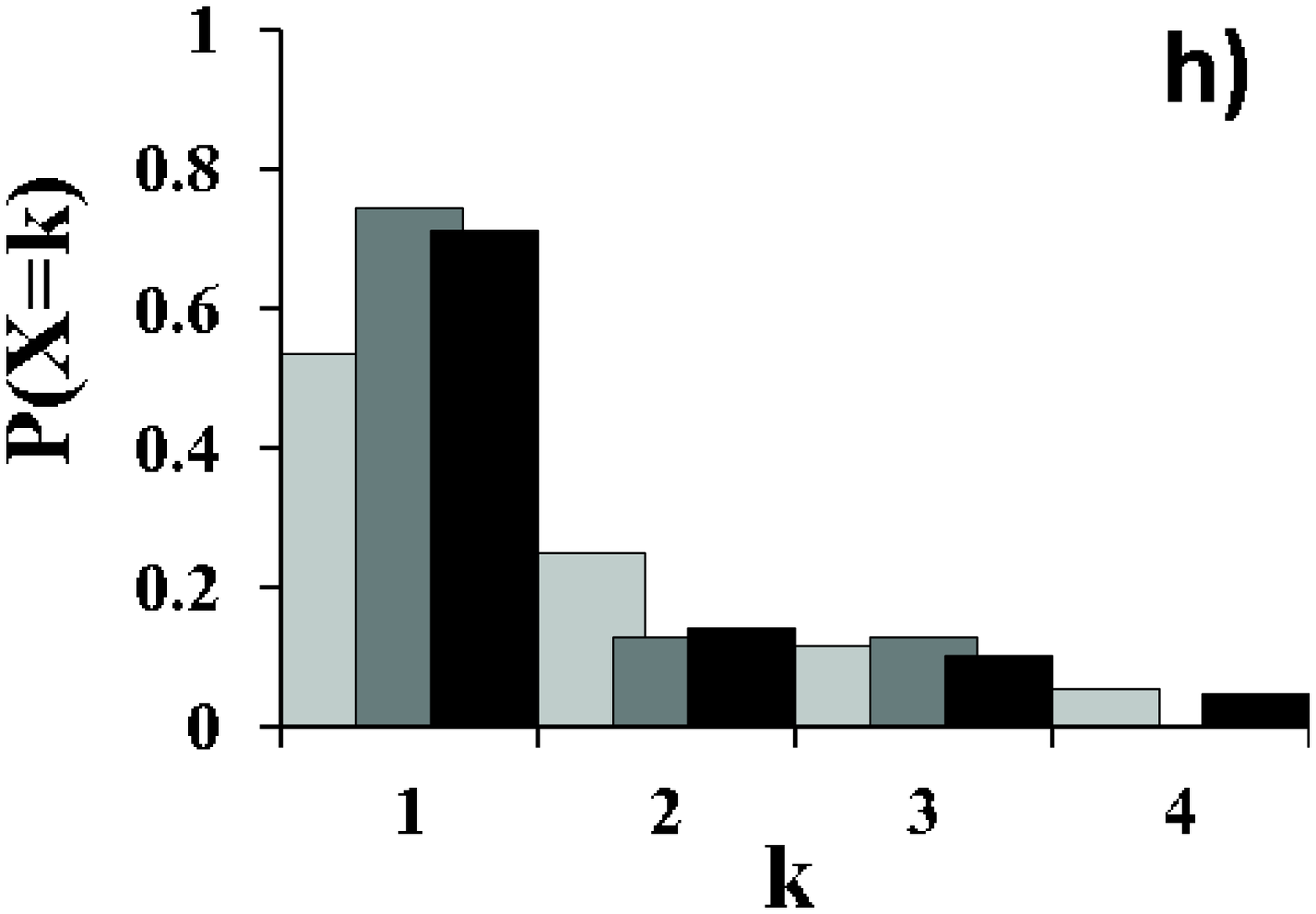}
               }
   \vspace{0.02\textwidth}
   \centerline{\includegraphics[width=0.43\textwidth]{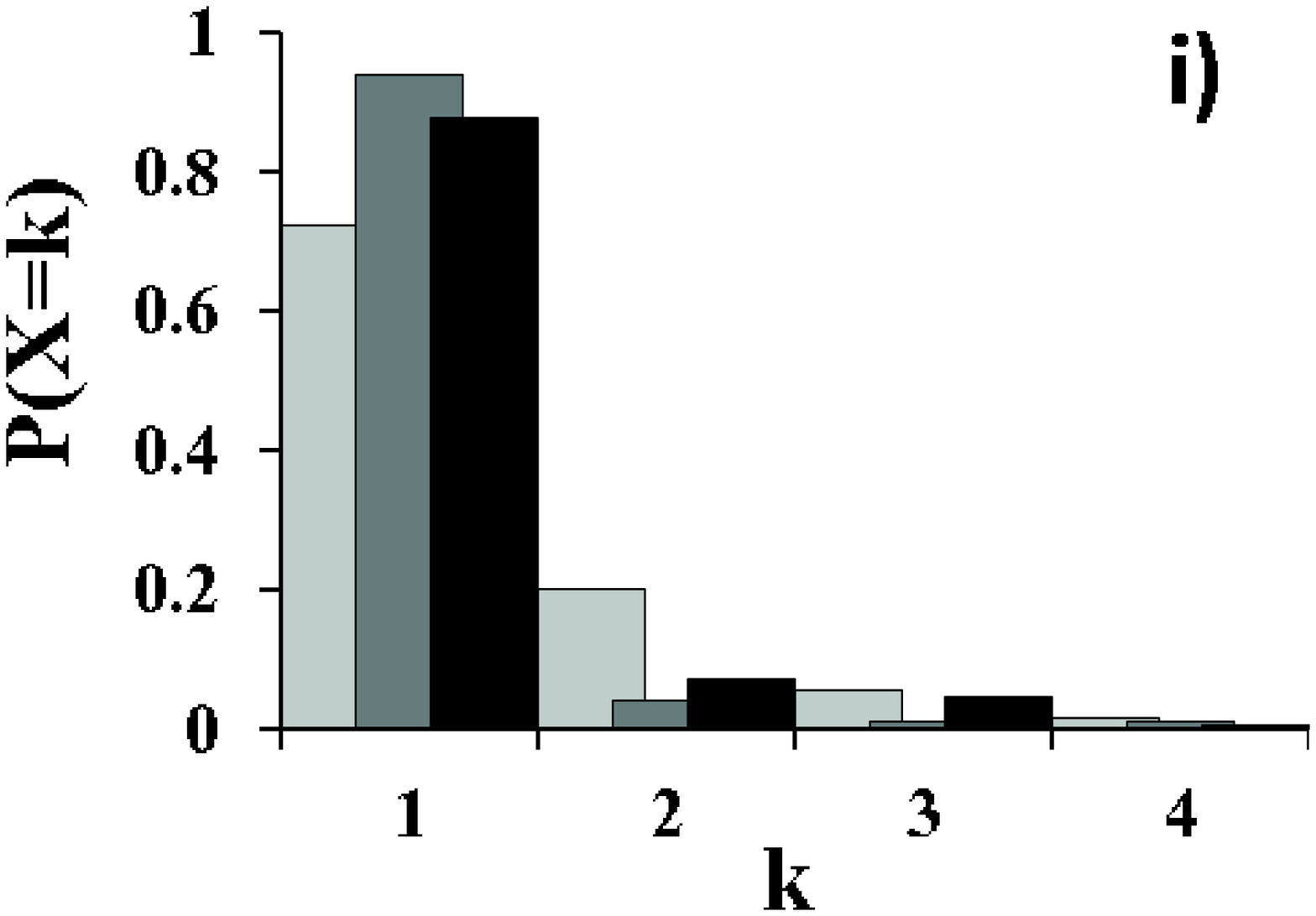}
               \includegraphics[width=0.43\textwidth]{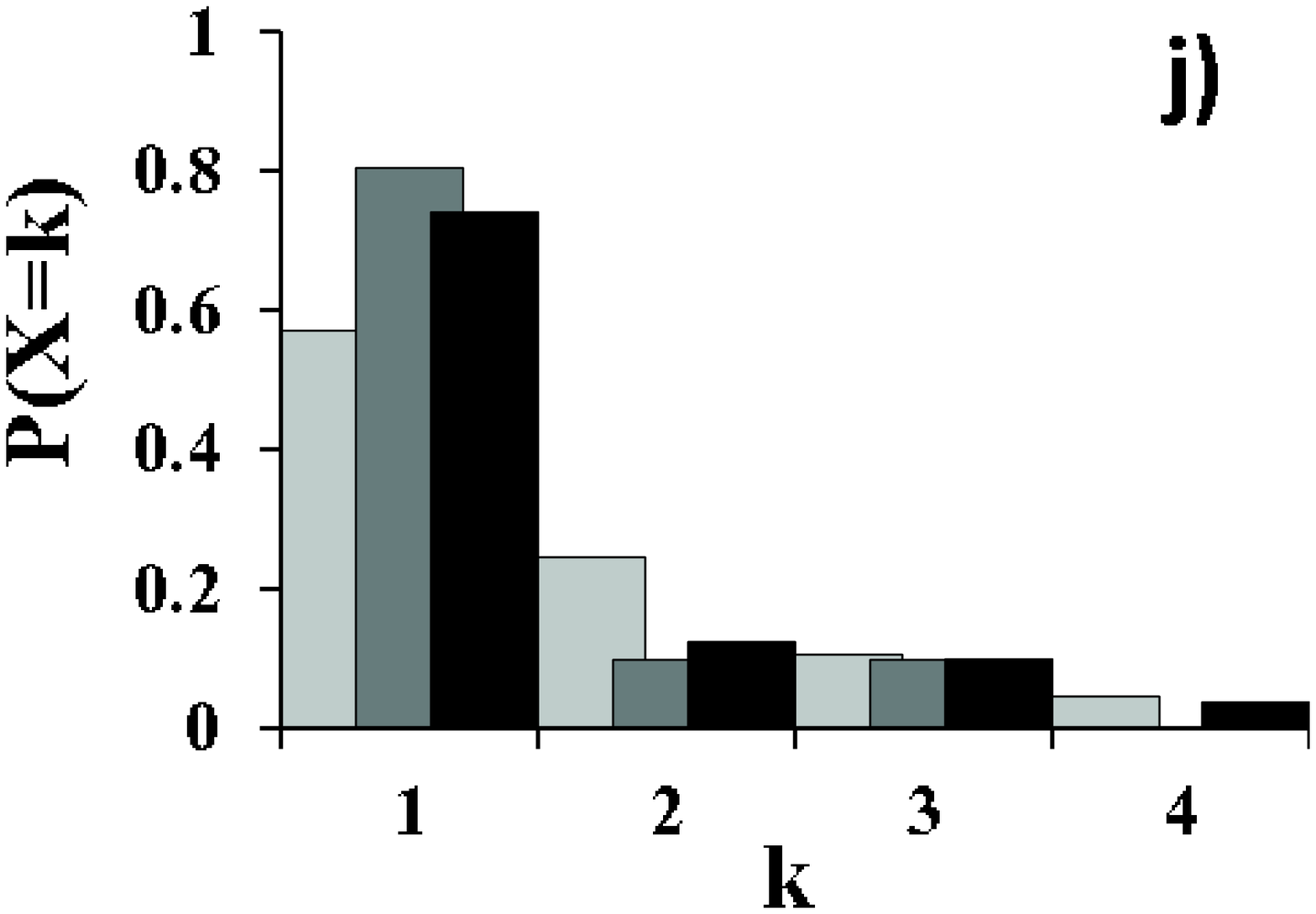}
               }
	 \caption{Geometric, observational and adjusted distributions for different ranges/values of key solar parameters:
	 					a) for the CME speed, 400 $\mathrm{km\,s^{-1}}$$<v<$600 $\mathrm{km\,s^{-1}}$;
	 					b) for the CME speed, $v>$1700 $\mathrm{km\,s^{-1}}$;
	 					c) for the CME source distance from the centre of the solar disc, $r>0.8$;
	 					d) for the CME source distance from the centre of the solar disc, $r<0.4$;
	 					e) for non-halo CMEs, $w<120^{\circ}$ ($w=1$);
	 					f) for halo CMEs, $w=360^{\circ}$ ($w=3$);
	 					g) for the associated flares of B\&C-class, $f=1$;
	 					h) for the associated flares of X-class, $f=3$;
	 					i) for the lowest interaction parameter, "S" (SINGLE), $i=1$;
	 					j) for the highest interaction parameter, "T" (TRAIN), $i=4$.
        }
   \label{fig11}
\end{figure}


The mathematically obtained geometric distribution underestimates the observed $|Dst|$ distribution for $k=1$, whereas it is overestimated for $k=2$. This can be seen in Figure \ref{fig11}, where the two are compared for a number of relationships. Therefore, new ``adjusted'' distributions for each of the key solar parameters were obtained by adding a specific constant to each bin to best fit the observed distribution in all the ranges, \emph{i.e.}, for all the values of key solar parameters. The constants are added so that the new distribution is also normalized (Table \ref{table6}) and are different for different $|Dst|$ bins, \emph{i.e.}, $k$ and different key solar parameters. It can be seen in Figure \ref{fig11} that the empirical distribution still slightly underestimates the observed $|Dst|$ distribution for $k=1$. However, the agreement between the two distributions for higher values of $k$ is substantially improved. For detailed mathematical formulations and procedures used to obtain probability distributions see Appendix.


\begin{table}
\caption{The constants added to geometric distribution to obtain adjusted distribution, for different $|Dst|$ bins, $k$, and different solar parameters.
				}
\label{table6}

\begin{tabular*}{0.99\textwidth}{ccccccccccc}
 		$k$	& &	$v$\tabnote{solar parameters: CME speed, $v$, CME source position distance from the centre of the solar disc, $r$, apparent width, $w$, associated flare class, $f$, and interaction parameter, $i$}		& &	$r$		& &	$w$		& &	$f$		& &	$i$\\
\hline
1	& &	0.13	& &	0.12	& &	0.14	& &	0.15	& &	0.15\\
2	& &	-0.10	& &	-0.12	& &	-0.12	& &	-0.12	& &	-0.13\\	
3	& &	-0.03	& &	0			& &	-0.02	& &	-0.02	& &	-0.01\\
4	& &	0			& &	0			& &	0			& &	-0.01	& &	-0.01\\
\end{tabular*}

\end{table}

The obtained empirical distributions are treated as probability distributions. For a specific solar parameter they provide the information on the probability for associating it a specific value of $k$, \emph{i.e.}, $|Dst|$ level. To combine the effect of the key solar parameters, \emph{i.e.}, to obtain a joint probability distribution, the key parameters were treated as mutually non-exclusive events, for which the following formula applies:


\begin{equation}
P(\mathrm{A}\cup \mathrm{B})=P(\mathrm{A})+P(\mathrm{B})-P(\mathrm{A}\cap \mathrm{B}).
\label{eq5}
\end{equation}


In general, $P(\mathrm{X})$, where $X=A,B$, is the (marginal) probability of the event X, $P(\mathrm{A}\cup \mathrm{B})$ is the probability that either event A or event B or both occur, and $P(\mathrm{A}\cap \mathrm{B})$ is their joint probability. Specifically, in our case $P(X)=P(X=k)$ is the probability that for a specific key solar parameter $X$ a specific $|Dst|$ level, $k$, will be observed. It should be noted that since a particular key parameters are tied to the same event, they should be regarded as mutually non-exclusive. In general, joint probability is given by:


\begin{equation}
P(\mathrm{A}\cap \mathrm{B})=P(\mathrm{A|B})\cdot P(\mathrm{B}).
\label{eq6}
\end{equation}


Where $P(\mathrm{A|B})$ is the conditional probability, \emph{i.e.} the probability for event A given that the event B occured. Assuming that the events are independent of each other, combining Equations (\ref{eq5}) and (\ref{eq6}) one gets:


\begin{equation}
P(\mathrm{A}\cup \mathrm{B})=P(\mathrm{A})+P(\mathrm{B})-P(\mathrm{A})\cdot P(\mathrm{B}).
\label{eq7}
\end{equation}


This assumption is not fully valid, due to the fact that not all key solar parameters are independent of each other, (\emph{e.g.}, CME speed and flare class, see \opencite{moon02}, \opencite{moon03}, \opencite{vrsnak05}, \opencite{maricic07}). Since the constructed geometric distribution directly depends on the key solar parameter for which it is constructed, the connection between two solar parameters leads to a relationship between two constructed geometric distributions. Moreover, positively correlated parameters will lead to conditional probability greater than the marginal probability for k=2,3,4 and \emph{vice versa} for k=1. Consequently, the assumption of independence redefines parameter space in a way that it will at worst underestimate the joint probability $P(\mathrm{A}\cap \mathrm{B})$, \emph{i.e.}, overestimate the probability $P(\mathrm{A}\cup \mathrm{B})$ for k=2,3,4 and \emph{vice versa} for k=1. Therefore, the constructed probability distribution will (slightly) overestimate geoeffectiveness, increasing to some extent the number of false alarms.

Finally, the probability of observing the $|Dst|$ value in a specific bin for a set of solar key parameters is then given by the formula derived from Equation (\ref{eq7}):


\begin{eqnarray}
P(|Dst|=k) = \sum_{\alpha}{P_{\alpha}}-\sum_{\alpha \ne \beta}{P_{\alpha}\cdot P_{\beta}}+\sum_{\alpha\ne \beta \ne \gamma}{P_{\alpha}\cdot P_{\beta}\cdot P_{\gamma}} - \nonumber \\
 -\sum_{\alpha \ne \beta \ne \gamma \ne \delta}{P_{\alpha}\cdot P_{\beta}\cdot P_{\gamma}\cdot P_{\delta}}+\sum_{\alpha \ne \beta \ne \gamma \ne \delta \ne \epsilon}{P_{\alpha}\cdot P_{\beta}\cdot P_{\gamma}\cdot P_{\delta} \cdot P_{\epsilon}}\,,
\label{eq8}
\end{eqnarray}

\noindent
where $P_{\alpha}$=$P(\alpha)$ represents the probability of a $|Dst|$ level $k$ for a specific solar key parameter $\alpha$ (CME speed, $v$, CME/flare source position distance from the centre of the solar disc, $r$, CME apparent width, $w$, flare class, $f$, and interaction parameter, $i$).


  \begin{figure}
   \centerline{\includegraphics[width=0.33\textwidth]{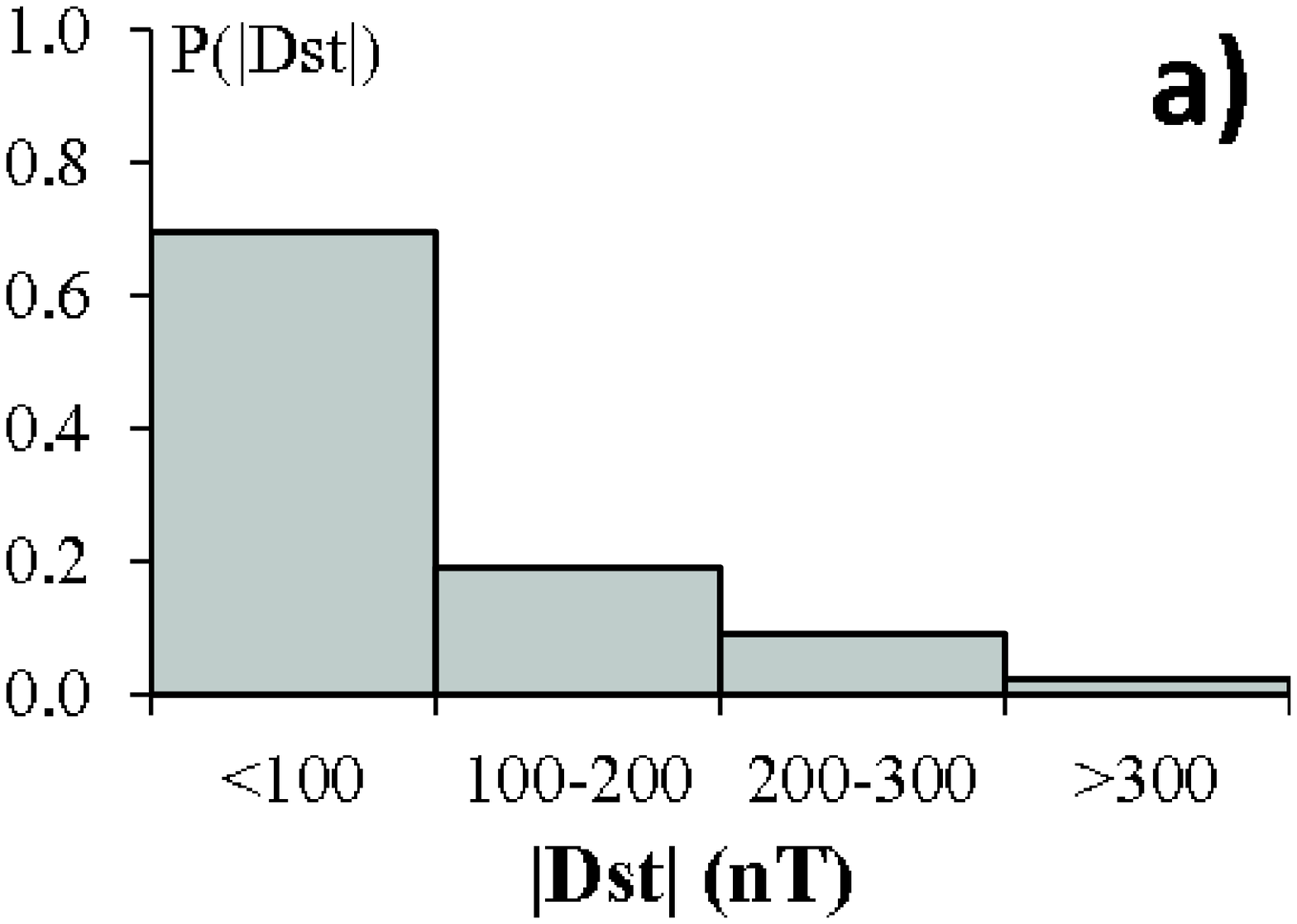}
               \includegraphics[width=0.33\textwidth]{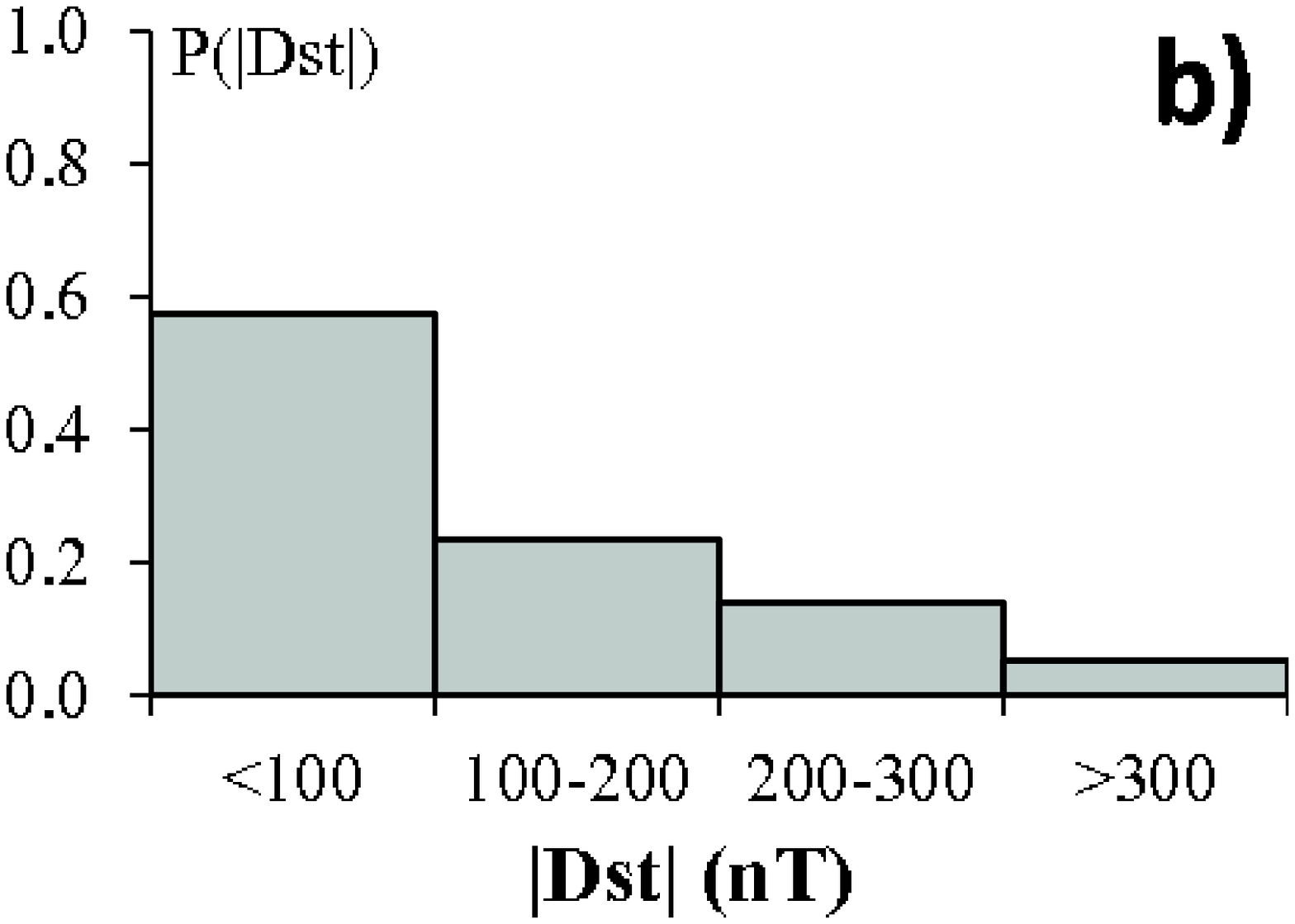}
               \includegraphics[width=0.33\textwidth]{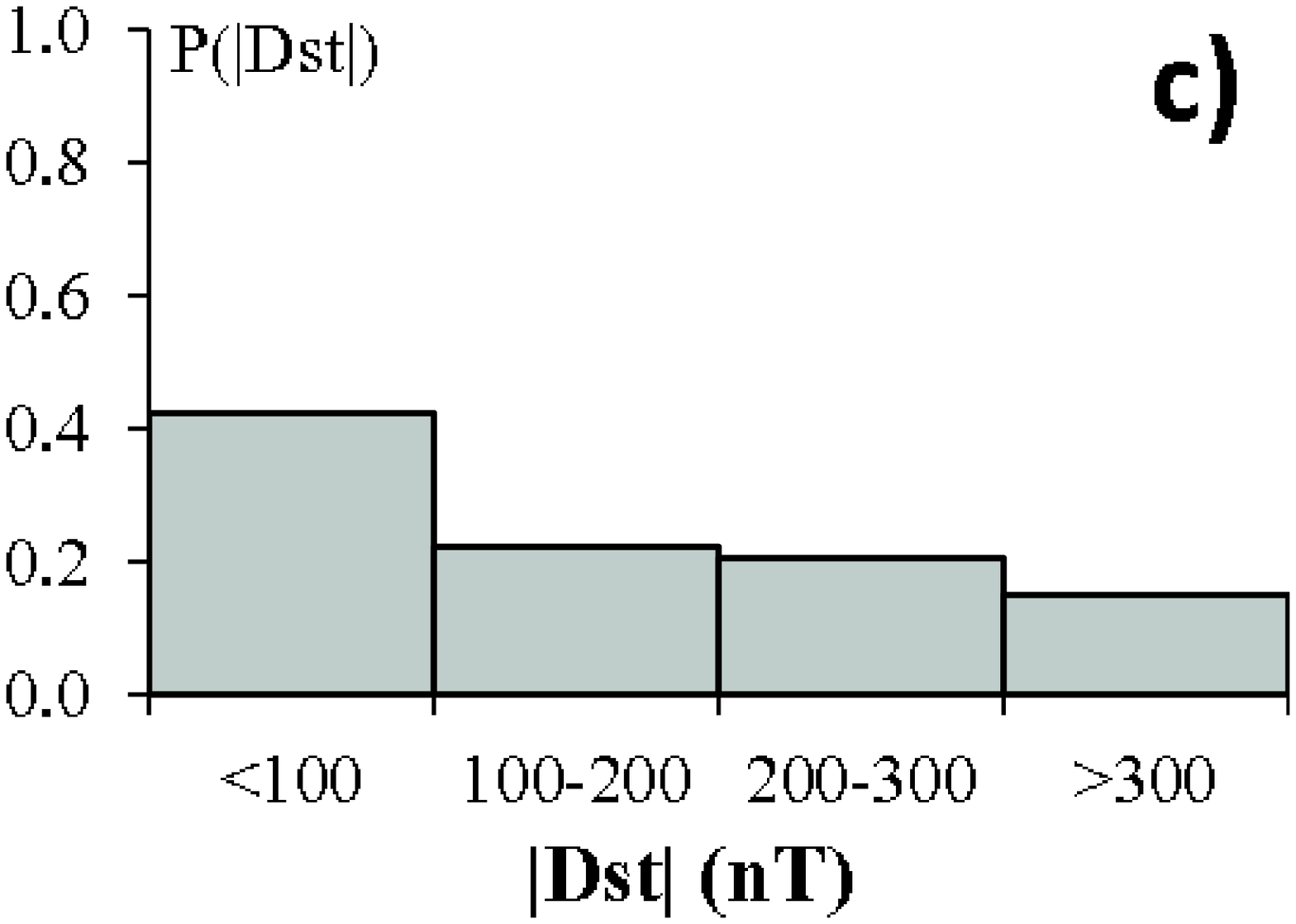}
               }
	 \caption{Probability distribution for observing $|Dst|$ in a specific $|Dst|$ bin for different sets of key solar parameters:
	 					a) $v=400$ km/s; $r=1$; $w=1$; $f=1$; $i=1$;
	 					b) $v=800$ km/s; $r=0.5$; $w=2$; $f=2$; $i=2$;
	 					c) $v=2000$ km/s; $r=0.01$; $w=3$; $f=3$; $i=4$.
        }
   \label{fig12}
\end{figure}


Based on the Equation (\ref{eq8}), probabilities of $|Dst|$ levels can be calculated for a specific set of key parameters $v$, $r$, $w$, $f$, and $i$. In Figure \ref{fig12} we present three different probability distributions obtained using Equation (\ref{eq8}) for three different key solar parameter sets. The results are also presented in Table \ref{table7}. It can be seen that the probabilities of large geomagnetic storms are higher for faster and wider CMEs which originate near the disc centre, are connected to more energetic flares and are likely to be involved in a CME--CME interaction.

This model constructs the geoeffectiveness probability distribution for a given CME. However, in its current form it cannot be used for forecasting. Since it is based on the distribution of CMEs in the SOHO era (a representative sample of 211 events) this probability distribution represents an ensemble of possible $|Dst|$ values for a given CME. Although the probability distribution changes with CME/flare parameters it is always highly asymmetric with greatest probability that CME will not be geoeffective. This depicts the general behavior of CMEs - a large majority of CMEs will never reach the Earth and/or will not have a favorable magnetic field orientation. Therefore, although the model produces a probability distribution it does not give a straightforward prediction of whether or not (and how strong) a geomagnetic storm will occur.

This is a different approach than used in previous models (\emph{e.g.} \opencite{srivastava05}; \opencite{valach09}; \opencite{uwamahoro12}), where prediction of geoeffectiveness is a direct output of the model. In \inlinecite{srivastava05} and \inlinecite{uwamahoro12} the threshold for the probability function is set to 0.5, \emph{i.e.} the prediction of the storm is based on the highest calculated probability.  Here, one cannot simply predict the $|Dst|$ level by stating that it has the largest probability, because the largest probability for all the CMEs is that they will not produce $|Dst|>100$ nT.  In order to derive the forecast based on this model, one will have to impose thresholds on the probability distribution, similar to that done by \inlinecite{valach09}. This, however, requires further study and will be reported elsewhere.

The differences between this model and other models mentioned above arise from the basic sample choice: our sample is based on CMEs, whereas in other studies samples were based on ICMEs or geomagnetic storms (\emph{i.e.} they presume the arrival of the CME at the Earth). In that sense, as explained in Section \ref{method} our model also indirectly takes into account false alarms, because they are incorporated into the distribution. That does not mean that it can avoid false alarms completely. Due to assumption of independence leading to Equation (\ref{eq7}), the effect of the false alarms may be increased, but this cannot be assessed at this point.  In the present form the model is not suitable for space weather forecast and additional calculations are needed to derive the expected $|Dst|$ level for a specific probability distribution. Once this is achieved and evaluation performed, results can be properly compared to other models and false alarms studied in more detail.


\begin{table}
\caption{Probabilities for observing a specific $|Dst|$ level for different sets of key solar parameters:
	 					I: $v=400$ km/s; $r=1 \mathrm{R_{SUN}}$; $w=1$; $f=1$; $i=1$;
	 					II: $v=800$ km/s; $r=0.5 \mathrm{R_{SUN}}$; $w=2$; $f=2$; $i=2$;
	 					III: $v=2000$ km/s; $r=0.01 \mathrm{R_{SUN}}$; $w=3$; $f=3$; $i=4$.
				}
\label{table7}

\begin{tabular*}{0.9\textwidth}{ccccccccccccc}
\hline
$|Dst|$	(nT)		& & & & & \multicolumn{7}{c}{$P(|Dst|)$(\%)}\\
 								& & & &	I			& & & &	II			& & & &	III\\
\hline
$<$100				  & & & &	70		& & & &	57			& & & &	42\\
100-200			    & & & &	19		& & & &	23			& & & &	22\\	
200-300			    & & & &	9			& & & &	14			& & & &	21\\
$>$100				  & & & &	30		& & & &	43			& & & &	58\\
$>$200				  & & & &	11		& & & &	19			& & & &	35\\
$>$300			    & & & &	2			& & & &	5			  & & & &	15\\
\end{tabular*}

\end{table}


\section{Summary and Conclusions}
				\label{summary}

From the presented statistical analysis we derived the key CME/flare parameters and quantified their influence on the probability of occurence of moderate and intense storms. Our results reflect some of relatively well-known relationships between remotely-observed solar properties and geomagnetic storms, namely the importance of CME initial speed, apparent width, source position and associated solar flare class. It also offers a quantification of these relationships and points out the significance in combining different solar parameters. It is shown that the CME--CME interaction is associated with a higher probability in causing intense storms. Moreover, it was shown that very slow, non-halo CMEs associated with B or C class flare are not expected to produce intense storms, unless involved in the CME--CME interaction with faster and wider CMEs, associated with stronger flares. The validity of the sample and of the results is confirmed by comparing the monthly CME/flare activity of the population (1392 CME--flare pairs in the SOHO era) with the monthly CME/flare and $Dst$ activity in our sample (211 $Dst$-associated CME--flare pairs).

The results of this statistical analysis can be used for prediction of the probability that a given event, observed by coronagraph, X-ray and EUV imagers at L1-point satellites (or even ground based instruments), will produce a major, intense or very intense geomagnetic storm at the Earth. An empirical statistical model for predicting geomagnetic storm levels was established that can be used as an early geomagnetic storm warning. It calculates the $|Dst|$ level probability distribution for a set of key solar parameters, based on the discrete probability distribution constructed by means of a geometric distribution. The distribution is shifted towards larger $|Dst|$ levels for faster and wider CMEs which originate near the centre of the disc, especially if they are connected to more energetic flares and are likely to be involved in a CME--CME interaction. However, the distribution is always highly asymmetric with the highest probability that a CME will not be geoeffective, reflecting the general behavior of CMEs (majority of them never reach the Earth and/or do not have favorable magnetic field orientation). Therefore, in order to forecast based on this model, further analysis is needed. The prediction at this stage is quite "crude" and does not provide a straightforward information whether or not a geomagnetic storm will occur and what would be its intensity. However, its advantage is that it offers an advance warning.

It should be noted that non-geoeffective CMEs were basically treated equally as the CMEs which never reached the Earth. This is due to the sampling, where the general idea was to observe solar sources and effects at the Earth, without interplanetary component. We found this view to be appropriate regarding the statistical aspect, where physical variables are treated as random variables. Although the false alarms are included in the sample which has been used for developing the model, it is not possible to distinguish whether or not they occur. This might represent a drawback of the model which will be analyzed in a future work.

%
\appendix

Hereafter follows a supplement to Section \ref{model}, providing detailed step-wise mathematical formulations and procedures used for estimating the probability distribution of geomagnetic storm level based on the remote solar observation of a CME and the associated solar flare. For this purpose an example-CME will be used with the following characteristics:
\begin{itemize}
\item First LASCO C2 appearance: 10 April 2001 05:30 UT.
\item Associated solar flare GOES peak time: 10 April 2001 05:26 UT.
\item First order LASCO catalog CME speed: 2411 km s$^{-1}$.
\item CME angular width: halo.
\item CME/flare source region location (distance from the centre of the solar disc, $r$): S23W09 ($r$=0.4067).
\item Flare X-ray class: X2.3.
\item CME--CME interaction level: train, T (very likely interacts with a halo CME that first appeared in the C2 field of view 9 April 2001 15:54 UT).
\end{itemize}

Based on the CME/flare characteristics the key parameters are defined in the following way:
\begin{itemize}
\item $v$ is a continuous parameter that is equal to the first order CME speed, measured in the LASCO field of view, expressed in km s$^{-1}$ and defined in a range $v > 400$. Therefore, $v = 2411$.
\item $r$ is a continuous parameter that is equal to the distance of a CME/flare position from the centre of the solar disc expressed in solar radii and defined in a range $0 < r \le 1$. Therefore, $r = 0.4067$.
\item $w$ is a discrete parameter with possible values 1 (non-halo CMEs), 2 (partial halo CMEs), and 3 (halo CMEs). Therefore, $w = 3$.
\item $f$ is a discrete parameter with possible values 1 (B or C flare), 2 (M flare), and 3 (X flare). Therefore, $f = 3$.
\item $i$ is a discrete parameter with possible values 1 (S, no interaction), 2 (S?, interaction not likely), 3 (T?, probable interaction), and 4 (T, interaction highly probable). Therefore, $i = 4$.
\end{itemize}


\section{The geometric probability distribution, $P(X=k)$}
\label{GD}

The geometric probability distribution is a probability distribution of a random variable $X$, where $X$ is a number of Bernoulli trials needed to get a success. There is an equal probability of success of each trial, $p$, and $X$ is defined on an endless set of discrete values $k=1,2,3,...$ (see e.g. \opencite{pitman}, and \opencite{stirzaker}). It is a discrete analogue of the exponential distribution. The probability density function for the geometric distribution is given by Equation (\ref{eq4}) in Section \ref{model} and an example is given in Figure \ref{fig13} for different probabilities of success in each trial, $p$.

\ifpdf
	bla
\else
  \begin{figure}
   \centerline{\includegraphics[width=0.5\textwidth]{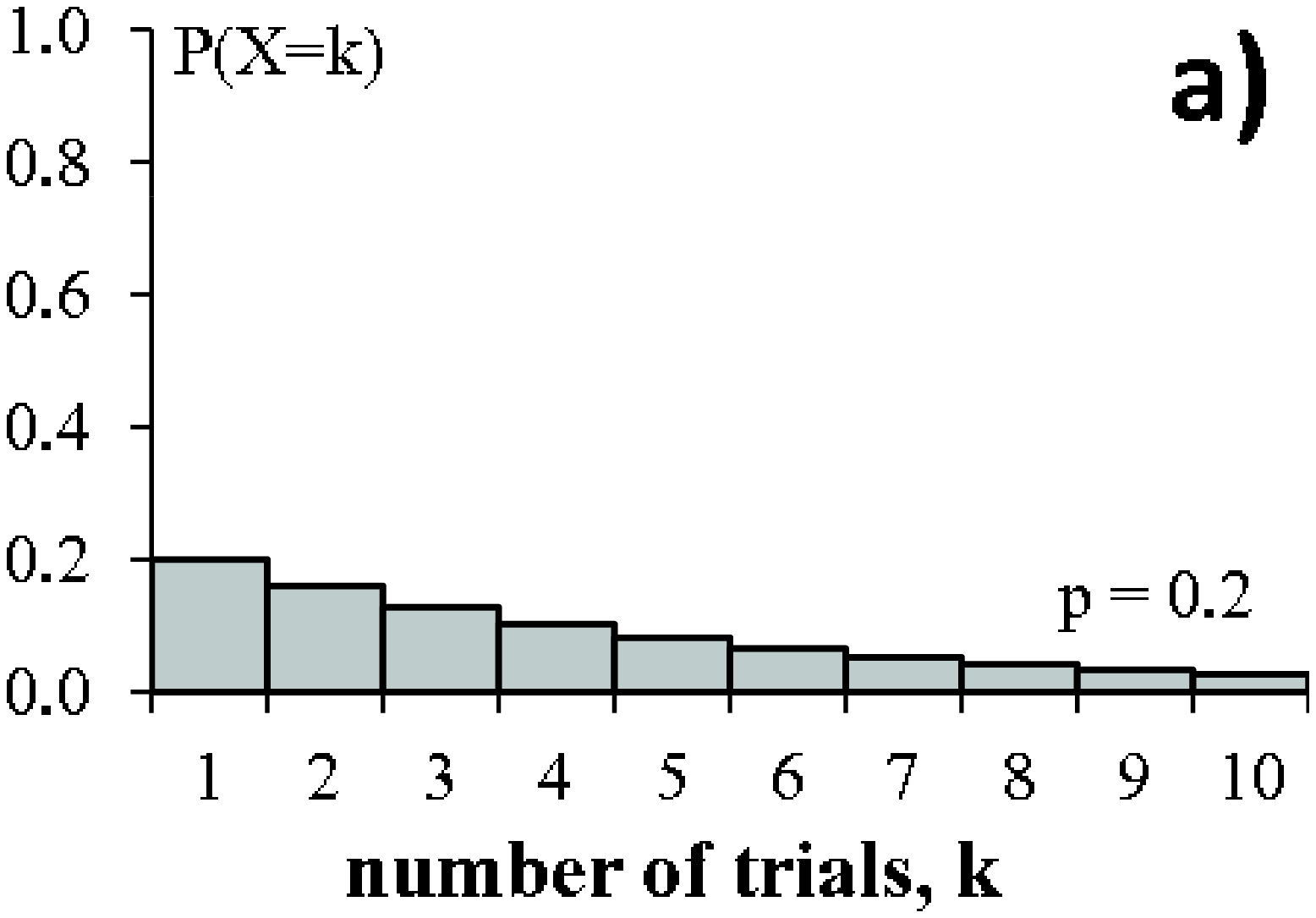}
               \includegraphics[width=0.5\textwidth]{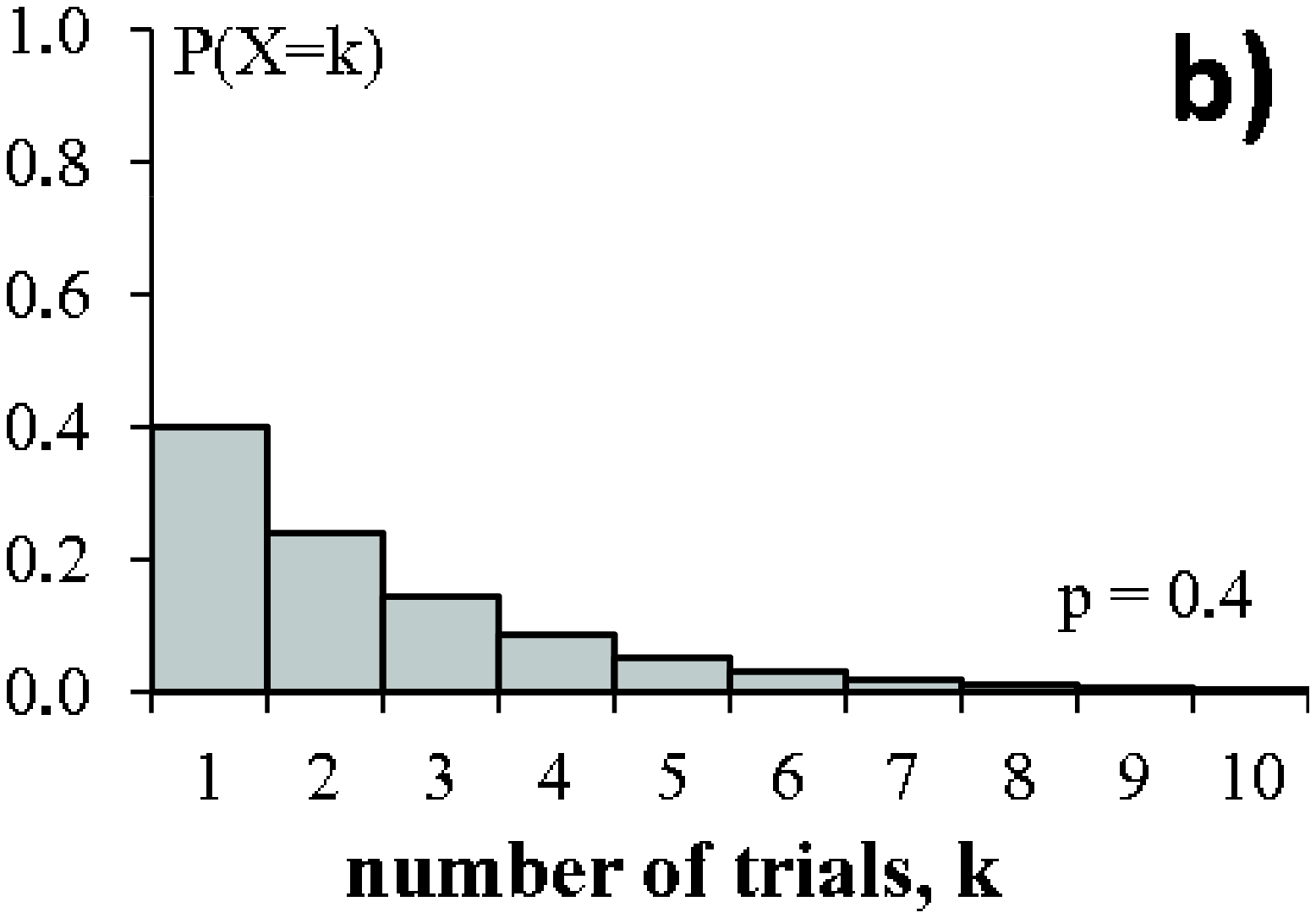}
               }
   \vspace{0.03\textwidth}
   \centerline{\includegraphics[width=0.5\textwidth]{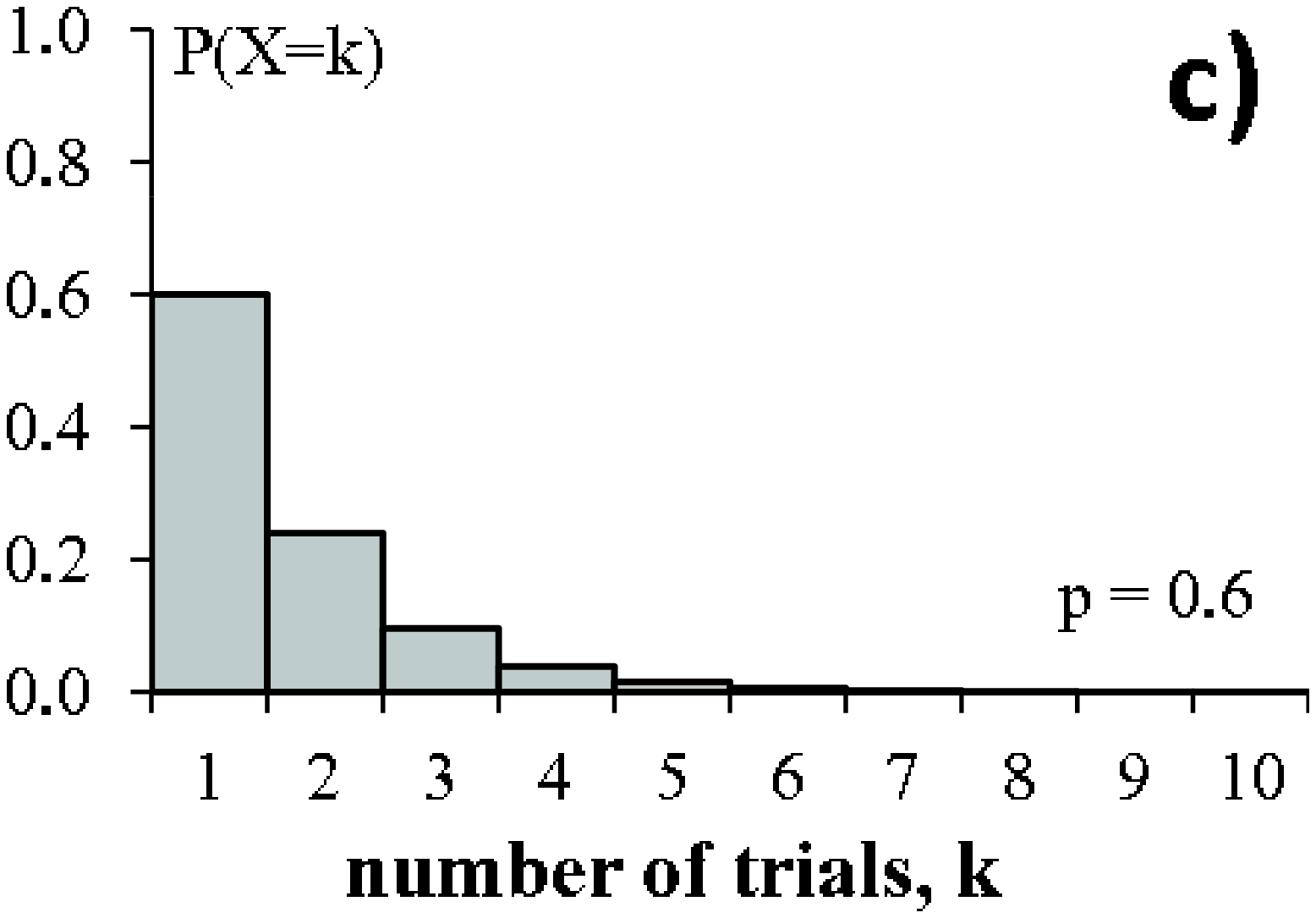}
               \includegraphics[width=0.5\textwidth]{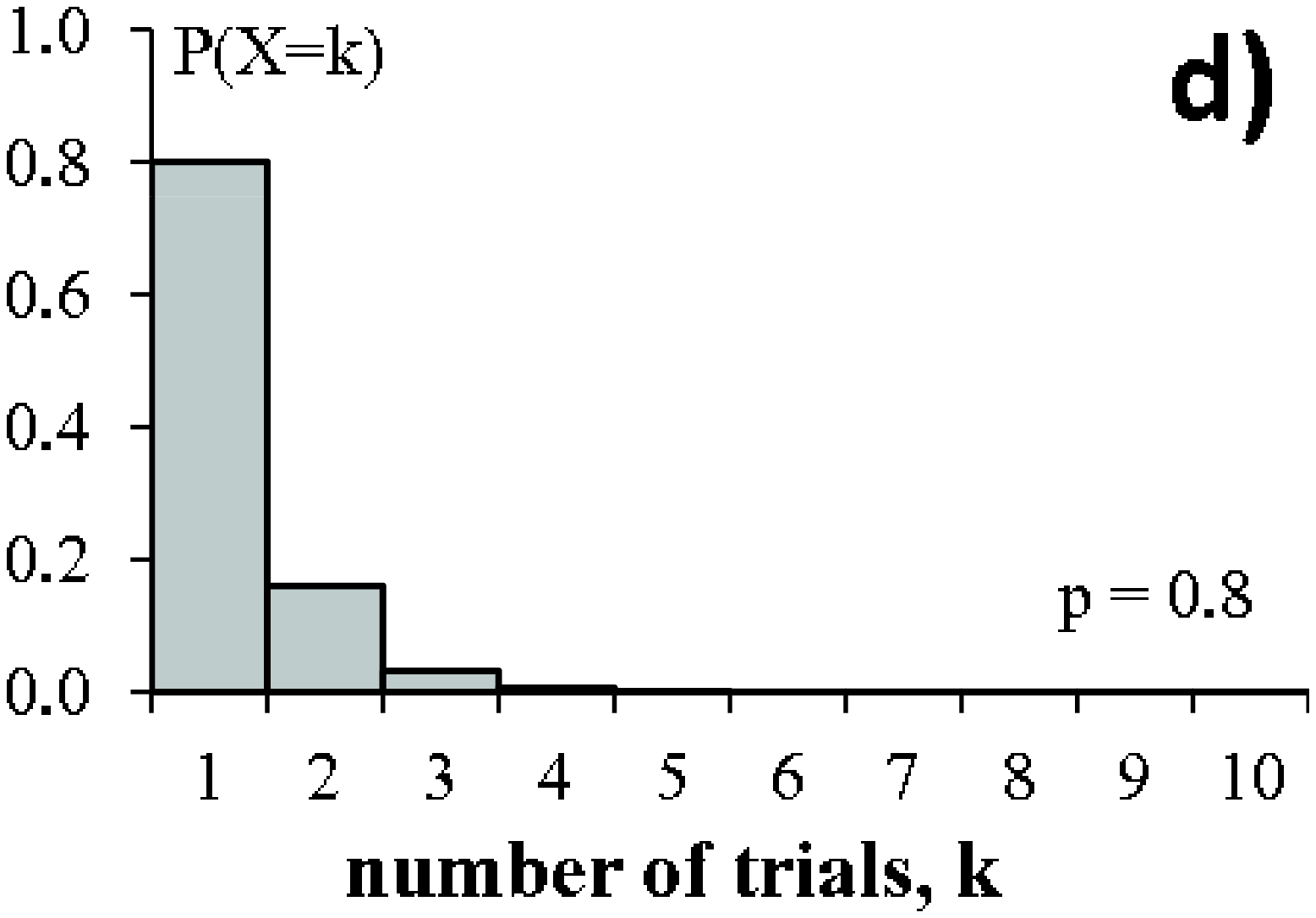}
               }	 
   \caption{The geometric probability distribution, $P(X=k)$ for different probabilities of success on each trial, $p$.
        }
   \label{fig13}
\end{figure}
\fi

It is easily found that the expected value of the geometrically distributed random variable $X$, i.e. the mean of the geometric distribution, is given by the following expression (for details see \opencite{stirzaker}):

\begin{equation}
m_{GD} = E(X) = \sum_{X \epsilon k}{X \cdot P(X)} = \sum_{k = 1}^{\infty}{k \cdot p \cdot (1-p)^{k-1}} = \frac{1}{p}.
\label{eq9}
\end{equation}

Therefore, the probability of the success in each trial, $p$ can be calculated if the mean of the geometric distribution, $m_{GD}$ is known:

\begin{equation}
p = \frac{1}{m_{GD}}.
\label{eq10}
\end{equation}

We use the formalism for geometric distribution to construct $|Dst|$ distributions observed throughout Section \ref{results}. For that purpose, different $|Dst|$ levels have to be associated with different numbers of trials ($k \longleftrightarrow |Dst|$) and the mean of $|Dst|$ distribution has to be associated with the mean of the geometric distribution ($m_{GD} \longleftrightarrow m_{DST}$).
We associate different $|Dst|$ levels with different number of trials, $k$ in the geometric distribution in the following way:
\begin{itemize}
\item $k=1 \longleftrightarrow |Dst|<100\,\mathrm{nT}$;
\item $k=2 \longleftrightarrow 100\,\mathrm{nT}<|Dst|<200\,\mathrm{nT}$;
\item $k=3 \longleftrightarrow 200\,\mathrm{nT}<|Dst|<300\,\mathrm{nT}$;
\item $k=4 \longleftrightarrow |Dst|>300\,\mathrm{nT}$.
\end{itemize}
Note that the value of $k$ is exactly 100 times smaller than the upper boundary for associated $|Dst|$ level, expressed in nT. It would be reasonable to assume that the mean of the geometric distributions would relate in a similar fashion to the mean of the $|Dst|$ distribution (\emph{i.e.} would be 100 times smaller). The mean of the $|Dst|$ distribution is in the first bin for all of the distributions troughout Section \ref{results}, \emph{i.e.} $m_{DST} < 100 \mathrm{nT}$. However, due to the fact that the geometric distribution is defined on a set $k=1,2,3,...$, the mean is always larger than 1. This is also seen from Equation \ref{eq9} ($p < 1$). Dividing $m_{DST}$ by 100 would not give a mathematically correct $m_{GD}$, but adding 1 to this relation solves this problem. Therefore:

\begin{equation}
m_{GD} = 1 + \frac{m_{DST}[nT]}{100}
\label{eq11}
\end{equation}

For simplicity, in further reading we will refer to $P(X=k)$ as $P(k)$. Note that for constructed distributions we will use the set of $k$ values $k =1,2,3,4$, based on the defined associations $k \longleftrightarrow |Dst|$.


\section{Probability distribution for CME speed ($v$), $P_{v}(k)$}
\label{v}

The change of the $|Dst|$ distribution mean with the CME speed, $v$, can be described with a linear function (see Figure \ref{fig5}):

\begin{equation}
m_{DST}(v) = a\cdot v + b.
\label{eq12}
\end{equation}

Here $a=0.04$,  $b=10.45$, and $|Dst|$ is expressed in nT.
For a given $v=2411$ Equation (\ref{eq12})  gives distribution mean $m_{DST}(v)=106.89$.
The geometric distribution mean is then calculated using Equation (\ref{eq11}), which in our example gives $m_{GD} = 2.07$.
The probability of the success in each trial, $p$  can be calculated using Equation (\ref{eq10}) and in our example equals $p = 0.48$.

For each $k=1,2,3,4$, a probability that the $k$-th trial is the first success, $P(k)$ can be calculated using Equation (\ref{eq4}) in Section \ref{model}.
In the example the results are as follows:
\begin{itemize}
\item $P(k=1) = 0.4833$;
\item $P(k=2) = 0.2497$;
\item $P(k=3) = 0.1290$;
\item $P(k=4) = 0.0667$.
\end{itemize}

Due to the fact that the geometric distribution does not stop at $k = 4$, this distribution is not normalized, \emph{i.e.} $\sum_{k=1}^{4}{P(k)} \ne 1$. Therefore, to define this distribution for $k=1,2,3,4$, it is necessary to renormalize the distribution:

\begin{equation}
P(k) = \frac{P(k)}{\sum_{k=1}^{4}{P(k)}}.
\label{eq13}
\end{equation}

In the example the results are as follows:
\begin{itemize}
\item $P(k=1) = 0.5204$;
\item $P(k=2) = 0.2689$;
\item $P(k=3) = 0.1389$;
\item $P(k=4) = 0.0718$.
\end{itemize}
Note that the ratio between different $P(k)$ does not change.

Finally, we construct an adjusted probability distribution adding constants shown in second column of Table \ref{table6}, as explained in Section \ref{model}. More specifically:
\begin{itemize}
\item $P_{v}(k=1) = P(k=1) + 0.13 = 0.6504$;
\item $P_{v}(k=2) = P(k=2) - 0.10 = 0.1689$;
\item $P_{v}(k=3) = P(k=3) - 0.03 = 0.1089$;
\item $P_{v}(k=4) = P(k=4) = 0.0718$.
\end{itemize}

Note that the adjusted probability distribution is normalized, since:

\begin{equation}
\sum_{k=1}^{4}{P_{v}(k)} = \sum_{k=1}^{4}{P(k)} = 1.
\label{eq14}
\end{equation}

$P_{v}(k)$ represents an empirically obtained probability distribution of $|Dst|$ level for a specific CME speed ($v = 2411$ km s$^{-1}$).


\section{Probability distribution for CME/flare position distance from the centre of the solar disc ($r$), $P_{r}(k)$}
\label{r}

The change of the $|Dst|$ distribution mean with CME/flare position distance from the centre of the solar disc, $r$, can be described with a power law function (see Figure \ref{fig9}):

\begin{equation}
m_{DST}(r) = a\cdot r^b.
\label{eq15}
\end{equation}

Here $a = 30.95$, $b = -0.83$, and $|Dst|$ is expressed in nT.
For a given $r = 0.4067$ Equation (\ref{eq15}) gives distribution mean $m_{DST}(r) = 65.31$.

The geometric distribution mean is calculated using Equation (\ref{eq11}). In the example $m_{GD} = 1.65$. The probability of the success in each trial, calculated using Equation (\ref{eq10}) is $p = 0.60$.

Probabilities calculated using Equation (\ref{eq4}) in Section \ref{model}, $P(k)$ are then:
\begin{itemize}
\item $P(k=1) = 0.6049$;
\item $P(k=2) = 0.2390$;
\item $P(k=3) = 0.0944$;
\item $P(k=4) = 0.0373$.
\end{itemize}

The renormalized distribution, calculated using Equation (\ref{eq13}) is:
\begin{itemize}
\item $P(k=1) = 0.6200$;
\item $P(k=2) = 0.2450$;
\item $P(k=3) = 0.0968$;
\item $P(k=4) = 0.0382$.
\end{itemize}

Finally, the adjusted probability distribution (adding constants shown in third column of Table \ref{table6}, as explained in Section \ref{model}) is:
\begin{itemize}
\item $P_{r}(k=1) = P(k=1) + 0.12 = 0.7400$;
\item $P_{r}(k=2) = P(k=2) - 0.12 = 0.1250$;
\item $P_{r}(k=3) = P(k=3) = 0.0968$;
\item $P_{r}(k=4) = P(k=4) = 0.0382$.
\end{itemize}

$P_{r}(k)$ represents an empirically obtained probability distribution of $|Dst|$ level for a specific CME/flare position distance from the centre of the solar disc ($r = 0.4067$ solar radius).


\section{Probability distribution for CME width ($w$), $P_{w}(k)$}
\label{w}

The change of the $|Dst|$ distribution mean with CME width, $w$, can be described with a quadratic function (see Figure \ref{fig9}):

\begin{equation}
m_{DST}(w) = a\cdot w^2 + b\cdot w + c.
\label{eq16}
\end{equation}

Here $a = 15.06$, $b = -34.60$, $c = 42.25$, and $|Dst|$ is expressed in nT.
For a given $w = 3$ this gives distribution mean $m_{DST}(w) = 73.99$.

The geometric distribution mean is calculated using Equation (\ref{eq11}). In the example $m_{GD} = 1.74$. The probability of the success in each trial, calculated using Equation (\ref{eq10}) is $p = 0.57$.

Probabilities calculated using Equation (\ref{eq4}) in Section \ref{model}, $P(k)$ are then:
\begin{itemize}
\item $P(k=1) = 0.5747$;
\item $P(k=2) = 0.2444$;
\item $P(k=3) = 0.1039$;
\item $P(k=4) = 0.0442$.
\end{itemize}

The renormalized distribution, calculated using Equation (\ref{eq13}) is:
\begin{itemize}
\item $P(k=1) = 0.5942$;
\item $P(k=2) = 0.2527$;
\item $P(k=3) = 0.1075$;
\item $P(k=4) = 0.0457$.
\end{itemize}

Finally, the adjusted probability distribution (adding constants shown in fourth column of Table \ref{table6}, as explained in Section \ref{model}) is:
\begin{itemize}
\item $P_{w}(k=1) = P(k=1) + 0.14 = 0.7342$;
\item $P_{w}(k=2) = P(k=2) - 0.12 = 0.1327$;
\item $P_{w}(k=3) = P(k=3) - 0.02 = 0.0875$;
\item $P_{w}(k=4) = P(k=4) = 0.0457$.
\end{itemize}

$P_{w}(k)$ represents an empirically obtained probability distribution of $|Dst|$ level for a specific CME width ($w = 360^{\circ}$, halo CME).


\section{Probability distribution for flare class ($f$), $P_{f}(k)$}
\label{f}

The change of the $|Dst|$ distribution mean with flare class, $f$, can be described with a quadratic function (see Figure \ref{fig9}):

\begin{equation}
m_{DST}(f) = a \cdot f^2 + b \cdot f + c.
\label{eq17}
\end{equation}

Here $a = 10.41$, $b = -17.90$, $c = 46.93$, and $|Dst|$ is expressed in nT.
For a given $f = 3$ this gives distribution mean $m_{DST}(f) = 86.92$.

The geometric distribution mean is calculated using Equation (\ref{eq11}). In the example $m_{GD} = 1.87$. The probability of the success in each trial, calculated using Equation (\ref{eq10}) is $p = 0.53$.

Probabilities calculated using Equation (\ref{eq4}) in Section \ref{model}, $P(k)$ are then:
\begin{itemize}
\item $P(k=1) = 0.5350$;
\item $P(k=2) = 0.2488$;
\item $P(k=3) = 0.1157$;
\item $P(k=4) = 0.0538$.
\end{itemize}

The renormalized distribution, calculated using Equation \ref{eq13} is:
\begin{itemize}
\item $P(k=1) = 0.5612$;
\item $P(k=2) = 0.2610$;
\item $P(k=3) = 0.1214$;
\item $P(k=4) = 0.0564$.
\end{itemize}

Finally, the adjusted probability distribution (adding constants shown in fifth column of Table \ref{table6}, as explained in Section \ref{model}) is:
\begin{itemize}
\item $P_{f}(k=1) = P(k=1) + 0.15 = 0.7112$;
\item $P_{f}(k=2) = P(k=2) - 0.12 = 0.1410$;
\item $P_{f}(k=3) = P(k=3) - 0.02 = 0.1014$;
\item $P_{f}(k=4) = P(k=4) - 0.01 = 0.0464$.
\end{itemize}

$P_{f}(k)$ represents an empirically obtained probability distribution of $|Dst|$ level for a specific flare class ($f = 3$, X class flare).


\section{Probability distribution for interaction level ($i$), $P_{i}(k)$}
\label{i}

The change of the $|Dst|$ distribution mean with interaction level, $i$, can be described with a power-law function (see Figure \ref{fig9}):

\begin{equation}
m_{DST}(i) = a \cdot i^b.
\label{eq18}
\end{equation}

Here $a = 38.39$, $b = 0.49$, and $|Dst|$ is expressed in nT.
For a given $i = 4$ this gives distribution mean $m_{DST}(i) = 65.77$.

The geometric distribution mean is calculated using Equation (\ref{eq11}). In the example $m_{GD} = 1.66$. The probability of the success in each trial, calculated using Equation (\ref{eq10}) is $p = 0.60$.

Probabilities calculated using Equation (\ref{eq4}) in Section \ref{model}, $P(k)$ are then:
\begin{itemize}
\item $P(k=1) = 0.5691$;
\item $P(k=2) = 0.2452$;
\item $P(k=3) = 0.1057$;
\item $P(k=4) = 0.0455$.
\end{itemize}

The renormalized distribution, calculated using Equation (\ref{eq13}) is:
\begin{itemize}
\item $P(k=1) = 0.5894$;
\item $P(k=2) = 0.2540$;
\item $P(k=3) = 0.1094$;
\item $P(k=4) = 0.0472$.
\end{itemize}

Finally, the adjusted probability distribution (adding constants shown in fifth column of Table \ref{table6}, as explained in Section \ref{model}) is:
\begin{itemize}
\item $P_{i}(k=1) = P(k=1) + 0.15 = 0.7394$;
\item $P_{i}(k=2) = P(k=2) - 0.13 = 0.1240$;
\item $P_{i}(k=3) = P(k=3) - 0.01 = 0.0994$;
\item $P_{i}(k=4) = P(k=4) - 0.01 = 0.0372$.
\end{itemize}

$P_{i}(k)$ represents an empirically obtained probability distribution of $|Dst|$ level for a specific interaction level ($i = 4$, interaction highly probable).


\section{Combined probability distribution for set of key parameters ($v,r,w,f,i$), $P(|Dst|)$}
\label{p}

Once we obtain the probability distribution of $|Dst|$ level for each of the key solar parameters ($v$, $r$, $w$, $f$, and $i$), their combined probability $P(k)=P(|Dst|)$ is calculated using Equation (\ref{eq8}) in Section \ref{model}. For our example this gives:
\begin{itemize}
\item $P(k=1) = P(|Dst|<100 nT) = 0.9982$;
\item $P(k=2) = P(100 nT<|Dst|<200 nT) = 0.5253$;
\item $P(k=3) = P(200 nT<|Dst|<300 nT) = 0.4056$;
\item $P(k=4) = P(|Dst|>300 nT) = 0.2178$.
\end{itemize}

Due to the fact that the set of parameters {$v$, $r$, $w$, $f$, and $i$} is not a complete set of independent variables for this distribution, this distribution is not normalized, i.e. $\sum P(|Dst|) \ne 1$. Therefore, it is necessary to renormalize the distribution (similarly to Equation (\ref{eq13})):
\begin{itemize}
\item $P(k=1) = P(|Dst|<100 nT) = 0.4649$;
\item $P(k=2) = P(100 nT<|Dst|<200 nT) = 0.2447$;
\item $P(k=3) = P(200 nT<|Dst|<300 nT) = 0.1889$;
\item $P(k=4) = P(|Dst|>300 nT) = 0.1014$.
\end{itemize}

Note that the ratio between different $P(|Dst|)$ does not change.
$P(|Dst|)$ represents an empirically obtained probability distribution of $|Dst|$ level for a specific set of key parameters ($v = 2411$, $r = 0.4067$, $w = 3$, $f = 3$, $i = 4$).


\begin{acks}

 The presented work has received funding from the European Union Seventh Framework Programme (FP7/2007-2013) under grant agreement n$^{\mathrm{o}}$ 263252 [COMESEP]. This work has been supported in part by Croatian Science Foundation under the project 6212 "Solar and Stellar Variability". This research has been funded by the Interuniversity Attraction Poles Programme initiated by the Belgian Science Policy Office (IAP P7/08 CHARM). L. Rodriguez acknowledges support from the Belgian Federal Science Policy Office through the ESA - PRODEX program. 
 We are grateful to the SOHO LASCO CME catalog team for providing the CME data. This CME catalog is generated and maintained at the CDAW Data Center by NASA and The Catholic University of America in cooperation with the Naval Research Laboratory. SOHO is a project of international cooperation between ESA and NASA.
 We are also grateful to the Solar-Terrestrial Physics (STP) Division of NOAA's (National Oceanic and Atmospheric Administration) National Geophysical Data Center (NGDC) for providing solar flare data.

\end{acks}


\bibliographystyle{spr-mp-sola}
\bibliography{REFs}


\end{article}
\end{document}